\pdfoutput=1
\documentclass[11pt]{article}
\usepackage[T1]{fontenc}
\usepackage{geometry} 
\usepackage[english]{babel}
\usepackage{cite}
\usepackage{amsmath}
\usepackage{amsthm}
\usepackage{amsfonts} 
\usepackage{amssymb} 
\usepackage{xspace}
\usepackage{mathtools} 
\usepackage{braket} 
\usepackage[linesnumbered,vlined]{algorithm2e}
\usepackage{xcolor} 
\usepackage{footnote} 
\usepackage{environ} 
\usepackage{suffix} 
\usepackage{apptools}
\usepackage{hyperref}
\usepackage{float}
\usepackage[capitalise]{cleveref} 

\makesavenoteenv{tabular}
\makesavenoteenv{table}

\makeatletter
\newtheorem*{rep@theorem}{\rep@title}
\newcommand{\newreptheorem}[2]{%
	\newenvironment{rep#1}[1]{%
		\def\rep@title{#2 \ref{##1}}%
		\begin{rep@theorem}}%
		{\end{rep@theorem}}}
\makeatother

\usepackage{thmtools} 
\usepackage{thm-restate} 

\def\eps{\varepsilon}

\newcommand{\size}[1]{\ensuremath{\left|#1\right|}}

\DeclareMathOperator{\poly}{poly}

\crefname{claim}{Claim}{Claims}
\crefname{property}{Property}{Properties}
\crefname{algocf}{Algorithm}{Algorithms}
\Crefname{algocf}{Algorithm}{Algorithms}

\crefname{propertyCarrierConfiguration}{Property}{Properties}
\creflabelformat{propertyCarrierConfiguration}{#2#1#3 of~\fullref{definition:carrierConfiguration}}
\crefmultiformat{propertyCarrierConfiguration}{Properties~#2#1#3}{ and~#2#1#3 of~\fullref{definition:carrierConfiguration}}{,~#2#1#3}{ and~#2#1#3 of~\fullref{definition:carrierConfiguration}}
\crefrangeformat{propertyCarrierConfiguration}{Properties~#3#1#4 to~#5#2#6 of~\fullref{definition:carrierConfiguration}}

\crefname{propertyWACCCarrierConfiguration}{Property}{Properties}
\creflabelformat{propertyWACCCarrierConfiguration}{#2#1#3 of~\fullref{wacc:def:carrierConfiguration}}
\crefmultiformat{propertyWACCCarrierConfiguration}{Properties~#2#1#3}{ and~#2#1#3~of~\fullref{wacc:def:carrierConfiguration}}{,~#2#1#3}{ and~#2#1#3~of~\fullref{wacc:def:carrierConfiguration}}
\crefrangeformat{propertyWACCCarrierConfiguration}{Properties~#3#1#4 to~#5#2#6 of~\fullref{wacc:def:carrierConfiguration}}

\newtheorem{theorem}{Theorem}[section]
\newreptheorem{theorem}{Theorem}
\newtheorem{lemma}[theorem]{Lemma}
\newtheorem{claim}[theorem]{Claim}
\newtheorem{definition}{Definition}

\newtheorem{corollary}[theorem]{Corollary}

\newenvironment{proofof}[1]{\begin{proof}[Proof of #1]}{\end{proof}}

\newcommand*{\fullrefsingle}[1]{\hyperref[{#1}]{\cref*{#1} (\emph{\nameref*{#1}})}} 

\makeatletter
\newcommand{\fullref}[1]{\IfAppendix{%
  \@tempswafalse
  \@for\next:=#1\do
    {\if@tempswa\ and \else\@tempswatrue\fi\fullrefsingle{\next}}%
}{\cref{#1}}}
\makeatother

\theoremstyle{remark}

\theoremstyle{definition}

\newcommand{\R}{\mathbb{R}}

\DeclarePairedDelimiter{\floor}{\lfloor}{\rfloor}
\DeclarePairedDelimiter{\ceil}{\lceil}{\rceil}
\newcommand{\ip}[1]{\left}

\newcommand{\hybrid}{\ensuremath{\mathsf{Hybrid}}\xspace}
\newcommand{\ncc}{\ensuremath{\mathsf{Node\text{-}Capacitated\ Clique }}\xspace}
\newcommand{\nccshort}{\ensuremath{\mathsf{NCC}}\xspace}
\newcommand{\ncczero}{\ensuremath{\nccshort_0}\xspace}
\newcommand{\congest}{\ensuremath{\mathsf{CONGEST}}\xspace}
\newcommand{\local}{\ensuremath{\mathsf{LOCAL}}\xspace}
\newcommand{\clique}{\ensuremath{\mathsf{Congested\ Clique}}\xspace}
\newcommand{\bcc}{\ensuremath{\mathsf{Broadcast\ Congested\ Clique}}\xspace}

\newcommand{\bccshort}{\ensuremath{\mathsf{BCC}}\xspace}
\newcommand{\wacc}[1][c]{\ensuremath{\mathsf{AC(#1)}}\xspace}
\newcommand{\waccfull}{\ensuremath{\mathsf{Anonymous\ Capacitated}}\xspace}

\newcommand{\knearest}{$k$\textsc{-nearest}\xspace}
\newcommand{\sdk}{$(S, d, k)$\textsc{-source detection}\xspace}

\newcommand{\ssp}{\textsc{-SSP}\xspace}

\newcommand\bigO[1]{\ensuremath{{O}(#1)}}
\WithSuffix\newcommand\bigO*[1]{\ensuremath{{O}\left(#1\right)}}
\newcommand\tildeBigO[1]{\ensuremath{{\tilde{{O}}}(#1)}}
\WithSuffix\newcommand\tildeBigO*[1]{\ensuremath{{\tilde{{O}}}\left(#1\right)}}
\newcommand\littleO[1]{\ensuremath{{o}(#1)}}
\WithSuffix\newcommand\littleO*[1]{\ensuremath{{o}\left(#1\right)}}
\newcommand\tildeLittleO[1]{\ensuremath{{\tilde{{o}}}(#1)}}
\WithSuffix\newcommand\tildeLittleO*[1]{\ensuremath{{\tilde{{o}}}\left(#1\right)}}
\newcommand\bigOmega[1]{\ensuremath{{\Omega}(#1)}}
\WithSuffix\newcommand\bigOmega*[1]{\ensuremath{{\Omega}\left(#1\right)}}
\newcommand\tildeBigOmega[1]{\ensuremath{{\tilde{{\Omega}}}(#1)}}
\WithSuffix\newcommand\tildeBigOmega*[1]{\ensuremath{{\tilde{{\Omega}}}\left(#1\right)}}
\newcommand\littleOmega[1]{\ensuremath{{\omega}(#1)}}
\WithSuffix\newcommand\littleOmega*[1]{\ensuremath{{\omega}\left(#1\right)}}
\newcommand\tildeLittleOmega[1]{\ensuremath{{\tilde{{\omega}}}(#1)}}
\WithSuffix\newcommand\tildeLittleOmega*[1]{\ensuremath{{\tilde{{\omega}}{\left(#1\right)}}}}
\newcommand\bigTheta[1]{\ensuremath{{\Theta}(#1)}}
\WithSuffix\newcommand\bigTheta*[1]{\ensuremath{{\Theta}\left(#1\right)}}
\newcommand\tildeTheta[1]{\ensuremath{{\tilde{{\Theta}}}(#1)}}
\WithSuffix\newcommand\tildeTheta*[1]{\ensuremath{{\tilde{{\Theta}}\left(#1\right)}}}

\makeatletter
\newcommand*{\whp}{%
    \@ifnextchar{.}%
        {w.h.p}%
        {w.h.p.\@\xspace}%
}
\makeatother
 
\newcommand{\interval}[1]{\ensuremath{\left[#1\right]}}
\newcommand{\polylog}{\poly\log}
\newcommand{\tmix}{\tau_{\text{mix}}}
\newcommand{\sqrtlgn}{\sqrt{\log{n}}}
\newcommand{\twosqlgnlglgn}{2^{\bigO{\sqrtlgn}}}
\newcommand{\tmixtwosqlgnlglgn}{\ensuremath{\tmix\cdot \twosqlgnlglgn}}

\RestyleAlgo{boxruled}
\LinesNumbered
\let\oldnl\nl
\newcommand{\nonl}{\renewcommand{\nl}{\let\nl\oldnl}}

\SetKw{Send}{send}

\SetKwFunction{AssignHelpers}{Assign-Helpers}
\SetKwFunction{TransmitSkeleton}{Transmit-Skeleton}
\SetKwFunction{LearnNeighborhood}{Learn-Neighborhood}
\SetKwFunction{LearnHelpers}{Learn-Helpers}
\SetKwFunction{ReassignSkeletons}{Reassign-Skeletons}

\NewEnviron{killcontents}{}

\title{On Sparsity Awareness in Distributed Computations}

\author{Keren Censor-Hillel
    \and Dean Leitersdorf
    \and Volodymyr Polosukhin
}

\date{Technion\footnote{\{ckeren, leitersdorf, po\}@cs.technion.ac.il}}

\begin{document}
\begin{titlepage}
\maketitle
\thispagestyle{empty}

\begin{abstract}
We extract a core principle that underlies seemingly different fundamental distributed settings, which is that sparsity awareness may induce faster algorithms for core problems in these settings. To leverage this, we establish a new framework by developing an intermediate auxiliary model which is weak enough to be successfully simulated in the classic \congest model given low mixing time, as well as in the recently introduced \hybrid model. We prove that despite imposing harsh restrictions, this artificial model allows balancing massive data transfers with a maximal utilization of bandwidth. We then exemplify the power we gain from our methods, by deriving fast shortest-paths algorithms which greatly improve upon the state-of-the-art. 

Specifically, we obtain the following end results for graphs of $n$ nodes:

\begin{itemize}
\item{A $(3+\eps)$ approximation for weighted, undirected APSP in $(n^{1/2} + n/\delta)\cdot \tau_{\text{mix}}\cdot 2^{O(\sqrt{\log{n}})}$ rounds in the \congest model, where $\delta$ is the minimum degree of the graph and $\tau_{\text{mix}}$ is its mixing time. For graphs with $\delta = \tau_{\text{mix}}\cdot 2^{\omega{(\sqrt{\log{n}})}}$, this takes $o(n)$ rounds, despite the $\Omega(n)$ known lower bound for general graphs [Nanongkai, STOC'14].}

\item{An ${(n^{7/6}/m^{1/2} + n^2/m)}\cdot \tau_{\text{mix}}\cdot 2^{O(\sqrt{\log{n}})}$-round exact SSSP algorithm in the \congest model, for graphs with $m$ edges and a mixing time of $\tau_{\text{mix}}$. This improves upon the previous algorithm of [Chechik and Mukhtar, PODC'20] for significant ranges of values of $m$ and $ \tau_{\text{mix}}$.}

\item{A \clique simulation in the \congest model which improves upon the previous state-of-the-art simulation of [Ghaffari, Kuhn, and SU, PODC'17] by a factor that is proportional to the average degree in the graph.}

\item{An $\tilde O(n^{5/17}/\eps^9)$-round algorithm for a $(1+\eps)$ approximation for SSSP in the \hybrid model. The only previous $o(n^{1/3})$ round algorithm for distance approximations in this model is for a much larger approximation factor of $(1/\eps)^{O(1/\eps)}$ in $\tilde O(n^\eps)$ rounds [Augustine, Hinnenthal, Kuhn, Scheideler, Schneider, SODA'20].}
\end{itemize}
\end{abstract}
\end{titlepage}

\tableofcontents
\thispagestyle{empty} 
\newpage

\setcounter{page}{1}
\section{Introduction}\label{sec:introduction}
The overarching theme of this paper is laying down an algorithmic infrastructure and employing it for designing fast algorithms in seemingly unrelated distributed settings, namely, the classic \congest model and the recently-introduced \hybrid model. 

The \congest model \cite{Peleg2000} abstracts a synchronous network of $n$ nodes, in which in each round of computation, each node can send a message of $O(\log n)$ bits on each of its links. A recent line of work addresses computing MST, distances, and data summarizaion in \congest with \emph{low-mixing time} \cite{Ghaffari2017DistributedMA,Ghaffari2018NewDA,Su2019DistributedDS} and finding small graphs in \congest benefits from efficient computation inside components of low-mixing time  \cite{Chang2019DistributedTD,ChangS19,ChangS20,Censor-HillelGL20,CensorHillelCLL20,EdenFFKO19,quantumCongestTriangles}. 
Low-mixing time is, in particular, a property of expander graphs, which have been shown to be useful for designing data centers \cite{Harsh2018ExpanderDF,DinitzSS20}.

The \hybrid model, recently introduced by \cite{AHKSS20}, abstracts networks supporting
high-bandwidth communication over \emph{local} edges, as well as very low-bandwidth communication over \emph{global} edges. Aligned with most previous work on this model, we assume here unbounded bandwidth for messages to neighbors, and $O(\log n)$-bit messages to a constant number of nodes anywhere in the network. This model in particular abstracts recent developments in hybrid data centers \cite{Cui2011ChannelAI,HanHLX15,HuangLLPLL13,WangAKPNKR10}.
Most research in the \hybrid model has been devoted to distance computation problems  \cite{AHKSS20,kuhn2020computing,feldmann2020fast,censorhillel2020distance}. 

While these settings highly differ, we pinpoint an approach that underlies computation in both settings, namely, \emph{sparsity and density awareness}. The key type of tasks which we tackle are tasks requiring transfer of massive amounts of data. 
Our general approach is to design load balancing mechanisms that leverage the full bandwidth of a given communication model.
Examples of tasks that enjoy our framework are matrix multiplication and distance computations.

\sloppy{For the purpose of our framework, as an auxiliary tool (not as a computational model in its own right), we define the \waccfull model with capacity $c$ (abbreviated \wacc), which is an all-to-all setting whose main characteristics are a limit on the bandwidth per node and the anonymity of nodes. To cope with the harsh nature of the \wacc model, which is needed in order to allow it to be efficiently simulated, we develop a distributed data structure and accompanying algorithms, dedicated towards load balancing and full utilization of the available bandwidth.}

We then show sparsity aware simulations of the \wacc model in the \congest and \hybrid settings. Specifically, the simulations focus on utilizing all the available bandwidth in the underlying models, even when given highly skewed inputs. Combined, these obtain our end results -- fast algorithms for distance computations in low-mixing time \congest and in the \hybrid model. 

~\\A first flavor of our end results: One of our main contributions is proving that the size $n$ of the graph is not fine-grained enough to capture the complexity of the all-pairs shortest-paths problem (APSP) in the \congest model. While there is an $\tilde{\Omega}(n)$ lower bound for general graphs, even when allowing large approximation factors \cite{Nanongkai2014DistributedAA}, and a matching randomized algorithm giving an exact solution \cite{Bernstein2019DistributedEW}, we show that one can go \emph{significantly} below this complexity, depending on the minimal degree in the graph and its mixing time.

\begin{restatable}[$(3+\eps)$-Approximation for APSP in \congest]{theorem}{sublinearAPSP}\label{cor:sublinearAPSP}
    For any constant $0 < \eps < 1$, and weighted, undirected graph ${G}$ with minimal degree $\delta$ and mixing time $\tmix$,
    there is an algorithm in the \congest model computing a $(3+\eps)$ approximation to APSP on $G$ within
    $(n^{1/2} + n/\delta)\cdot \tmixtwosqlgnlglgn$ rounds\footnote{\label{footnote1}Note that there is a typo in the complexity stated for this theorem in the SPAA'21 proceedings.}, \whp
\end{restatable}

For any constant $0 < \gamma < 1$, consider a graph $G$ with $\delta = n^\gamma \cdot \tmix\cdot 2^{\littleOmega{\sqrtlgn}}$. Using \cref{cor:sublinearAPSP}, it is possible to approximate weighted, undirected APSP on $G$ in $n^{1/2}\cdot \tmixtwosqlgnlglgn + O(n^{1-\gamma})$ rounds, \whp, in the \congest model, which is a major improvement over the linear general case. This approach is aligned with a  conclusion that is obtained by the single-source shortest paths (SSSP) algorithm of \cite{Ghaffari2018NewDA} that reflects that $n$ and the diameter $D$ are insufficient for capturing the complexity of SSSP. Our result suggests that for APSP the  dependence parameters could be  $n,\delta$ and $\tmix$, and this opens the complexity landscape of APSP in the \congest model to much further exciting research.
\subsection{Our Contributions}
\subsubsection{Fast Algorithms for the \hybrid Model}

The pioneering works of \cite{AHKSS20, kuhn2020computing} lay down technical foundations that show that utilizing \emph{both} the local and global edges in the \hybrid model allows solutions which are much faster than algorithms which only use the local or global edges. One of their prime contributions is showing that
the complexity of exact and approximate APSP is $\tilde \Theta(n^{1/2})$ rounds.

The $\tilde \Theta(n^{1/3})$-round regime is also of importance in this model, as \cite{kuhn2020computing,censorhillel2020distance} show a variety of algorithms with this complexity. For diameter, a lower bound of $\tilde \Omega(n^{1/3})$ rounds for exact unweighted diameter and for a $(2-\eps)$ approximation for weighted diameter is shown (the first lower bound in the \hybrid model for a problem with a small output), matched with a $\tilde O(n^{1/3})$ round algorithm for a $(1+\eps)$ approximation for unweighted diameter, and a $2$ approximation for weighted diameter. For weighted distances, $\tilde O(n^{1/3})$-round algorithms are shown for approximations from polynomially many sources, in addition to exact distances from $\tilde O(n^{1/3})$ sources.

Our main contribution in the \hybrid model is breaking below $o(n^{1/3})$ rounds for a $(1+\eps)$ approximation for weighted single source shortest paths (SSSP), which also implies a $(2+\eps)$ approximation for weighted diameter. As we elaborate upon in the following subsections, this requires establishing an entire foundation of techniques which are a core contribution of the paper. 
We show the following algorithm which completes in $\tilde O(n^{5/17})$ rounds, \whp (as common,
\emph{high probability} is at least $1-n^{-c}$, for a constant $c>1$).

\begin{restatable}[$(1+\eps)$-Approximation for SSSP in \hybrid]{theorem}{approxSSSP}
\label{theorem:approxSSSP}
Given a weighted, undirected graph $G = (V, E)$, with $n = |V|$ and $m = |E|$, a value $0 < \eps < 1$, and a source $s \in V$, there is a \hybrid model algorithm computing a $(1+\eps)$-approximation of SSSP from $s$, in $\tildeBigO{n^{5/17} / \eps^9}$ rounds, \whp
\end{restatable}

Our results provide an interesting open question: what is the best round complexity for a $2$ approximation of the diameter? We show that for a $(2+\eps)$ approximation, $o(n^{1/3})$ rounds suffice, while $(2-\eps)$ requires $\tilde \Omega(n^{1/3})$ rounds, as stated above.

\subsubsection{Fast Algorithms for the \congest Model}

As a warm-up, we illustrate the power of our \wacc model and its simulation in the \congest model, by showing how to simulate the \clique model (a synchronous model where every two nodes can exchange messages of $O(\log n)$ bits in every round) in \wacc and hence in \congest. Simulation of algorithms from the \clique model may both significantly simplify algorithm design as well as improve results in other distributed models, as seen in previous works \cite{Ghaffari2017DistributedMA,Chang2019DistributedTD,ChangS19,ChangS20,Censor-HillelGL20,CensorHillelCLL20,EdenFFKO19,kuhn2020computing}.

We get the following result, from which one can already show new algorithms which beat the state-of-the-art in the \congest model by simulating existing \clique algorithms. 

\begin{restatable}[\clique Simulation in \congest]{theorem}{cliqueInCongest}\label{cor:hc:cliqueSimulation}
    Consider a graph $G$. Let $A$  be an algorithm which runs in the \clique model over $G$ in $t$ rounds. If each node $v$ has $\bigO{\deg{v}\cdot \log{n}}$ bits of input and output, then there is an algorithm which simulates $A$ in $(t\cdot {{n^2}/{m}})\cdot\tmixtwosqlgnlglgn$ rounds of the \congest model over $G$, \whp.
\end{restatable}

Here, $\tmix$ is the \emph{mixing time} of the graph, which is roughly the number of rounds required for a lazy random walk to reach the stationary distribution (a precise definition is not needed for understanding our results). A previous non trivial simulation of the \clique model in the \congest model is due to~\cite{Ghaffari2017DistributedMA}. It emulates one \clique round in $O({n\cdot \tmix\cdot(1 + \frac{\Delta^2\log{n}\cdot\tmix}{n})\log{n}\log^*{n}})$ rounds of the \congest model, where $\Delta$ is the maximum degree in the graph. Moreover, for certain graphs, this is improved to $\tilde O(n\cdot \tmix)$.
Thus, our simulation improves upon both previous algorithms for all graphs with 
$m=n\cdot 2^{\littleOmega{\sqrtlgn}}$edges by a factor that is roughly proportional to the average degree, namely, faster by a factor of
$m / (n \cdot 2^{\alpha\cdot \sqrtlgn})$ for some constant $\alpha \geq 1$.
Intuitively, in \cite{Ghaffari2017DistributedMA}, if every node in \congest desires to message every other node then this requires many rounds. We circumvent this by sending the input of low degree nodes to high degree ones, which then simulate the \clique model and send back the output to the low degree nodes.

\sloppy{Finally, such a simulation implies a relation between lower bounds in the \congest and \clique models.} Specifically, simulating one \clique round in $T$ rounds of the \congest model shows that a lower bound of $R$ rounds in the \congest model implies a lower bound of $R/T$ rounds in the \clique model. 
Due to \cite[Theorem 4]{DuckerPowerClique}, we know that lower bounds for some problems in the \clique model imply lower bounds in bounded-depth circuit complexity, and are therefore considered hard to obtain.
Plugging our results from \Cref{cor:hc:cliqueSimulation} in the $R/T$ lower bound for the \clique model shows that if one constructs a family of graphs $\mathcal{G}$ with $m$ edges and $\tmix$ mixing time, for which solving some problem $P$ in the \congest model requires $R$ rounds, then $P$ has a lower bound of $(R\cdot m/(n^2\cdot \tmix\cdot 2^{\alpha{\sqrtlgn}}))$
rounds in the \clique model for some constant $\alpha \geq 1$. This means that any value of $\tmix$ that is below $R\cdot m/(n^2\cdot 2^{\alpha\sqrtlgn})$ implies a lower bound that is considered hard in the \clique model. 

\paragraph{SSSP\@.}
The current state-of-the-art exact SSSP algorithm in the \clique model, due to~\cite{SparseMMPODC2019}, runs in $O(n^{1/6})$ rounds. Using our result from \cref{cor:hc:cliqueSimulation}, a
simulation of this algorithm in the \congest model runs in  $({n^{13/6}}/{m})\tmixtwosqlgnlglgn$ rounds, \whp. We further improve this result by presenting a solution that is faster for any graph which is not extremely dense, namely, for which $m=\littleO{n^2}$. In a nutshell, our speed-up is due to the fact that the \clique algorithm does not use all of the $\Omega(n^2)$ bandwidth available to it in every round, and so it is inefficient to simulate directly in \congest. Thus, we instead simulate our \wacc model, which better reflects the complexity of such algorithms, giving us the following. 
\begin{restatable}[Exact SSSP in \congest]{theorem}{exactSSSPInHC}\label{theorem:hc:exactSSSP}
    Given a weighted, undirected graph $G$ and a source node $s \in G$, there is an algorithm in the \congest model that ensures that every node $v \in G$ knows the value $d_G(s, v)$, within ${(n^{7/6}/m^{1/2} + n^2/m)}\cdot\tmixtwosqlgnlglgn$ rounds, \whp.
\end{restatable}

Consider graphs with a 
mixing time $\tmix=\tildeBigO{1}=\polylog(n)$. The diameter $D$ of such graphs is also $\tildeBigO{1}$. If such a graph has at least $m = n^{3/2}\cdot 2^{\littleOmega{\sqrtlgn}}$ edges, then our SSSP algorithm runs asymptotically faster than the state-of-the-art $\tildeBigO{n^{1/2}D^{1/4}}$-round algorithm of~\cite{ChechikM20}.

\paragraph{APSP\@.}
We now turn our attention to the APSP problem. As observed by Nanongkai \cite[Observation 1.4]{Nanongkai2014DistributedAA}, to solve APSP in the \congest model, a node $v$ is required to learn $\tildeBigO{n}$ bits of information. In the worst case, for a node with a constant degree, this takes $\tildeBigOmega{n}$ rounds. For this reason,  slightly modified requirements for the output have been considered.

We consider a \emph{shortest-path query problem}, in which we separate the computation of shortest paths into two phases, one in which the input graph is pre-processed, and another in which a query set of pairs of nodes is given, $Q \subseteq V \times V$, and every node $v$ is required to learn the distance to every node $u$ such that $(v, u) \in Q$. 
The round complexity of this problem is thus bi-criteria, measured both in terms of pre-processing time and in terms of query time. We analyze the round complexity of the query in terms of the \emph{query load}, where given a node $v$, $q_v = \size{\{u\ |\ (v, u) \in Q \lor (u, v) \in Q\}}$ denotes the number of queries which $v$ is a part of, and the total, normalized query load is $\ell = \max_{v \in V} q_v / \deg{(v)}$. 
\begin{restatable}[$(3+\eps)$-Approximation for Shortest-Path Query in \congest]{theorem}{apspQuery}\label{claim:hc:apspQuery}
    For any constant $0 < \eps < 1$ and a weighted, undirected graph $G$ with $m$ edges,
    there is a \congest algorithm which after 
    $(n^{1/2} + {n^{11/6}}/{m})\cdot\tmixtwosqlgnlglgn$ rounds
     of pre-processing, solves any instance of the $(3+\eps)$-approximate shortest path query problem with a known load $\ell$, in $\ell\cdot \tmixtwosqlgnlglgn $ rounds, \whp.
\end{restatable}

By denoting $\delta$ as the minimal degree in the graph, one gets $\ell \leq n/\delta$ and $m \geq n \cdot \delta$, which implies \cref{cor:sublinearAPSP} stated above.

Finally, we consider a version of APSP, which we call \emph{Scattered APSP}, where the distance between every pair of nodes is known to \emph{some} node, not necessarily the endpoints themselves. That is, we require that every node $u$ knows, for every node $v$, the identity of a node $w_{uv}$, which stores the distance $d_{G}(u,v)$.
\begin{restatable}[$(3+\eps)$-Approximation for Scattered APSP in \congest]{theorem}{SAPSPInHC}\label{theorem:hc:sapsp}
    There exists an algorithm in the \congest model, that given a weighted, undirected
    input graph $G=(V, E)$ with $n = |V|$ and $m = |E|$, and some constant $0 < \eps < 1$, solves the $(3+\eps)$-approximate Scattered APSP problem on $G$,
    within $((n^{11/6}/m + n^{1/3} + m / n)/\eps + n^{1/2})\tmixtwosqlgnlglgn$ rounds\footnote{See footnote~\ref{footnote1}.}, \whp 
\end{restatable}

\paragraph{Roadmap.} After a survey of additional related work, \Cref{sec:prelim} provides required definitions. Additional preliminaries appear in \Cref{app:prelim}. \Cref{sec:wacc} is dedicated to the definition of carrier configurations in the \wacc model, and for a flavor of a sample proof, namely, sparse matrix multiplication. The bulk of our infrastructure for \wacc is deferred to \Cref{app:sec:wacc}. Finally, \Cref{sec:subCubeRootAlgorithms} and \Cref{sec:congest} provide proofs of our end results in the \hybrid and \congest models, respectively. Both sections have some content deferred to \Cref{app:littleO} and \Cref{app:sec:congest}, respectively. We conclude with a discussion in \Cref{sec:discussion}.

\subsection{Additional Related Work}

\paragraph{\hybrid.}

There is growing interest in the recently introduced \hybrid network model \cite{AHKSS20}. It is further studied by \cite{kuhn2020computing,feldmann2020fast,censorhillel2020distance}. \cite{AHKSS20,kuhn2020computing,censorhillel2020distance} consider the same values for the local bandwidth $\lambda$ and global bandwidth $\gamma$ as we do, and address mainly distance computations. 
The complexity of exact weighted APSP is $\tildeTheta{n^{1/2}}$ rounds, by combining the upper bound of \cite{kuhn2020computing}, and the lower bound of \cite{AHKSS20}. The complexity of $3$ approximate weighted $n^{2/3}$-SSP is $\tildeTheta{n^{1/3}}$ rounds, by combining the upper bound of \cite{censorhillel2020distance}, and the lower bound of \cite{kuhn2020computing}.
\cite{AHKSS20,kuhn2020computing} show how to simulate one round of the \bcc (a synchronous distributed model where every node can broadcast a (same) $O(\log n)$-bit message to all nodes per round) and \clique models on the skeleton graph with $\tildeBigO{n^{2/3}}$ nodes in $\tildeBigO{n^{1/3}}$ rounds of the \hybrid model and obtain various distance related results using them. For example \cite{kuhn2020computing} show $(7+\eps)$ and $(3+\eps)$ weighted $n^{x}$-SSP approximation in $O(n^{1/3}/\eps)$ and $\tildeBigO{n^{0.397 + n^{x/2}}}$ rounds  \whp, respectively. \cite{censorhillel2020distance} improve on some of those results by simulating ad-hoc models which exploit additional communication abilities of the \hybrid model. For instance, they show $3$ approximate weighted $n^{x}$-SSP approximation in $O(n^{1/3}/\eps)$ rounds \whp.
\cite{feldmann2020fast} solve distance problems for sparse families of graphs using local congestion bounded by $\log{n}$. Another special case of the \hybrid model is the \ncc model, which contains only global edges
\cite{AugustineCCPSS20,AGGHKL19}.

\paragraph{\congest.}
Distance problems are extensively studied in the \congest model due to being fundamental tasks. One of the important problems in the \congest model is $(1+\eps)$-approximate SSSP \cite{Sarma2011DistributedVA,Becker2017NearOptimalAS,Henzinger2019ADA}. The state-of-the-art randomized algorithm  \cite{Becker2017NearOptimalAS} solves it in $\tildeBigO{(\sqrt{n} + D)\eps^{-3}}$ rounds \whp, even in the more restricted Broadcast \congest model. This is close to the $\tilde{\Omega}{(\sqrt{n} + D)}$ lower bound  by \cite{Sarma2011DistributedVA}. The state-of-the-art deterministic algorithm \cite{Henzinger2019ADA} completes in $n^{1/2 + o(1)} + D^{1 + o(1)}$ rounds.  The complexity of exact SSSP is still open, with algorithms  given in \cite{Elkin2017DistributedES,Ghaffari2018ImprovedDA,Forster2018AFD,ChechikM20}. The current best known algorithm is of \cite{ChechikM20} runs in $\tildeBigO{n^{1/2}D^{1/4} + D}$ rounds, \whp.

There is a lower bound of $\tildeBigOmega{n}$ rounds to compute APSP \cite{Nanongkai2014DistributedAA} which was then tweaked to give a $\bigOmega{n}$ lower bound for the weighted case \cite{CensorHillelKP17}. Over the years there was much progress in understanding the complexity of this problem \cite{Peleg2012DistributedAF,Lenzen2013EfficientDS,Lenzen2015FastPD,Agarwal2018ADD,Elkin2017DistributedES,Bernstein2019DistributedEW,Agarwal2019DistributedWA,Ramachandran20FasterDA}. For  weighted exact APSP, the best known  randomized algorithm \cite{Bernstein2019DistributedEW} is optimal up to polylogarithmic terms. The best deterministic algorithm to  date \cite{Ramachandran20FasterDA} completes in $\tildeBigO{n^{4/3}}$ rounds. For unweighted exact APSP, $\tildeBigO{n}$ rounds algorithms are known \cite{Peleg2012DistributedAF,Lenzen2013EfficientDS,HolzerW12}.

To go below the $\bigOmega{n}$ lower bound, \cite{Lenzen2013FastRT,Lenzen2015FastPD} propose to build name-dependent routing tables and APSP. This means that the algorithm is allowed to choose small labels and output results that are consistent with those labels. Choosing the labels carefully thus overcomes the need to send too many bits of information to low-degree nodes. There is also a line of work which breaks below the lower bound for some certain graph families \cite{GhaffariH16PODC,GhaffariH16SODA,LiP19,Parter20a}.
\cite{Peleg2012DistributedAF,Frischknecht2012NetworksCC,Abboud2016NearLinearLB,GrossmanKP20,Ancona2020DistributedDA} study exact and approximate diameter, eccentricities, girth and other problems. 

Recently, \cite{Ghaffari2017DistributedMA,Ghaffari2018NewDA,Su2019DistributedDS} noticed that it is possible to develop faster \congest algorithms when the underlying graph has low mixing time. This was shown for MST \cite{Ghaffari2017DistributedMA}, maximum flow, SSSP and transshipment \cite{Ghaffari2018NewDA} and frequency moments \cite{Su2019DistributedDS}. Algorithms for subgraph-freeness and related variants also enjoy fast computations on graphs with low mixing time \cite{Chang2019DistributedTD,quantumCongestTriangles,ChangS19,Chang2019DistributedTD,ChangS19,ChangS20,Censor-HillelGL20,CensorHillelCLL20,EdenFFKO19}.

\paragraph{\clique and \bcc.}
In the \clique and \bcc models, distance related problems like APSP, $k$-SSP and diameter and their approximation are studied in \cite{Nanongkai2014DistributedAA,CensorHillel2016AlgebraicMI,Gall2016FurtherAA,CensorHillel2020SparseMM,SparseMMPODC2019,Dory2020ExponentiallyFS,Holzer2015ApproximationOD,Becker2017NearOptimalAS}.
We mention that models related to the \bcc and \clique models have been studied in~\cite{Becker2015BriefAA,Becker2016TheEffectOf}, who explore the regime between broadcast and unicast.

\addtocontents{toc}{\protect\setcounter{tocdepth}{2}}

\section{Preliminaries}\label{sec:prelim}
The following are some required definitions, while \Cref{app:prelim} contains additional definitions and basic claims.
We begin with a variant of the \hybrid model, introduced in \cite{AHKSS20}.

\begin{definition}[\hybrid Model]

In the \hybrid model, a synchronous network of $n$ nodes with identifiers in \interval{n} is given by a graph $G=(V,E)$. In each round, every node can send and receive $\lambda$ messages of $O(\log n)$ bits to/from each of its neighbors (over \emph{local edges} $E$) and an additional $\gamma$ messages in total that are $\bigO{\log{n}}$ bits long to/from any other nodes in the network (over \emph{global edges}). If in some round more than $\gamma$ messages are sent via global edges to/from a node, only $\gamma$ messages selected adversarially are delivered.
\end{definition}

All of \cite{AHKSS20,kuhn2020computing,censorhillel2020distance} use $\lambda=\infty,\gamma=\log{n}$. To our knowledge only \cite{feldmann2020fast} considers more restrictive settings of $\lambda=1,\gamma=\log{n}$.

We introduce the \waccfull model with capacity $c$ (abbreviated \wacc), which we show is powerful for extracting core distributed principles. The model has two defining characteristics which are \emph{anonymity} and \emph{restricted message capacity}. 

\begin{definition}[The \wacc Model]
\label{def:wacc}
    The \waccfull model with capacity $c$ is a distributed synchronous communication model, over a graph $G=(V, E)$ with $n$ nodes, where each node $v$ can send and receive $c$ messages in every round, each of size $O(\log n)$ bits, to and from any node in the graph. Nodes have identifiers in $[n]$, however, every node has a unique $O(\log n)$-bit, \emph{communication token}, initially known only to itself. Node $v$ can send each message either to some node $w$ whose communication token is already known to $v$, or, to a node selected uniformly, independently at random from the entire graph.
    \end{definition}

\wacc[c] bears some similarity to the $\ncczero$ model \cite{AugustineCCPSS20} with an empty \emph{initial knowledge graph}. Unlike the $\ncczero$ model, \wacc[c] has additional capacity to send $c$ messages from each node and ability to send messages to random nodes. Our motivation for defining \wacc is to provide a setting which is both strong enough in order to solve challenging problems, yet, at the same time is weak enough in order to be simulated efficiently in the settings of interest. 
We note the importance of having both identifiers and communication tokens. An identifier is chosen from a given, hardcoded set and thus can be used for assigning tasks to specific nodes. Communication tokens both assist in dealing with anonymity, and enhance the ability of the \wacc model to be easily simulated in other distributed settings -- \emph{a simulating algorithm can encode routing information of the underlying model in the tokens}.

~\\
Many of our results hold for weighted graphs $G=(V, E, w)$, where $w\colon E\mapsto \set{1, 2, \dots, W}$ for a $W$ which is polynomial in $n$. Whenever we \emph{send} an edge $e$ as part of a message, we assume $w(e)$ is sent as well. We assume that all graphs that we deal with are connected.

Given a graph $G=(V, E)$ and a pair of nodes $u, v\in V$, we denote by $hop(u,v)$ the hop distance between $u$ and $v$, by $N^k_G(v)$ a subset of the $k$ closest nodes to $v$ with ties broken arbitrarily,
and by $d_G^h(u, v)$ the weight of the lightest path between $u$ and $v$ of at most $h$-hops, where if there is no path of at most $h$-hops then $d_G^h(u, v)=\infty$. In the special case of $h=\infty$,
we denote by $d_G(u, v)$ the weight of a lightest path between $u$ and $v$. We also denote by $\deg_{G}{(v)}$ the degree of $v$ in $G$, and, in the directed case, $\deg_{G}^{in}{(v)},\ \deg_{G}^{out}{(v)}$ denote the in-degree and out-degree of $v$ in $G$, respectively. When it is clear from the context we sometimes drop the subscript $G$.

\begin{definition}[$k$-Source Shortest Paths (k\ssp)]
    Given a graph $G = (V, E)$, in the \emph{$k$-source shortest path problem}, we are given a set $S\subseteq V$ of $k$ sources. Every $u\in V$ is required to learn the distance $d_G(u, s)$ for each source $s\in S$. The cases where $k=1$, $k=n$ are called the \emph{single source shortest paths} problem (SSSP), and \emph{all pairs shortest paths} problem (APSP), respectively.
\end{definition}

\begin{definition}[Scattered APSP]
\label{def:scatteredAPSP}
Given a graph $G = (V, E)$, in the \emph{Scattered APSP} problem, for every pair of nodes $u, v \in V$, there exist nodes $w_{uv}, w_{vu} \in V$ (potentially $w_{uv} = w_{vu}$), such that $w_{uv}$ and $w_{vu}$ know $d_G(u, v)$, $u$ knows the identifier of $w_{uv}$, and $v$ knows the identifier of $w_{vu}$. 
\end{definition}

In the approximate versions of these problems, each $u\in V$ is required to learn an $\left(\alpha,\beta\right)$-approximate distance $\widetilde{d}(u, v)$ which satisfies $d(u, v)\leq \widetilde{d}(u, v) \leq \alpha \cdot d(u, v) + \beta$, and in case $\beta=0$, $\widetilde{d}(u, v)$ is called an $\alpha$-approximate distance.

\begin{definition}[Diameter]
    Given a graph $G=(V, E)$, the diameter $D=\max_{u,v\in V}\set{d(u,v)}$ is the maximum distance in the graph. An $\alpha$-approximation of the diameter $\widetilde{D}$ satisfies $D/\alpha\leq\widetilde{D}\leq D$. 
\end{definition}

\section{The \waccfull Model}
\label{sec:wacc}

The role of defining the \wacc model is its power given by our ability to efficiently simulate it 
in the \hybrid and \congest models. Applications of this strength are exemplified by improved algorithms for distance computation problems in these models.

To this end, we design fast algorithms in the \wacc model for the useful tools of sparse matrix multiplication and hopset construction (\cref{wacc:main:results}). Such algorithms already exist in all-to-all communication models, such as the \clique \cite{SparseMMPODC2019,Dory2020ExponentiallyFS}. However, fundamental load balancing and synchronization steps that are simple to implement when assuming a bandwidth of $\Theta(n^2)$ messages per round as in the \clique model, provide formidable challenges when nodes cannot receive $\Theta(n)$ messages each. For instance, when multiplying two matrices, while the number of finite elements in row $v$ of both input matrices might be small, in the output matrix, row $v$ might be very dense, so that node $v$ would not be able to learn all of this information. This even implies that it is not always the case that every node can even know about all the edges incident to it in some overlay graph (e.g., a hopset). 
The crux in the way we overcome these challenges is in introducing the \emph{carrier configuration} distributed data structure that performs automatic load balancing (\cref{subsec:carrierConfiguration}). 
Missing proofs are deferred to \cref{app:sec:wacc}.

\subsection{Carrier Configurations}\label{subsec:carrierConfiguration}

A carrier configuration is a distributed data structure for holding graphs and matrices, whose main objective is to provide a unified framework for load balancing in situations where substantial amounts of data need to be transferred. The key is that when using the carrier configurations data structure, an algorithm does not need to address many load balancing issues, as those are dealt with under-the-hood by the data structure itself. Therefore, this data structure allows us to focus on the core concepts of each algorithm and abstract away challenges that arise due to skewed inputs.

The data structure crucially enables our algorithms to enjoy sparsity awareness, by \emph{yielding complexities that depend on 
the average degree in an input graph rather than its maximal degree}.
This allows one to eventually deal with data which is \emph{highly skewed} and which would otherwise cause a slow-down by having some nodes send and receive significantly more messages than others.

We stress that the manner in which we implement carrier configurations is inherently distributed in nature. In many cases, when two nodes $u,\ v$ store data, $D(u),\ D(v)$, respectively, in a carrier configuration, the data is dispersed among many other nodes, and when operations are performed using both $D(u)$ and $D(v)$, the nodes which store the data perform direct communication between themselves, without necessarily involving $u$ or $v$.

In more detail, the carrier configuration data structure is based on three key parts. (1) \emph{Carrier Nodes:} every node $v$ gets a set $C_v$ of \emph{carrier nodes}, where $|C_v| = \Theta(\deg(v)/k)$ and $k$ is the average degree in the graph, which \emph{help} $v$ to carry information and store its input edges in a distributed fashion. A key insight is that it is possible to create such an assignment of carrier nodes and also maintain that each node itself is not a carrier for too many other nodes, thus avoiding congestion. (2) \emph{Ordered Data Storage:} the data of $v$ is stored in a sorted fashion across $C_v$ in order to enable its efficient usage. In particular, it takes $O(\log n)$ bits in order to describe what ranges of data are stored in a specific carrier node. (3) \emph{Communication Structure:} the nodes $C_v \cup \{v\}$ are connected using a communication tree (\cref{definition:communicationTree}), which enables fast broadcasting and aggregation of information between $v$ and $C_v$.

A formal definition of a \emph{carrier configuration data structure} (\cref{definition:carrierConfiguration}), as well as an extended \wacc-specific definition (\cref{wacc:def:carrierConfiguration}) are given in \cref{subsec:waccAppCarconf}.

\textbf{Carrier Configuration Toolbox.} Typically, data is converted from a \emph{classical} storage setting (every node knows the edges incident to it) into being stored in a carrier configuration, and then operations are applied to change the state of the configuration.
Thus, we show in the \wacc model how to convert a classical graph representation to a carrier configuration. Then, we provide basic tools, e.g., given two matrices held in carrier configurations, produce a third configuration holding the matrix resulting from computing the point-wise minimum of the two input matrices. The descriptions and implementations of these are deferred to \cref{app:wacc:carrierconfiguration}.

\subsection{Sparsity Aware Distance Computations}
\label{wacc:main:results}

In order to give a taste of the type of algorithms which we construct in the \wacc model, we present an outline of our \emph{sparse matrix multiplication} algorithm.

We build a foundation in the \wacc model
which enables us to eventually implement sparse matrix multiplication and hopset construction algorithms, as inspired by the \clique model algorithms in \cite{SparseMMPODC2019},\footnote{We note that while some results of \cite{SparseMMPODC2019} were improved in \cite{Dory2020ExponentiallyFS}, the improvements focus on reducing the amount of synchronous rounds of the algorithm, yet do not decrease the message complexity in way that assists us.}  in order to solve distance related problems. Ultimately, compared to \cite{SparseMMPODC2019}, our main contribution is significantly reducing the \emph{message} complexity overhead of the load balancing mechanisms. In the \clique implementation, various load balancing steps require $\Theta(n^2)$ messages, which is trivial in the \clique model, yet is highly problematic in the \wacc model (and in the \congest and \hybrid models which ultimately simulate it). Interestingly, for the relevant sparsity used in distance computations, the sparse matrix multiplication itself requires $o(n^2)$ messages for actual transfer of matrix elements, and in the \clique algorithm, the message complexity is dominated by the $\Theta(n^2)$ overhead. We reduce this overhead significantly such that the message complexity of the algorithm is dominated by the messages which actually transfer matrix elements.

We show how to perform sparse matrix multiplication in the \wacc model. To simplify our proofs, we assume that the two input matrices and the output matrix have the same average number of finite elements per row.
We note that it is possible to use the same ideas we show here in order to prove the general case where the matrices have different densities.

\begin{restatable}[Sparse Matrix Multiplication]{theorem}{SparseMM}\label{theorem:wacc:sparseMM}
Given two $n \times n$ input matrices $S, T$, both with an average number of \emph{finite} elements per row of at most $k$, and stored in carrier configurations, $A$ and $B$, it is possible to output the matrix $\hat P$, in carrier configuration $C$, where $\hat P$ is $P = S \cdot T$, with only the smallest $k$ elements of each row computed, breaking ties arbitrarily.
This takes $\tilde O(k \cdot n^{1/3}/c + n/(k\cdot c) + 1)$ rounds, in the \wacc model, \whp \end{restatable}

\begin{proofof} {\cref{theorem:wacc:sparseMM}}
Throughout the algorithm, we assume that every piece of data \emph{sent directly} to a node is also known by all its carrier nodes. This is possible to do with only a $\tilde O(1)$ multiplicative factor to the round complexity, due to \fullref{lemma:carrierConfiguration:broadcastAggregate}. Further, as the input matrices are stored in carrier configuration, due to \cref{itm:carrierConfiguration:communicationToken}, every entry $S[i][j]$ is stored in $A$ alongside the communication tokens of both $i$ and $j$ (likewise for entries of $T$ stored in $B$). Thus, whenever a value is sent in the algorithm from one of the input matrices, we assume that it is sent alongside these communication tokens.

Denote by $\Delta$ the maximal number of finite elements in a row of $S$ or $T$. In the proof below, we show a round complexity of $\tilde O(k \cdot n^{1/3}/c + n/(k\cdot c) + \Delta / (n^{1/3} \cdot c) + n^{2/3}/c + 1)$, which is bounded by the claimed round complexity of $\tilde O(k \cdot n^{1/3}/c + n/(k\cdot c) + 1)$.\footnote{Notice that $\tilde O(k \cdot n^{1/3}/c + n/(k\cdot c) + \Delta / (n^{1/3} \cdot c) + n^{2/3}/c + 1) = \tilde O(k \cdot n^{1/3}/c + n/(k\cdot c) + 1)$, since: 1) It always holds that $\Delta \leq n$, and so $\Delta / (n^{1/3}\cdot c) = O(n^{2/3}/c)$, and 2) If $k < n^{1/3}$, then $n^{2/3}/c = O(n/(k \cdot c))$, yet if $k \geq n^{1/3}$, then $n^{2/3}/c = O(k \cdot n^{1/3}/c)$.}

Throughout the proof, we construct various sets of nodes (for instance, $V_{i}$) where every node in the set knows the communication tokens and identifiers of all the other nodes in the set ($Tokens(V_i)$), and thus we assume that the nodes in the each set locally compute some arbitrary node (some fixed $v_i \in V_i$) which is the \emph{leader} of the set. 
~\\

\noindent\textbf{Step: Partitioning the Input -- Sets $V_i$}\ \\
Denote by $\deg_S(v),\ \deg_T(v)$ the number of finite elements in row $v$ of $S$, and column $v$ of $T$, respectively.

Denote $W_1, \dots, W_{n^{1/3}/2}$, a hardcoded partition of $V$ into equally sized sets. Using \fullref{wacc:tools:routing} and \fullref{wacc:tools:grouping}, every $v \in W_i$ broadcasts $\deg_S(v),\ \deg_T(v)$ to $W_i$, within $O(n^{2/3}/c + 1)$ rounds. The nodes in $W_i$ locally partition $W_i$ into $W_{i, 1}, \dots, W_{i, j_i}$, for some $j_i$, where for each $k \in [j_i]$, $\sum_{v \in W_{i, k}} \deg_S(v) + \deg_T(v) \leq 4k \cdot n^{2/3} + 4\Delta$. Since $\sum_{v \in V}\deg_S(v) + \deg_T(v) \leq 2nk$, there exists a way to create this partitioning such that $\sum_{i \in [n^{1/3}]} j_i \leq n^{1/3}$. 

From here on, we refer to the sets $W_{i, k}$ as $V_j$. We denote by $S[V_j]$ the rows of $S$ corresponding to the nodes $V_j$, and by $T[V_j]$ the columns of $T$ corresponding to the nodes $V_j$. As $\sum_{i \in [n^{1/3}]} j_i \leq n^{1/3}$, there are at most $n^{1/3}$ sets $V_i$ -- the sets $V_1, \dots, V_{n^{1/3}}$.
For each set $V_{i}$, an arbitrary $v_{i} \in V_{i}$ is designated as its \emph{leader}. The identifiers and communication tokens of all the leaders are broadcast using \fullref{wacc:tools:broadcasting}, within $\tilde O(n^{1/3}/c + 1)$ rounds.

We partition the information held by the sets $V_i$. We compute $C_i = \{0 = c_0, c_1, \dots c_{n^{1/3}/4-1}, c_{n^{1/3}/4} = n\}$, where the total number of finite entries in\footnote{The notation $S[X][a:b]$ denotes all rows of $S$ with indices in $X$, from column $a$ to column $b$.} $S[V_i][c_j+1:c_{j+1}]$ is at most $16k \cdot n^{1/3} + 16\Delta/n^{1/3} + |V_i| \leq 16k \cdot n^{1/3} + 16\Delta/n^{1/3} + 2n^{2/3}$. Recall that for each $V_i$, it holds that $\sum_{v \in V_i} \deg_S(v) + \deg_T(v) \leq 4k \cdot n^{2/3} + 4\Delta$, and also $|V_i| \leq 2n^{2/3}$, implying the existence of such a set $C_i$. 

The nodes in $V_i$ compute the values in $C_i$ using binary search. To see this, given any $p_1, \dots, p_c$, the number of finite entries in each of $S[V_i][0:p_1], \dots, S[V_i][0:p_c]$ can be computed by the nodes in $V_i$ in $\tilde O(1)$ rounds, using \fullref{wacc:tools:groupBroadcasting}. Thus, each $c_j \in C_i$ can be found using binary search, with $c$ binary searches run in parallel in $\tilde O(1)$ rounds. In total $\tilde O(|C_i|/c + 1) = \tilde O(n^{1/3}/c + 1)$ rounds are required to compute $C_i$.
~\\

\noindent\textbf{Step: Creating Intermediate Representations  -- Sets $U_{i, j}$, $P_{i, j, \ell}$ Matrices}\ \\
Let $U_{i, j}$, for $i, j \in [n^{1/3}]$, be a hard-coded partition of $V$ into equally sized sets. The goal of this step is for nodes $U_{i, j}$ to compute an \emph{intermediate representation} of the product $S[V_i]\times T[V_j]$. Therefore, we desire that each $V_i$ sends all its data to $U_{i, j}, U_{j, i}$, for each $j \in [n^{1/3}]$, in some load-balanced manner. We show how, for each $i \in [n^{1/3}]$, $V_i$ sends $S[V_i]$ to $U_{i, j}$, for each $j \in [n^{1/3}]$, and in a symmetric way (with matrix $T$ instead of $S$) this can be done for $U_{j, i}$.

In $\tilde O(n^{2/3}/c + 1)$ rounds, allow communication within each $U_{i, j}$ by invoking \fullref{wacc:tools:grouping}. Let some $u_{i, j} \in U_{i, j}$, be denoted \emph{leader}, and broadcast the leaders of all the sets using \fullref{wacc:tools:broadcasting} within $\tilde O(n^{2/3}/c + 1)$ rounds. Node $u_{i, j}$ sends to both $v_i, v_j$ all of $Tokens(U_{i, j})$. Each node $v_i$ receives $n^{2/3}$ messages, thus this takes $\tilde O(n^{2/3}/c + 1)$ rounds using \fullref{wacc:tools:routing}. Now, each node $v_i$ broadcasts to $V_i$ all the tokens it receives, using \fullref{wacc:tools:groupBroadcasting}, within $\tilde O(n^{2/3}/c + 1)$ rounds.

Leader $v_i$ sends the contents of $C_i$ to all the leader nodes $u_{i, j}, u_{j, i}$, for any $j \in [n^{1/3}]$, in $\tilde O(n^{2/3}/c + 1)$ rounds using \fullref{wacc:tools:routing}. Each $u_{i, j}$ broadcasts to $U_{i, j}$ the contents of $C_i$ and $C_j$, within $\tilde O(n^{1/3}/c + 1)$ rounds using \fullref{wacc:tools:groupBroadcasting}.

We send information from $V_i$ to $U_{i, j}$. The $\ell$-th node in $U_{i, j}$ learns the finite elements in $S[V_i][c_\ell+1:c_{\ell+1}]$. By definition of $C_i$, for any $\ell$, the number of finite elements in $S[V_i][c_\ell+1:c_{\ell+1}]$ is at most $16k \cdot n^{1/3} + 16\Delta/n^{1/3} + 2n^{2/3}$, bounding the number of messages each node desires to receive. Every finite element held by a node $v \in V_i$ needs to be sent to $O(n^{1/3})$ nodes in the graph, in total.
Further, since the finite elements of $v$ are stored in its carrier nodes, and each carrier node of $v$ stores $O(k)$ elements, each node sends at most a total of $O(k \cdot n^{1/3})$ messages. Thus, this routing can be accomplished in $\tilde O(k\cdot n^{1/3}/c + \Delta/(n^{1/3} \cdot c) + n^{2/3}/c + 1)$ rounds, using \fullref{wacc:tools:routing}.\footnote{To perform the routing, it is required that the carrier nodes of $v \in V_i$ know the communication tokens of the nodes in $U_{i, j}$ which receive the messages. All the nodes in $V_i$ received $Tokens(U_{i,j})$, and so $v$ knows the required tokens. At the start of the proof, we state that we assume that every message a node receives is also broadcast from it to its carriers, and so also the carriers of $v$ know the required communication tokens.}.

We shuffle data within $U_{i, j}$. Recall that $|U_{i, j}| = n^{1/3}$. The first $|C_i| = |C_j| = n^{1/3}/4$ nodes of $U_{i, j}$ received data from $S[V_i],\ T[V_j]$, according to
$C_i, C_j$. Denote $C_{i, j} = C_i \cup C_j$.
We desire that for each interval, $[c_\ell + 1, c_{\ell+1}]$, for $c_\ell \in C_{i, j}$, node $\ell \in U_{i, j}$ knows all the elements $S[V_i][c_\ell + 1, c_{\ell+1}]$ and $T[c_\ell + 1, c_{\ell+1}][V_j]$. To do so, notice that $S[V_i][c_\ell + 1, c_{\ell+1}]$ (likewise with $T$) is fully contained in the data which some node in $q \in U_{i, j}$, where $q \in [n^{1/3}/4]$ already knows.
Thus, each node $\ell \in [n^{1/3}/4]$ of $U_{i, j}$ denotes by $val(\ell)$ how many other nodes in $U_{i, j}$ are reliant on the data which it itself received from $V_i$.\footnote{This can be done since all the nodes in $U_{i, j}$ know all of $C_{i, j}$.} Then, we invoke \fullref{wacc:tools:groupMulticasting} in $\tilde O(k\cdot n^{1/3}/c + \Delta/(n^{1/3} \cdot c + n^{2/3}/c + 1)$ rounds in order to route the required information. Finally, each node $\ell \in [n^{1/3}/2]$ from $U_{i, j}$ knows $S[V_i][c_\ell + 1, c_{\ell+1}]$ and $T[c_\ell + 1, c_{\ell+1}][V_j]$, denoted $S_{i, j, \ell}, T_{i, j, \ell}$, respectively, and can compute the product $P_{i, j, \ell} = S_{i, j, \ell} \times T_{i, j, \ell}$.

We are at a state where for each $U_{i, j}$, every $\ell \in [n^{1/3}/2]$ of $U_{i, j}$ knows some matrix $P_{i, j, \ell}$ such that $S[V_i]T[V_j] = \sum_{\ell \in [n^{1/3}/2]}P_{i, j, \ell}$.\ \\

\noindent\textbf{Step: Sparsification -- $\hat{P}_{i, j, \ell}$ Matrices}\ \\
We sparsify the $P_{i, j, \ell}$ matrices. Recall that we desire to output $\hat P$ which is $P = S\times T$, with only the $k$ smallest entries on each row.
Fix $i \in [n^{1/3}]$, $\ell \in [n^{1/3}/2]$, and denote by $Q_{i, \ell}$ as the matrix of size $|V_i| \times n$ created by concatenating $P_{i, 0, \ell}, \dots, P_{i, n^{1/3}, \ell}$.
As shown in \cite{SparseMMPODC2019},
we are allowed to keep only the $k$ smallest entries on each row of $Q_{i, \ell}$, without damaging the final output $\hat{P}$. That is, throwing out elements at this stage is guaranteed to throw out elements which are only in $P$ and not in $\hat{P}$.

Fix $i \in [n^{1/3}], \ell \in [n^{1/3}/2]$, and denote by $N_{i, \ell}$ the nodes numbered $\ell$ in each $U_{i, j}$. The nodes in  $N_{i, \ell}$ perform a binary search for each row of $Q_{i, n^{1/3}, \ell}$, to determine the cutoff for the $k^{th}$ smallest element on that row.
The leaders $u_{i, 0}, \dots, u_{i, n^{1/3}}$ broadcast to each other $Tokens(U_{i, j}), \dots, Tokens(U_{i, n^{1/3}})$, using \fullref{wacc:tools:routing}, taking $O(n^{2/3}/c + 1)$ rounds. Each leader $u_{i, j}$ broadcasts the $O(n^{2/3})$
tokens it received to $U_{i, j}$, using \fullref{wacc:tools:groupBroadcasting}, within $\tilde O(n^{2/3}/c + 1)$ rounds. Now, the nodes $N_{i, \ell}$ all know $Tokens(N_{i, \ell})$.

The nodes $N_{i, \ell}$ proceed in $\tilde O(1)$ iterations to perform concurrent binary searches on the threshold values for each row in $Q_{i, \ell}$.
Let $n_{i, \ell}$ be an arbitrarily chosen leader node for $N_{i, \ell}$. In every iteration, node $n_{i, \ell}$ broadcasts $p_1, \dots, p_{|V_i|}$ values, each a tentative threshold value for a row in $Q_{i, \ell}$, using \fullref{wacc:tools:groupBroadcasting}, and in response, aggregates from the nodes of $N_{i, \ell}$, using \fullref{wacc:tools:groupBroadcasting} the total number of entries in each row of $Q_{i, \ell}$ below, equal to, and above the queried threshold. Each such iteration takes $\tilde O(|V_i|/c + 1) = \tilde O(n^{2/3}/c + 1)$ rounds, and after $\tilde O(1)$ iterations of this procedure, all the nodes in $N_{i, \ell}$ know a threshold for every row they posses, informing them which values in $Q_{i, n^{1/3}, \ell}$ can be thrown out.

Define the matrices $\hat P_{i, j, \ell}$ by removing from $P_{i, j, \ell}$ the entries which are thrown away due to the thresholds.\ \\

\noindent\textbf{Step: Balancing the Intermediate Representation}\ \\
The nodes computed the matrices $\hat P_{i, j, \ell}$, yet,
some matrices $\hat P_{i, j, \ell}$ may be too dense in order to transport out of the nodes which locally hold them. Even though we sparsified the matrices above, the sparsification steps perform were on several $\hat P_{i, j, \ell}$ matrices at once, and thus we can still have single $\hat P_{i, j, \ell}$ matrices which remains very dense. Let $\hat P_{x, y, z}$ be such a very dense matrix. We overcome this challenge by having more nodes compute $\hat P_{x, y, z}$ from scratch, allowing each node to only take responsibility for retaining a part of $\hat P_{x, y, z}$.

For each $i\in [n^{1/3}]$, we pool the nodes $U_i = \bigcup_{j \in [n^{1/3}]} U_{i, j}$ and redistribute them such that areas in the matrix which are too dense get more nodes.

Each node $\ell \in [n^{1/3}/2]$ from  $U_{i, j}$ computes the number of finite values in $\hat P_{i, j, \ell}$, denoted as $p_{i, j, \ell}$. Node $u_{i, j}$ computes $p_{i, j} = \sum_{\ell \in [n^{1/3}/2]} p_{i, j, \ell}$ using
on $U_{i, j}$ within $\tilde O(n^{1/3}/c + 1)$ rounds due to \fullref{wacc:tools:groupBroadcasting}. Finally, $u_{i, j}$ broadcasts to the other leader nodes $u_{i, 0}, \dots, u_{i, n^{1/3}}$ the values $p_{i, j, 0}, \dots, p_{i, j, n^{1/3}/2}$, and then broadcasts to $U_{i, j}$ all the $O(n^{2/3})$ values that it received -- taking $O(n^{2/3}/c + 1)$ rounds due to \fullref{wacc:tools:routing} and \fullref{wacc:tools:groupBroadcasting}.

Due to the fact that each row of $\hat P$ has at most $k$ elements in total, and currently $\hat P$ is distributed across $n^{1/3}/2$ matrices which need to be summed, all the nodes $U_i$ hold at most $k \cdot |V_i| \cdot n^{1/3}/2 \leq k \cdot n$ elements. That is, the sum $p_i = \sum_{j\in[n^{1/3}]}p_{i, j}$ is at most $k \cdot n$. Each node $\ell \in [n^{1/3}/2]$ from  $U_{i, j}$ observes $\hat P_{i, j, \ell}$ and locally breaks it up into $t(i, j, \ell) = \ceil{p_{i, j, \ell} / (2k \cdot n^{1/3})}$ matrices $\hat P_{i, j, \ell, 0}, \dots, \hat P_{i, j, \ell, t(i, j, \ell)}$ which sum up to $\hat P_{i, j, \ell}$ and each have at most $2k\cdot n^{1/3}$ finite elements. This creates at most $n^{2/3}$ such matrices. That is, $\sum_{j\in[n^{1/3}], \ell\in[n^{1/3}/2]} t(i, j, \ell) \leq \sum_{j\in[n^{1/3}], \ell\in[n^{1/3}/2]} \ceil{p_{i, j, \ell} / (2k \cdot n^{1/3})}  \leq  n^{2/3}$. Thus, the total number of intermediate matrices, spread over the nodes $U_i$, is at most $n^{2/3}$. Each $\ell \in [n^{1/3}/2]$ from $U_{i, j}$ is allocated $t(i, j, \ell) - 1$ other nodes from $U_i$, called \emph{auxiliary nodes} in order to send them the matrices $\hat P_{i, j, \ell, 1}, \dots, \hat P_{i, j, \ell, t(i, j, \ell)}$. Notice that since all nodes in $U_i$ know all the values $p_{i, j, \ell}$, all the nodes can locally know which node in $U_i$ is allocated to help which other node in $U_i$. However, node $\ell$ cannot send to all its auxiliary nodes all of this information, instead it sends to each of its auxiliary nodes the data it received from $S$ and $T$ with which it computed the matrix $P_{i, j, \ell}$, and the $O(n^{2/3})$ thresholds which it used to turn $P_{i, j, \ell}$ into $\hat P_{i, j, \ell}$. This can be accomplished via \fullref{wacc:tools:groupMulticasting} within $\tilde O(k \cdot n^{1/3}/c + n^{2/3}/c + 1)$ rounds, since each node wishes to multicast at most $\tilde O(k \cdot n^{1/3})$ data (as this is the bound on the total data each node received from $S$ and $T$), to at most $|U_i| = O(n^{2/3})$ other nodes, and each node desires to receive multicast messages only from one other node.\ \\

\noindent\textbf{Step: Summation}\ \\
We send data from $U_i$ to $V_i$ to perform the final summation step. Node $v \in V_i$ learns all of the information held in the nodes in $U_i$ which pertains to $\hat{P}[v]$.
All the nodes in $U_i$ know $Tokens(V_i)$, and vice versa and can thus communicate. Each node in $V_i$ wishes to receive $O(k \cdot n^{1/3})$ data from $U_i$, and likewise, every node in $U_i$  wishes to send $O(k \cdot n^{1/3})$ data. Thus, this communication is executed using \fullref{wacc:tools:routing} in $\tilde O(k \cdot n^{1/3} / c + 1)$ rounds. Upon receiving the data, each node $v \in V_i$ computes the at most smallest $k$ entries of $\hat P[v]$. Now, $\hat P$ is stored in a \emph{partial carrier configuration} (with every node $v$ having $C_v^{out} = \{v\}$), and thus we invoke \fullref{wacc:tools:transpose}, within $\tilde O(\sqrt{nk}/c + n/(k \cdot c) + 1)$ rounds, to ensure that $\hat P$ is stored in a carrier configuration $C$. Notice that $\tilde O(\sqrt{nk}/c + 1) = \tilde O(k \cdot n^{1/3}/c + n^{2/3}/c + 1)$, and so we are within the stated round complexity.
\end{proofof}
\section{Breaking Below \texorpdfstring{$o(n^{1/3})$}{n in 1/3} in  \hybrid }\label{sec:subCubeRootAlgorithms}
We show a $(1+\eps)$ approximation for weighted SSSP in the \hybrid model within $\tilde O(n^{5/17}/\eps^9)$ rounds, further implying that it is possible to compute a $(2+\eps)$ approximation for the weighted diameter in this number of rounds.
We achieve this by combining a simulation of our \wacc model and of the \bcc model. Roughly speaking, this incorporates \emph{density} awareness in addition to the \emph{sparsity} awareness discussed thus far.

A key approach in previous distance computations in the \hybrid model~\cite{AHKSS20, kuhn2020computing,censorhillel2020distance} is to construct an overlay \emph{skeleton graph}, and show that solving distance problems on such skeleton graphs can be extended to solutions on the entire graph. In a nutshell, given a graph $G = (V, E)$ and some constant $0 < x < 1$, a \emph{skeleton graph} $S_x = (M, E_S)$ is generated by letting every node in $V$ independently join $M$ with probability $n^{x-1}$. Two skeleton nodes in $M$ have an edge in $E_S$ if there exists a path between them in $G$ of at most $\tilde O(n^{1-x})$ hops. 
In particular, the nodes of $S_x$ are \emph{well spaced} in the graph and satisfy a variety of useful properties. The central distance related property is that pair of far enough nodes in $G$ have skeleton nodes at \emph{predictable intervals} on some shortest path between them.

Given such a skeleton graph, $S_x = (M, E_s)$, with $|M| = \Theta(n^x)$, previous work showed that it is possible in the \hybrid model to let the nodes in $M$ \emph{take control} over the other nodes in the graph and use \emph{all} the global bandwidth available to perform messaging between nodes in $M$. In essence, after $\tilde \Theta(n^{1-x})$ pre-processing rounds, the $\tilde \Theta(n)$ global bandwidth available to the \emph{entire} graph in each round of the \hybrid model is utilized such that every node in $M$ can send and receive $\tilde \Theta (n/|M|) = \tilde \Theta (n^{1-x})$ messages per round from any other node in $M$, in an amortized manner.\footnote{In $\tilde \Theta(n^{1-x})$ rounds, each node in $M$ can send and receive $\tilde \Theta(n^{2-2x})$ messages to other nodes in $M$.} For a given $v \in M$, we denote by $H_v$ the set of \emph{helper} nodes of $v$ which contribute their globally communication capacity to $v$; it is guaranteed that all nodes in $H_v$ are at most $\tilde O(n^{1-x})$ hops away from $v$ in $G$.

A formal definition of skeleton graphs encapsulating all of the above, is in \cref{def:skeleton} in \cref{app:skelConst}.
These skeleton graphs are built upon nodes randomly selected from the input graph $G$, and thus the number of edges in $S_x$ correlates to the density of neighborhoods in $G$ -- the graph $S_x$ is either sparse or dense, depending on $G$. We split into cases according to the sparsity of $S_x$, which can be computed using known techniques from the \local and \ncc models.

\textbf{Sparse $S_x$:} A hurdle that stands in the way for going
below $o(n^{1/3})$ rounds is that one must choose $x>2/3$, in order for the $\tilde \Theta(n^{1-x})$-round pre-processing step to not exceed the goal complexity.
However, in order to use the previously shown routing techniques, the identifiers of the nodes in $M$ must be globally known, a task which can be shown to take $\tilde \Omega(n^{x/2}) = \omega (n^{1/3})$ rounds. This leads to an \emph{anonymity} issue -- letting nodes in $M$ communicate with one another although no node in the graph knows the identifiers of all the nodes in $M$.

We overcome this anonymity problem by showing a routing algorithm which allows messaging over $S_x$ without assuming that the identifiers of the nodes in $M$ are globally known. This allows us to simulate the \wacc model over $S_x$.
By simulating algorithms from the \wacc model on $S_x$, we directly get a $(1+\eps)$-approximation for SSSP in $o(n^{1/3})$ rounds, with the exact round complexity dependent on the sparsity of $S_x$.
However, as $S_x$ approaches having $|M|^2 = \Theta(n^{2x})$ edges, the round complexity of the simulated \wacc model algorithms approaches $\tilde \Theta(n^{1/3})$. As such, using the techniques so far, we solve all cases except for \emph{very dense} skeleton graphs $S_x$.

\textbf{Dense $S_x$:} To tackle a \emph{dense} $S_x$, we present a \emph{density aware} simulation of the \bccshort model over $S_x$. The \bccshort model is simulated over $S_x$ in~\cite{AHKSS20} within $\tilde{O}(n^{1/3})$ rounds. Our observation is that as $S_x$ is more dense, broadcasting messages in the input graph can be made more efficient. In essence, as $S_x$ is more dense, neighborhoods in the original input graph $G$ are closely packed, and so when a node receives some message, it can efficiently share it with many nodes in its neighborhood. With this in hand, we can simulate the \bccshort algorithm from \cite[Theorem 8]{Becker2017NearOptimalAS} for approximate SSSP very quickly on \emph{dense} skeleton graphs.

\textbf{Tying up the pieces:} Each simulation result by itself is insufficient, as in the extreme cases each solution takes $\tilde O(n^{1/3})$ rounds. Yet, by combining them and using each when it is better, based on the sparsity of $S_x$, we achieve the resulting $\tilde O(n^{5/17}) = o(n^{1/3})$ round algorithm, for all graphs.

~\\
The outline of the rest of this section:
We first perform routing over skeleton graphs where the receivers of messages do not know which nodes desire to send them messages (\Cref{subsec:routingHybrid}). In \Cref{subsec:simHybrid}, we simulate the \wacc model in the \hybrid model. Next, we state that the \bccshort model can be simulated in the the \hybrid model, and defer the proof to \cref{app:littleO}.
Finally, in \Cref{subsec:SSSPHybrid}, we combine our various algorithms to yield the SSSP approximation result from which the weighted diameter approximation result also follows.

\subsection{Oblivious Token Routing}
\label{subsec:routingHybrid}

The following claim shows how to route unicast messages inside a skeleton graph, and is based on \fullref{thr:tokenRouting} shown in \Cref{app:ObliviousTR}. 

\begin{restatable}[Skeleton Unicast]{claim}{SklUnicast}\label{skeletons:unicast}
Given a graph $G$, a skeleton graph $S_x=(M, E_S)$, and a set of messages between the nodes of $M$, s.t. each $v \in M$ is the sender and receiver of at most $k = \tilde O(n^{2-2x})$ messages, and each message is initially known to its sender, it is possible to route all given messages within $\tilde O(n^{1-x})$ rounds of the \hybrid model.
\end{restatable}

\fullref{skeletons:unicast} is an extremely important requirement for showing our $o(n^{1/3})$ round algorithms. Previously, \cite{kuhn2020computing} required that every node knows how many messages every other node intends to send to it. In turn, this would require that for each $v \in M$, all the other nodes in $M$ know the identifier of $v$. But the latter can be shown to take $\omega(n^{x/2}) = \omega(n^{1/3})$ rounds, since $x > 2/3$ (as elaborated above). Therefore, the necessity of our strengthened claim follows.

\subsection{\texorpdfstring{\wacc}{AC(c)} and \bccshort Simulations in \hybrid}
\label{subsec:simHybrid}
\begin{restatable}[\wacc Simulation in  \hybrid]{theorem}{WaccSim}
\label{claim:wacc:simulation}
    Consider a graph $G=(V, E)$, and a skeleton graph $S_x=(M, E_S)$ of $G$, for some \emph{constant} $2/3 < x < 1$. Let $ALG_{AC}$ be an algorithm which runs in the \wacc model with capacity $c=\tildeTheta{n^{2-2x}}$ over $S_x$ in $t$ rounds. Then there exists an algorithm which simulates $ALG_{AC}$ within $\tildeTheta{t \cdot n^{1-x}}$ rounds of the \hybrid model over $G$, \whp Further, it is ensured that at the start of the simulation, every node in $M$ knows the \emph{communication tokens} of all its neighbors in $S_x$.
\end{restatable}

\begin{proofof}{\cref{claim:wacc:simulation}}We show that we can instantiate the \wacc model over $S_x$ and then show how to simulate each round.

In order to instantiate the \wacc model over the nodes $M$, the \wacc model definition asserts that every node $v \in M$ has an identifier in $[|M|]$ as well as a communication token whose knowledge enables other nodes to communicate with $v$. In order to ensure the condition related to identifiers is satisfied, we invoke \fullref{hybrid:treeConstruction} over $G$ and $S_x$ in $\tilde O(1)$ rounds. For the second condition, each node $v \in M$ uses its original identifier in the \hybrid model as its communication token in the \wacc model, which enables us to use \fullref{skeletons:unicast} in order to later simulate the rounds of the \wacc model. Finally, notice that it holds that every node in $M$ knows the communication tokens of all its neighbors in $S_x$, as the communication tokens are the \hybrid model identifiers.

In each round of the \wacc model, each node sends/receives at most $c$ messages to/from any other node, such that for each message it either knows the communication token of the recipient or the recipient is chosen independently, uniformly at random. We split each round of the \wacc model into two phases. First, we route the messages which use communication tokens, and second, we route those messages destined to random nodes. Since the communication tokens in the \wacc model are the identifiers from the \hybrid model, the first phase is implemented straightforwardly using \fullref{skeletons:unicast}, taking $\tildeBigO{n^{1-x}}$ rounds of the \hybrid model over $G$, as required.

For the second phase, denote by $R_v$ the messages which node $v$ desires to send to random targets. Node $v$ selects, uniformly and independently, $|R_v| \leq c$ random nodes from $G$, where each node is the target for a different message in $R_v$ -- from here on, we assume that a message in $R_v$ has the identifier of a random node in $G$ attached to it. For each helper $u \in H_v$, node $v$ assigns $c / |H_v| = \tilde O(n^{1-x})$ messages from $R_v$, denoted by $R_v^u$, to $u$. Within $\tilde O(n^{1-x})$ rounds, node $v$ uses the local edges of the \hybrid model in order to inform $u \in H_v$ about $R_v^u$. Next, each node $u$ uses the global edges of the \hybrid model in order to send the messages $R_v^u$ to their targets in $G$, within $\tilde O(n^{1-x})$ rounds, due to \fullref{thr:sendingToUniform}. Finally, given that a node $u\in V$ received a message in this way, node $u$ selects uniformly and independently at random a node such that $u \in H_v$ and forwards that message $v$. Notice that it can be the case that a certain $u \in V$ does \emph{not} help any node, and thus there does \emph{not} exist a node $v \in M$ such that $u \in H_v$ -- in such a case, $u$ will report back to whichever node sent it the message saying that it cannot forward it. However, the definition of a skeleton graph promises that the total number of helper nodes of nodes in $M$ is at least $\tilde\Omega{(n)}$ (Property~\ref{itm:helper:many} of \fullref{def:skeleton}). As every node assists at most $\tilde O(1)$ other nodes, this implies that at least a poly-logarithmic fraction of the nodes assist other nodes, and so if we repeat the above process and resend messages that \emph{bounced back}, within $\tilde O(1)$ iterations of the above algorithm, \whp, all messages are forwarded.

Notice that the above methodology does not produce a uniform distribution of receivers over the nodes $M$, since there might be some node $v \in M$ such that all of $H_v$ \emph{only} help $v$, and a node $u \in M$ such that all the nodes $H_u$ also help other nodes in $M$ (thus $u$ is less likely than $v$ to receive a random message). This happens even though each node only helps at most $\tilde O(1)$ other nodes and the number of nodes in $H_v$ is \emph{exactly the same} for each $v \in M$. Thus, we augment the probabilities with which a node $v \in M$ \emph{accepts} a message. Each node $v \in M$ observes its helpers $H_v$ and computes the probability, denoted $p_v$, that given that a uniformly chosen node $u \in H_v$ received a message, $u$ forwards the message to $v$. Now, the nodes $M$ utilize an Aggregate and Broadcast tool of~\cite[Theorem 2.2]{AGGHKL19} (see \fullref{thr:aggregateAndBroadcast}) in order to compute $p = \min_{v\in M}p_v$. Every node $v \in M$ now \emph{accepts} each message it received with probability $p/p_v$, independently. In the case that a node $v \in M$ rejects a message, it notifies the original sender of the message that the message was rejected -- this is done by sending messages in the reverse direction to the way they were sent previously. In the case that a node $v \in M$ hears that a message it sent was rejected, it will attempt to resend it by repeating the entire algorithm above. As every node $u \in V$ helps at most $\tilde O(1)$ nodes $v \in M$, it holds that $p = \tilde \Omega(1)$, and so within $\tilde O(1)$ iterations of the above algorithm, \whp, every message will be successfully delivered.
\end{proofof}

The following is proven in \Cref{app:littleO}.
\begin{restatable}[\bccshort Simulation in \hybrid]{theorem}{BCCSim}
\label{thr:bccInHybrid}
    Given a graph $G=(V, E)$, a skeleton graph $S_x=(M, E_S)$, for some \emph{constant} $2/3 < x < 1$, with average degree $k = \Theta(|E_S|/|M|)$, and an algorithm $ALG_{BCC}$ in the \bcc model which runs on $S_x$ in $t$ rounds, it is possible to simulate $ALG_{BCC}$ by executing $\tilde O(t \cdot (n^{2x-1}/\sqrt{k} + n^{1-x}))$ rounds of the \hybrid model on $G$. This assumes that prior to running $ALG_{BCC}$, each node $v \in S_x$ has at most $\tilde O(\deg_G(v))$ bits of input used in $ALG_{BCC}$, including, potentially, the incident edges of $v$ in $S_x$, and that the output of each node in $ALG_{BCC}$ is at most $O(t \log n)$ bits. 
\end{restatable}

\subsection{A \texorpdfstring{$(1+\eps)$}{1+epsilon}-Approximation for SSSP}
\label{subsec:SSSPHybrid}
We show an $\tilde O(n^{5/17}/\eps^9)$ round algorithm for a $(1+\eps)$-approximation of weighted SSSP in the \hybrid model. We begin by showing how to use the \wacc algorithm from \fullref{theorem:wacc:approxSSSP} in the \hybrid model for sparse skeleton graphs, with low maximal degree and even lower average degree, we then show an algorithm in the \hybrid model for graphs with low average degree, yet high maximal degree, and then finally we show how to use the \bcc algorithm from \cite[Theorem 8]{Becker2017NearOptimalAS} in the \hybrid model for dense graphs, with high average degree. Combining these claims using \fullref{theorem:approxSSSP} gives the desired result.

\begin{lemma}[SSSP with Low Average and Maximal Degrees]
\label{theorem:hybrid:approxSSSP:wacc}
Given a \emph{weighted} input graph $G = (V, E)$, and a skeleton graph $S_x=(M, E_S)$, s.t. the average and maximal degrees in $S_x$ are $k = \tilde O(n^{x/2})$ and $\Delta_{S_x} = O(n^{2-2x})$, respectively, a value $0 < \eps < 1$, and a source node $s \in M$, ensures that every node in $M$ knows a $(1+\eps)$-approximation to its distance from $s$ over the edges $E_S$, within $\tilde O((n^{11x/6 - 1} + n^{1-x})/\eps)$ rounds in the \hybrid model, \whp
\end{lemma}

\begin{proofof}{\cref{theorem:hybrid:approxSSSP:wacc}}
    Use \fullref{claim:wacc:simulation} to simulate in the \hybrid model over $G$ the SSSP algorithm from \fullref{theorem:wacc:approxSSSP} over $S_x$. Set capacity $c = \tilde \Theta(n^{2-2x})$ for a round complexity  of 
    $\tilde O(n^{1-x} \cdot (((n^x)^{5/6} + k \cdot (n^x)^{1/3})/(c\cdot \eps) + 1/\eps + \Delta_{S_x}/c)) = \tilde O(n^{1-x} \cdot (((n^x)^{5/6} + n^{x/2} \cdot (n^x)^{1/3})/(c\cdot \eps) + 1/\eps + \Delta_{S_x}/c)) = \tilde O((n^{5x/6 - 2 + 2x + 1 - x} + n^{1-x})/\eps + \Delta_{S_x} / n^{1-x} ) = \tilde O((n^{11x/6 - 1} + n^{1-x})/\eps)$.\end{proofof}

\fullref{theorem:wacc:approxSSSP} depends on the maximal degree, and so applying it to a skeleton graph with high maximal degree is inefficient. Thus, for sparse skeleton graphs with a high maximal degree, we show a \hybrid algorithm ensuring one node in the skeleton graph can learn all of the skeleton graph and inform the nodes of their desired outputs. The proof of the following appears in \Cref{app:littleO}.

\begin{restatable}[SSSP with Low Average and High Maximal Degrees]{lemma}{SSSPLAHM}
\label{theorem:hybrid:approxSSSP:highMaximalDegree}
Given a \emph{weighted} input graph $G = (V, E)$, and a skeleton graph $S_x=(M, E_S)$, for some \emph{constant} $2/3 < x \leq 12/17$, such that the average and maximal degrees in $S_x$ are at most $\tilde O(n^{x/2})$ and at least $\Delta_{S_x} = \omega(n^{2-2x})$, respectively, and a source node $s \in M$, ensures that every node in $M$ knows its distance from $s$ over the edges $E_S$, within $\tilde O(n^{1-x})$ rounds in the \hybrid model, \whp

\end{restatable}

We use our efficient \bccshort simulation for \emph{dense} skeleton graphs.

\begin{lemma}[SSSP with High Average Degree]
\label{thr:ssspByBccInHybrid}
    Given a weighted, undirected input graph $G = (V, E)$, a skeleton graph $S_x=(M, E_S)$, for some \emph{constant} $2/3 < x < 1$, such that the average degree in $S_x$ is at least $k=\tildeBigOmega{n^{x/2}}$, a value $0 < \eps < 1$, and a source node $s\in M$, ensures that every node in $M$ knows a $(1+\eps)$-approximation to its distance from $s$ over the edges $E_S$, within  $\tildeBigO{(n^{7x/4-1} + n^{1-x})/\eps^9}$ rounds in the \hybrid model, \whp
\end{lemma}
\begin{proofof}{\cref{thr:ssspByBccInHybrid}}
    \fullref{thr:bccInHybrid} simulates the SSSP approximation algorithm of \cite[Theorem 8]{Becker2017NearOptimalAS} on $S_x$. As the complexity of \cite[Theorem 8]{Becker2017NearOptimalAS} is $\tilde O(\eps^{-9})$ rounds of the \bccshort model, the simulation takes $\tilde O(\eps^{-9} \cdot (n^{2x-1}/\sqrt{k} + n^{1-x})) = \tilde O(\eps^{-9} \cdot (n^{2x-1 - (x/2)/2} + n^{1-x})) =  \tildeBigO{(n^{7x/4-1} + n^{1-x})/\eps^9}$ rounds of the \hybrid model.
\end{proofof}

Finally, we combine \fullref{theorem:hybrid:approxSSSP:wacc,theorem:hybrid:approxSSSP:highMaximalDegree,thr:ssspByBccInHybrid}.

\approxSSSP*

\begin{proofof}{\cref{theorem:approxSSSP}}
    Denote $x=12/17$. Construct a skeleton graph $S_x=(M, E_S)$ in $\tilde O{(n^{1-x})} = \tilde O(n^{5/17})$ rounds using \fullref{claim:skeleton}. Notice that \fullref{claim:skeleton} can ensure that source $s$ is also in $M$. Using \fullref{thr:aggregateAndBroadcast}, compute $k = \Theta(|E_S|/|M|)$ and the maximal degree in $S_x$, the value $\Delta_{S_x}$, in $\tilde O(1)$ rounds.
    
    First, ensure every node in $S_x$ knows a $(1+\eps)$-approximation to its distance from $s$ over the edge set $E_S$ only. Thus, selectively deploy \cref{thr:ssspByBccInHybrid,theorem:hybrid:approxSSSP:wacc,theorem:hybrid:approxSSSP:highMaximalDegree}, as follows. If $k = \tilde \Omega(n^{x/2})$, invoke \cref{thr:ssspByBccInHybrid} in $\tildeBigO{(n^{7x/4-1} + n^{1-x})/\eps^9} = \tildeBigO{(n^{(7/4)\cdot(12/17)-1} + n^{1-12/17})/\eps^9} = \tildeBigO{(n^{4/17} + n^{5/17})/\eps^9} = \tildeBigO{n^{5/17}/\eps^9}$ rounds. Otherwise, $k= O(n^{x/2})$, and so split the algorithm into two cases, according to $\Delta_{S_x}$. If $\Delta_{S_x} = O(n^{2-2x})$, invoke \cref{theorem:hybrid:approxSSSP:wacc}, in $\tilde O((n^{11x/6 - 1} + n^{1-x})/\eps) = \tilde O((n^{(11/6) \cdot (12/17) - 1} + n^{1-12/17})/\eps) = \tildeBigO{n^{5/17}/\eps}$ rounds, and otherwise invoke \cref{theorem:hybrid:approxSSSP:highMaximalDegree} in $\tilde O(n^{1-x}) = \tildeBigO{n^{5/17}}$ rounds.
    
    Finally, due to Property~\ref{itm:skeleton:distance} of the definition of a skeleton graph, we invoke \fullref{thr:computeMSSPFromMSSPOnSkeleton}, in $\tilde O(n^{5/17})$ rounds, to ensure every $v \in V$ knows a $(1+\eps)$-approximation to its distance from $s$ in $G$.
\end{proofof}

\fullref{theorem:approxSSSP} with $\alpha=1+\eps/2$ and \fullref{thr:ssspToDiameter} give the following.
\begin{restatable}[$(2+\eps)$-Approximation for Weighted Diameter in \hybrid]{theorem}{HybridWeightedDiameter}\label{thr:weightedDiameter2PlusEps}
    It is possible to compute a $(2+\eps)$-approximation for weighted diameter in $\tilde O(n^{5/17}/\eps^9)$ rounds in the \hybrid model, \whp
\end{restatable}

\section{Fast Distance Computations in the \congest Model}\label{sec:congest}
In \Cref{subsec:simCongest} we show how to simulate the \wacc model in \congest.
We then employ our simulation to simulate the \clique model in \congest (\Cref{subsec:CCsimCongest}), and to derive our distance computations in \congest (\Cref{subsec:distancesCongest}).

\subsection{\texorpdfstring{\wacc}{AC(c)} Simulation in \congest}
\label{subsec:simCongest}

\noindent\textbf{Key Principle.}
In the \congest model, each node $v$ can send or receive $\deg_G{(v)}$ messages in each round, implying a total bandwidth of $2m$ messages per round. We aim to build efficient distance tools which utilize this bandwidth \emph{completely}. Consider the \wacc[m/n] model, which also has a bandwidth of $2m$ messages per round. Potentially, by comparing the bandwidths, it could be possible to simulate one round of the \wacc model in single round of the \congest model. However, in the \congest model, nodes with degree $\littleO{m / n}$ can not send $m / n$ messages in a single round, regardless of how an algorithm tries to do this. 

To overcome this problem, we notice that the bandwidth of the nodes with degree at least $m / (4n)$ is at least $7m / 4$. Thus, the key principle we show is that the high degree nodes, denoted by $H$, can \emph{learn} all of the input of the low degree nodes, denote by $L$. The higher the degree of a node, the more nodes it simulates. Formally, we create an assignment $\rho \colon L\mapsto H$ where for every $\ell \in L$, the node $\rho(\ell)\in H$ simulates $\ell$, and we ensure that $\rho$ is \emph{globally known}.

We then simulate the \wacc[m/n] model using only the nodes in $H$, and finally we send back the resulting output to the nodes in $L$. To allow all-to-all communication between the nodes in $H$, we use the routing algorithms developed in \cite{Ghaffari2017DistributedMA,Chang2019DistributedTD} and stated in \cref{app:congest} and pay an overhead of $\tmixtwosqlgnlglgn$ rounds for the simulation.

Some problems, such as Scattered APSP, require the output to be stored on some node, but do not specify on which. We adapt for this case by allowing nodes to produce an \emph{auxiliary output}. After the simulation is over, every node $u$, given the communication token of any node $v$  can compute the identifier of node $w$ where the auxiliary output of $v$ is stored.

~\\
\noindent\textbf{Carrier Configuration and Communication Tokens.}
In addition to simulating the \wacc model in the \congest model, we let every node know the communication token of its neighbors as well as construct a carrier configuration directly in the \congest model. This benefits some graph problems greatly.

Note that in the simulation in the \hybrid model, the carrier configuration is constructed in the \wacc model itself. However, in the case of the \congest model, we cannot delegate this task efficiently to the \wacc model since building a carrier configuration requires every node to be able to send its incident edges to arbitrary nodes in the graph. Doing so takes $\Omega(\Delta \cdot n/m)$ rounds in the \wacc[m/n] model, yet only $O(\tmix \cdot n^{o(1)})$ rounds in \congest.

However, building a carrier configuration in the \congest model is also not directly possible, as a low degree carrier might learn $\bigOmega{m / n}$ edges of other nodes, which could take $\bigOmega{m / n}$ rounds. Therefore, instead of each node sending its edges directly to its carriers, it sends them to the nodes which simulate its carriers.

Overall this might sound like a back and forth process, as we simulate low degree nodes by high degree nodes and then split the edges of the high degree nodes among nodes which may be simulated nodes. 
However, using the \wacc model grants us the modular approach we aim for.

~\\
\noindent\textbf{Supergraphs.}
\fullref{theorem:wacc:hopsets} constructs hopsets and \cref{theorem:wacc:approxAPSP} computes Scattered APSP in the \wacc model, and both require the input graph to have $\bigOmega{n^{3/2}}$ edges.\footnote{This assumption can be removed at the expense of more complicated proofs, yet would not imply any speed-up for our end-results.} We aim to apply those results on general graphs and so augment $G$ with $\bigO{n}$ added nodes and $\bigTheta{n^{3/2}}$ added edges 
while preserving distances between the nodes of $V$ and ensuring that all added edges are globally known. We call the resulting graph $G'$ an \emph{$n^{3/2}$-supergraph}, build a carrier configuration holding $G'$, and apply the \wacc algorithms on $G'$.

\begin{definition}
    Given a weighted graph $G=(V, E, w)$ with $n=\size{V}$ and $m=\size{E}$, and a number $0\leq m'\leq n^2$, a weighted graph $G'=(V', E', w')$ is an $m'$-\emph{supergraph} of $G$, if $G'=(V', E', w')$ is obtained from $G$ by adding $\ceil{m' / n}$ new nodes and adding an infinite weight edge between every added node and every original node. 
\end{definition}

Clearly, a supergraph preserves the original distances.

We now simulate the \wacc model in the \congest model.

\begin{restatable}[\wacc Simulation in \congest]{theorem}{hcWACCSimulation}\label{lemma:hc:waccSimulation}
    Consider a graph $G$ and some constant $c$. Let $m'$ be some number s.t. $0\leq m'\leq n^2$.
    Let $k=m/n$ be the average degree of $G$ and let $A$ be an algorithm which runs in the \wacc model over $G$ in $t$ rounds. For each $v$ denote by $i_v\log{n}$ and $o_v\log{n}$ the number of bits in the input and output of the node $v$, respectively. Let $i_c=\max_{v\in V}\set{1 + i_v/\deg_G(v)}$ and $o_c=\max_{v\in V}\set{1 + {o_v/\deg_G(v)}}$ be the input and output \emph{capacities}: the minimum number of rounds required for any node to send or receive its input or output, respectively.
    
    There exists an algorithm which simulates $A$ within $(m'/m + i_c + \ceil{c/k}\cdot t + o_c)\tmixtwosqlgnlglgn$ rounds of the \congest model over $G$. 
    The above works even if $A$ requires a carrier configuration of $G'$ (an $m'$-supergraph of $G$) or communication tokens of neighbors as an input.
    Furthermore, each node $u$ might produce some unbounded auxiliary output, in which case the output is known to some (not necessary same) node $v$ such that each node $w$ can compute the identifier of $v$ given the communication token of $u$.

\end{restatable}

\begin{proofof}{\cref{lemma:hc:waccSimulation}}

\textbf{Identifiers:}
    First, compute $k=m/n$ in $O(D) = O(\tmix)$ rounds.
    By the definition of the \congest model, each node $v$ has an identifier $v\in\interval{n}$, denoted the \emph{original} identifier. Use \fullref{cor:hc:identifiers}, to compute a set of \emph{new} identifiers $ID_{new}\colon V\mapsto \interval{n}$. We abuse notation and denote by $\deg_G(i)$ the degree of node $v$ with identifier $ID_{new}(v)=i$. Each node $v$ locally computes for each new identifier $i\in \interval{n}$ its value $\floor{\log\deg_G{(i)}}$. A node $v\in V$ with degree that is less than $2^{\floor{\log{k}}-2}$ is a \emph{low} degree node $v\in L$, otherwise it is a \emph{high} degree node $v\in H$. According to the properties of $ID_{new}$, the new identifiers of nodes in $L$ are smaller than those of nodes in $H$. We abuse the notation and treat $L$ and $H$ as set of new identifiers.
    For each $i\in L$ denote $x_i=\floor{\log\deg_G{(i)}}$ and for each $j\in H$ denote $y_j=\floor{\log\deg_G{(i)}}$. 
    
    \textbf{Simulation Assignment:} The numbers $\size{L}, \size{H}$ and the sets $\set{x_i}_{i\in L}, \set{y_j}_{j\in H}$ satisfy the conditions of \fullref{lemma:hc:assignment}, since $\sum_{i\in L}{2^{x_i}} + \sum_{j\in H}{2^{y_j}}=\sum_{v\in V}{2^{\floor{\deg_G{(v)}}}}>\sum_{v\in V}{2^{\deg_G{(v)} - 1}}=2m/2=m$. 
    Hence, there is a partition of $L$ into $\size{H}$ sets $\set{I_j}$, satisfying $\size{I_j}\leq 4\cdot \floor{2^{y_j}/k}\leq 4\floor{\deg_G{(v)}/k}$. The node $j\in H$ simulates the nodes in $I_j$, and for simplicity it also simulates itself. Denote by $\rho(i)$ the new identifier of the node simulating the node with new identifier $i$.
    
    \textbf{Input:} For every $j \in H$, we now deliver the information node $j$ requires to simulate nodes in $I_j$. Each $u$ sends to its neighbors its new identifier $ID_{new}(u)$. As $u$ knows its new identifier $ID_{new}(u)$, it also knows the new identifier of $j$. Using \fullref{claim:hc:routing}, $u\in L$ sends to the node $\rho(u)$ its new and original identifier together with the new and original identifiers of its neighbors and input.
    A node $v\in H$, for each neighbor $w$ of node $u\in I_{v}$,can now locally compute the new identifier of $\rho(w)$.
    This invokes the algorithm from \fullref{claim:hc:routing} at most $\bigO{i_c}$ times. In each invocation, each $u\in L$ sends at most $\deg_G(u)$ messages and each node $v\in H$ receives at most $4\cdot (\deg_G(v)/k)\cdot k=\deg_G(v)\cdot \twosqlgnlglgn$ messages, as required.
    \textbf{Instantiation:} We now instantiate the \wacc model. As a communication token of node $v$ in the \wacc model, we use the concatenation of $v$, $\rho(v)$ and $ID_{new}(v)$. Clearly, identifiers are unique in $\interval{n}$ and communication tokens are unique of size $\bigO{\log{n}}$ bits. While pre-possessing, we already guaranteed that the node which simulates $u$ knows the communication token of all neighbors of $u$.
    Now the new identifier assignment $ID_{new}$ and simulation assignment $\rho$ satisfy the demands of \fullref{lemma:hc:waccSimulation:carrierConfiguration} and we use it to build carrier configuration.
    
    \textbf{Round Simulation:} During one round of the \wacc model, each node can send and receive at most $c$ messages. Each message is sent either to a random node or a node with a known communication token. We split into two phases. First, sending the messages to nodes with known communication tokens, and second sending the messages to random nodes.
    
    In the first phase, we use the fact that the new identifier of the destination is a part of the communication token. Each node $v\in H$ has to send and receive messages on behalf of all $I_v$. Thus, node $v\in H$ has to send and receive at most $(4\cdot \floor{\deg{(v)}/k} + 1)\cdot c\leq (4\cdot\deg{(v)}+1)\ceil{c/k}$ messages. It is therefore enough to invoke the algorithm from \fullref{claim:hc:routing} for $\tildeBigO{\ceil{c/k}}$ times to deliver all the messages \whp.
    
    In the second phase, for each message, we chose a new identifier independently and uniformly. Since for node $u$ we know the new identifier of $\rho(u)$, we also know where to route the message to. As $n$ nodes sample uniformly at most $c$ messages, each new identifier is sampled at most $\tildeBigO{c}$ times, \whp. Thus, the number of messages that a high degree node $v\in H$ has to send or receive is at most $\tildeBigO{\deg{(v)}c/k}$. Thus, \whp, $\tildeBigO{\ceil{{c/k}}}$ invocations of algorithm from \fullref{claim:hc:routing} suffice.
    
    \textbf{Main Output:} Finally, we send outputs back to the simulated nodes. This works in a similar manner, with a node $v\in H$ splitting the output of each node $u\in I_{v}$ into $\ceil{{o_u/\deg{(u)}}}$ batches of size at most $\deg{(u)}$. Notice that since $v$ receives the identifiers of all neighbors of $u$, it knows $\deg{(u)}$. Then, for $o_c$ rounds, each node uses \fullref{claim:hc:routing} to send one batch to each node it simulates.
    
    \textbf{Auxiliary Output:} The auxiliary output that some node $u$ produces is stored in the node $v=\rho(u)$ which simulates it. Since $\rho(u)=v$ is a part of the communication token of $u$, each node $w$ which knows the communication token of $u$, knows also $v$.

    \textbf{Round Complexity:} By \fullref{cor:hc:identifiers}, computing new identifiers takes $\bigO{\tmix + \log{n}}$ rounds. 
    By \fullref{claim:hc:routing}, sending the input requires $i_c\cdot \tmixtwosqlgnlglgn$ rounds, the simulation of the $t$ rounds of the \wacc model requires $\ceil{\frac{c}{k}}\cdot t \cdot \tmixtwosqlgnlglgn$ rounds, and sending the output back takes $o_c\cdot\tmixtwosqlgnlglgn$ rounds, \whp By \fullref{lemma:hc:waccSimulation:carrierConfiguration}, building the carrier configuration takes $\ceil{m'/m}\tmixtwosqlgnlglgn $ rounds \whp. Thus, the execution terminates in $(m'/m + i_c + \ceil{c/k}\cdot t + o_c)\tmixtwosqlgnlglgn$ rounds, \whp.
\end{proofof}

\subsection{Faster \clique Simulation}
\label{subsec:CCsimCongest}

The \wacc model is, in a sense, a generalization of the \clique model, directly implying the following.

\begin{claim}[\clique Simulation in \wacc]\label{lemma:wacc:clique}
There is an algorithm, which executes one round of the \clique model in the \wacc[c] in $\tildeBigO{\frac{n}{c}}$ rounds \whp 
\end{claim}
\begin{proofof}{\cref{lemma:wacc:clique}}
    Initially, for $\tildeBigO{\frac{n}{c}}$ rounds, each node sends its communication token and identifier to $c$ (not necessary distinct) randomly sampled nodes. By Chernoff and union bounds, all nodes receive the identifiers and communication tokens of all other nodes \whp
    For an additional $\tildeBigO{\frac{n}{c}}$ rounds, the nodes use the learned communication tokens to deliver the messages they have.
\end{proofof}

Combining \fullref{lemma:wacc:clique,lemma:hc:waccSimulation} implies a \emph{density aware} simulation of the \clique model in the \congest model -- the more edges the input graph has, the faster the simulation.

\begin{reptheorem}{cor:hc:cliqueSimulation}[\clique Simulation in \congest]Consider a graph $G=(V, E)$, and $A$ an algorithm which runs in the \clique model over $G$ in $t$ rounds. For each $v$ denote by $i_v\log{n}$ and $o_v\log{n}$ its number of input and output bits, respectively. Let $i_c=\max_{v\in V}\set{1 + i_v/\deg_G(v)}$ and $o_c=\max_{v\in V}\set{1 + {o_v/\deg_G(v)}}$ be the input and output \emph{capacities}: rounds required for any node to send or receive its input or output, respectively.
    Then $A$ can be simulated within $(i_c + {{n^2}/{m}}\cdot t + o_c)\cdot\tmixtwosqlgnlglgn$ rounds of the \congest model over $G$, \whp.
\end{reptheorem}

\subsection{Improved Distance Computations}
\label{subsec:distancesCongest}

We show an exact SSSP algorithm via a simulation of the algorithm from \fullref{lemma:hc:waccSimulation}.
\exactSSSPInHC*
\begin{proofof}{\cref{theorem:hc:exactSSSP}}
    The claim follows from simulating \fullref{theorem:wacc:exactSSSP} of \wacc[m/n] over $G$ using \fullref{lemma:hc:waccSimulation}, in $ (1 + \tilde O(m^{1/2}n^{1/6}/c + n/c + n^{7/6}/m^{1/2})\cdot\ceil{c / k})\tmixtwosqlgnlglgn={(n^{7/6}/m^{1/2} + n^2/m)}\cdot\tmixtwosqlgnlglgn$ rounds, \whp
\end{proofof}

Now, we define and approximate our first APSP relaxation.

\begin{definition}[Shortest Path Query Problem]\label{def:SPquery}
Given an input graph $G$, a \emph{query set} is a set of $q$ source-destination pairs $Q=\set{(s_i, t_i)}_{i=1}^{q}$ called \emph{queries}.
For each node $u$, the \emph{source and 
destination loads}, $u_s$ and $u_t$, respectively, are the number of times $u$ appears as a source and destination in $S$ divided by its degree. The maximum $\ell$ over all source and destination loads is the query set \emph{load}.

A \emph{shortest path query problem} is a query set of size $q$ and load $\ell$, s.t. every
$s_i$ knows the identifier of
$t_i$. The goal is to answer all queries, that is,
$s_i$ computes or approximates $d_G(s_i, t_i)$.

Given an input graph $G$, an algorithm is a \emph{shortest path query algorithm} if, after some pre-processing of $T_{pre-processing}$ rounds, given any query set of size $q$ and load $\ell$, the algorithm solves shortest path query problem within an additional number of $T_{query}$ rounds.
\end{definition}

We follow the approach of \cite{SparseMMPODC2019} in order to design a $(3+\eps)$-approximate shortest path query algorithm, using our methods from the \wacc model. For this we use the following important tool whose proof deferred to \cref{app:hc:knearest}.

\begin{restatable}[\knearest in \congest]{lemma}{hcKnearest}\label{claim:hc:knearest}
    Given a graph $G$, it is possible in the \congest model, within ${(k \cdot n^{4/3}/m + n^{5/3}/m + 1)}\cdot\tmixtwosqlgnlglgn$ rounds, \whp, to compute the distance $d_G(v, u)$ from every node $v$ to every node $u$ which is one of the closest $k$ nodes to $v$ (with ties broken arbitrarily).
\end{restatable}

\apspQuery*
\begin{proofof}{\cref{claim:hc:apspQuery}}
    To approximate the distance $d_G(s, t)$, we compute $\min\set{d_G^{n^{1/2}}(s, t), d_G(s, p(s)) + d_G(p(s), t)}$, where $p(s)$ is the closest node to $s$ in some globally known hitting set $A$ of all $N^{n^{1/2}}_G(s)$. Thus, while pre-possessing, we verify that each node $s$ knows $d_G(s, v)$ for each $v\in A$.

    \textbf{Pre-processing:}
    First, we execute the algorithm from \cref{claim:hc:knearest} with $k=n^{1/2}$ to get the distance to each node in $N^{n^{1/2}}_G(v)$. Now, each node enters $A$ independently with probability $\tildeBigO{n^{-1/2}}$. W.h.p., the set $A$ is of size $\tildeBigO{n^{1/2}}$ and is a hitting set for each $N^{n^{1/2}}_G(v)$. Let $\eps'=\eps/3$. We compute $(1+\eps')$-approximate MSSP from $A$ using
    $\tildeBigO{n^{1/2}}$ invocations of the $(1+\eps)$ approximate SSSP algorithm \cite[Corollary 5]{Ghaffari2018NewDA}. Each $v$ computes $p(v)\in A\cap N_{G}^{n^{1/2}}(v)$, which is the sampled node closest to $v$, which exists in the set of its $n^{1/2}$ nearest neighbors \whp.
    
   Computing  $n^{1/2}$-nearest takes ${(k \cdot n^{4/3}/m + n^{5/3}/m + 1)}\cdot\tmixtwosqlgnlglgn$ rounds, \whp The complexity of solving $n^{1/2}$-SSP is $n^{1/2}\cdot\tmix \cdot 2^{\bigO{\sqrt{log{n}}}}$ rounds \whp. Overall complexity of the pre-processing thus $(n^{1/2} + {n^{11/6}}/{m})\cdot\tmixtwosqlgnlglgn$ rounds \whp. 

    \textbf{Query:}
    Whenever node $s_i$ needs to approximate its distance to $t_i$, the node $s_i$ requests from $t_i$ the distance to $p(s_i)$, for which $t_i$ knows a $(1+\eps')$ approximation, $\tilde{d}(p(s_i), t_i)$, due to \cite[Corollary 5]{Ghaffari2018NewDA}.
    The node $s$ approximates its distance to $t_i$ as $\tilde{d}(s_i, t_i)=d(s_i, p(s_i)) + \tilde{d}(p(s_i), t_i)$. The approximation factor follows from \fullref{claim:3approximation}.
    To execute the routing, the nodes invoke \cref{claim:hc:routing}. The number of rounds for solving a query with load $\ell$ is  $\ell\cdot \tmixtwosqlgnlglgn $ \whp
\end{proofof}

By denoting $\delta$ as the minimal degree in the graph, one gets load $\ell \leq n/\delta$ and $m \geq n \cdot \delta$, which implies our main result \cref{cor:sublinearAPSP}.

{\renewcommand\footnote[1]{}\sublinearAPSP*}

Finally, we use our \wacc simulation together with \fullref{theorem:wacc:approxAPSP} to obtain our Scattered APSP algorithm in the \congest model.

{\renewcommand\footnote[1]{}\SAPSPInHC*}
\begin{proofof}{\cref{theorem:hc:sapsp}}
    Simulate \fullref{theorem:wacc:approxAPSP} using \fullref{lemma:hc:waccSimulation}. Split the output of each node $u$ to two parts. The first part, $s_u$, is $\tilde O(n^{1/2})$ bits encoding the communication tokens of nodes which store the distances from $u$. Thus, the output capacity is $o_c=\tildeBigO{n^{1/2}}$. The communication tokens decoded from $s_u$, allow $u$ to know where its distances are stored. The second part of the output of $u$ is the auxiliary output which is the distances it knows. Thus, the algorithm completes in 
    $((n^{11/6}/m + n^{1/3} + m / n)/\eps + n^{1/2})\tmixtwosqlgnlglgn$
    rounds, \whp 
\end{proofof}

\section{Discussion}
\label{sec:discussion}
We believe that additional problems in various fundamental distributed settings could be solvable using our infrastructure for sparsity aware computation. This is a broad open direction for further research.

With respect to the specific results shown here, a major goal would be to construct a  more sparse hopset in the \wacc model. 
Further, one could attempt to show sparse matrix multiplication algorithms which relax the assumption that the input matrices and output matrix are bounded by the same number of finite elements, as this could directly improve the complexity of our 
$k$-SSP algorithm.
Either of these improvements is likely to significantly reduce the round complexity of many of our end results, in both the \hybrid and \congest models.

\section*{Acknowledgements}
This project was partially supported by the European Union’s Horizon 2020 Research and Innovation Programme under grant agreement no. 755839. The authors would like to thank Michal Dory and Yuval Efron for various helpful conversations. We also thank Fabian Kuhn for sharing a preprint of \cite{kuhn2020computing} with us.

\bibliographystyle{plainurl}
\bibliography{References}

\newpage
\appendix

\section{Preliminaries -- Extended}\label{app:prelim}

\subsection{Definitions}
\textbf{Additional Models.} \sloppy{Throughout the paper, we also refer to the \clique and \bcc models. Both are synchronous models, where every node can communicate in each round with every other node by sending messages of $O(\log n)$ bits. In the \clique model, the messages between every pair of nodes can be unique, while in the \bcc model each node sends the same message to all the other nodes.}

Similarly to \cite{AGGHKL19} we define a \emph{distributive aggregate function} and an \emph{Aggregation-Problem}.
\begin{definition}[Aggregate Function]
An \emph{aggregate function} $f$ maps a multiset $S = \set{x_1, . . . , x_N }$ of input values to some value $f (S)$. An aggregate function $f$ is called \emph{distributive} if there is an aggregate function $g$ such that for any multiset $S$ and any partition $S_1,\cdots ,S_\ell$ of $S$, it holds that $f(S)=g(f(S_1),...,f(S_\ell))$.
\end{definition}
\begin{definition}[Aggregate-and-Broadcast Problem]
In the \emph{Aggregate-and-Broadcast Problem} we are given a distributive aggregate function $f$ and each node $u$ stores exactly one input value $val(u)$. The goal is to let every node learn $f(\Set{val(u)}_{u\in V})$.
\end{definition}

\subsection{Mathematical Tools}

Similarly to \cite{kuhn2020computing}, we will use families of $k$-wise independent functions.
\begin{definition}[Family of $k$-wise Independent Random Functions]\label{def:pseudoRandom}
    For finite sets $A, B$, let $\mathcal{H}=\set{h\colon A\mapsto B}$ be a family of hash functions. Then $\mathcal{H}$ is called \emph{$k$-wise independent} if for a random function $h\in H$ and for any $k$ distinct keys $\set{a_i}_{i=1}^{k}\subseteq A$, we have that $\set{b_i}_{i=1}^{k}\subseteq B$ are independent and uniformly distributed random variables in $B$.
\end{definition}

In particular, we are interested in a hash function on which nodes can agree within a small amount of communication.
\begin{claim}[Seed]{{\cite{Salil12}\cite[Lemma D.1]{kuhn2020computing}}}\label{lemma:pseudoRandom}
    For $A=\set{0,1}^a$ and $B=\set{0,1}^b$, there is a family of $k$-wise independent hash functions $\mathcal{H} = \set{h \colon A \mapsto B}$ such that selecting a function from $\mathcal{H}$ requires $k\cdot\max\set{a,b}$ random bits and computing $h(x)$ for any $x \in A$ can be done in $\poly(a,b,k)$ time.
\end{claim}

Unlike \cite{kuhn2020computing} we do not necessarily apply the hash function $h\in\mathcal{H}$ sampled from the family of $k$-wise independent random functions on distinct sets of arguments, but rather on a multiset of arguments where each argument appears at most $\tildeBigO{1}$ times. So, we show in \fullref{claim:pseudoHashOnMultiSet} the property similar to \cite[Lemma D.2]{kuhn2020computing}, which bounds the number of collisions.

\begin{claim}[Conflicts]\label{claim:pseudoHashOnMultiSet}
    There exists a value $k=\tildeTheta{1}$ such that for a sufficiently large $n$, given a function $h\colon A\mapsto B\in \mathcal{H}$ (with $\size{A},\size{B}=\tildeTheta{n}$) sampled from a family of $k$-wise independent hash functions, and a multi-set of keys $A'=\set{a_i}_{i=1}^{\tildeBigO{n}}$, in which each key appears in $A'$ at most $\tildeBigO{1}$ times, each value $b\in B$ appears in the multiset of values $B'=h(A')=\set{h(a_i)}_{i=1}^{\tildeBigO{n}}$ at most \tildeBigO{1} times \whp
\end{claim}
\begin{proofof}{\cref{claim:pseudoHashOnMultiSet}}
    Split $A'$ greedily into $\tildeBigO{1}$ sets of distinct keys $\set{A'_j}_{j=1}^{\tildeBigO{1}}$. 
    Consider some $A'_j$. Let $\mathcal{H}$ be a family of $k$-wide independent hash functions, for some $k$ to be determined. By the definition of  $h\in\mathcal{H}$, the random variables $B'_j=h(A'_j)=\set{h(a)\colon a\in A'_j}\subseteq B'$ are $k$-wise independent and uniformly distributed in $B$. Thus the probability to sample some particular $b\in B$ is $\frac{1}{\size{B}}=\tildeTheta{\frac{1}{n}}$. By a Chernoff Bound for variables with bounded independence \cite[Theorem 2]{SSS95} and a union bound over all $b\in B$ and $A'_j$, there is a large enough $k=\tildeTheta{1}$, such that each value $b\in B$ appears in $B'_j$ at most \tildeBigO{1} times \whp Thus, each value $b$ appears in $B'=\bigcup_{j=1}^{\tildeBigO{1}} B'_j$ at most $\tildeBigO{1}\cdot\tildeBigO{1}=\tildeBigO{1}$ times \whp
\end{proofof}

\subsection{Distance Tools}
\begin{claim}[APSP using \knearest and MSSP]\label{claim:3approximation} (see e.g. \cite
{SparseMMPODC2019})
    Let $G=(V, E, w)$ be a weighted graph, let $c$ be a constant, let $k$ be a value, and let $A$ be a set of nodes marked independently with probability at least $k^{-(c + 1)\log{n}}$.

    With probability at least $1 - n^{-c}$, it holds that $N^k(v)\cap A\ne\emptyset$. Denote by $p(s)\in A\cap N^k(v)$ one of the closest nodes to $s$ in $A\cap N^k(v)$. Denote by $\tilde{d}\colon A\times V\mapsto \R$ the $\alpha$-approximate distance from $A$ to other nodes for some $\alpha$. With probability at least $1 - n^{-c}$, for any pair of nodes $s, t\in V$ it holds that $\hat{d}(s, t)=d(s, p(s)) + \tilde{d}(p(s), t)$ is a $3\alpha$-approximate weighted distance between $s$ and $t$.

\end{claim}

\section{The \waccfull Model -- Extended}\label{app:sec:wacc}

\Cref{subsec:waccAppTools} contains proofs of various technical tools for routing information in the \wacc model -- we note that if taken as black-boxes, its contents can be skipped without harming the understanding of the main contributions of this section. Then, we show how to build carrier configurations and how to work with them, in \Cref{subsec:waccAppCarconf}. We use sparse matrix multiplication \cref{theorem:wacc:sparseMM} to construct hopsets in \Cref{subsec:waccAppHopset}, which eventually allows us to obtain our fast algorithms for SSSP and MSSP in
\Cref{subsec:waccAppSSSP} and
\Cref{subsec:waccAppSSP}, respectively.

\subsection{General Tools}
\label{subsec:waccAppTools}
We show basic tools which are useful in the \wacc model, for overcoming the anonymity challenges, as well as for solving problems related to communication with limited bandwidth.

We introduce the following notation. Given a set of nodes $W \subseteq V$, denote by $Tokens(W)$ 
the set of pairs of communication tokens and identifiers of the nodes in $W$.

\subsubsection{Basic Message Routing}
\begin{lemma}[Routing]
\label{wacc:tools:routing}
Given a set of messages and a globally known value $k$, if each node desires to send at most $k$ messages and knows the communication tokens of their recipients, and each node is the recipient of at most $k$ messages, then it is possible to deliver the messages in $\tildeBigO{k / c + 1}$ rounds of the \wacc model, \whp 
\end{lemma}
\begin{proofof}{\cref{wacc:tools:routing}}
    Denote by $m_v$ the messages that node $v$ desires to send. We proceed in $\Theta{(\ceil*{k\log^2{n}/c})}$ rounds, where in each round each node $v$ samples messages from $m_v$ that are not yet sent, independently with probability $\Theta{(\frac{c}{k \log n})}$, and sends them to their destinations.
    The probability that some message is not sampled during this procedure is $\left(1 - \Theta{(\frac{c}{k \log n})}\right)^{\Theta{(\ceil*{k\log^2{n}/c})}}=\bigO{1/\poly{n}}$. Thus, by applying a union bound over all messages, each message is sent \whp
    
    For any given round, the probability for a specific message to be sent or received by some node is at most $\Theta{(\frac{c}{k \log n})}$ (it is zero for rounds after the one in which it has been sent). Thus, due to the independence between messages, by a Chernoff Bound and a union bound over senders, receivers and rounds, on each round, each node sends or receives at most $c$ messages \whp

\end{proofof}

\subsubsection{Anonymous Communication Primitives}

\begin{definition}[Communication Tree]\label{definition:communicationTree}
Given a graph $G=(V, E)$ and a node $s \in V$, a \emph{communication tree rooted at $s$} in $G$ is a $\tilde O(1)$-depth directed tree which is rooted at $s$ and spans $V$, such that each node has at most 2 edges directed away from it.
A \emph{communication tree over $W \subseteq V$}, satisfies the conditions above, yet, only spans $W$ and not $V$.
\end{definition}

A communication tree rooted at $s$ allows to efficiently broadcast 
messages from $s$ to the entire graph as well as compute aggregation functions.

We show it is possible to build many communication trees in parallel.

\begin{lemma}[Constructing Communication Trees]
\label{wacc:tools:treeConstruction}
Given a set of nodes $S$, it is possible to construct for each $s \in S$ a communication tree rooted at $s$, $T_s$, such that each node in the graph knows the edges incident to it in each tree. This takes $\tildeBigO{|S|/c + 1}$ rounds of the \wacc model, \whp
\end{lemma}

\begin{proofof}{\cref{wacc:tools:treeConstruction}}
Consider the task of constructing $T_s$ for a single node $s$. Node $s$ randomly samples two nodes, $v_1, v_2$, and tells them it is their parent in $T_s$. Nodes $v_1, v_2$ each sample two nodes and repeat this process. At each step, a node $v_i$ might sample a node $v_j$ which is already in $T_s$. In such a case, $v_j$ rejects the demand of $v_i$ to add it as a child. Thus, when building the next level of the tree, we repeat the choosing step $\tilde O(1)$ times, ensuring, \whp, that each node has two nodes as its children. Notice that this ensures, \whp, that at every level in $T_s$, except for the last, each node has exactly two children, and thus the depth of $T_s$ is $\tilde O(1)$ \whp Thus, in $\tilde O(1)$ rounds, a communication tree from $s$ which spans the entire graph is constructed. 

In each round, every node sends and receives $\tilde O(1)$ messages, \whp, thus we can perform this for $\tilde O(c)$ nodes in parallel, taking $\tilde O(|S|/c)$ rounds overall to build such trees for all $s \in S$.
\end{proofof}

\begin{lemma}[Message Doubling on Communication Trees]
    \label{wacc:tools:messageDoublingOnTrees}
    Let $W \subseteq V$ be a set of nodes, and $T_s$ a communication tree rooted at $s \in W$, which spans $W$. It is possible for $s$ to broadcast a set of $M$ messages to the entire set $W$ within $\tilde O(|M|/c+ 1)$ rounds of the \wacc model, \whp, while utilizing only the communication bandwidth of the nodes in $W$. Likewise, it is possible to compute $k$ aggregation functions on values of the nodes in $W$, in $\tilde O(k/c+ 1)$ rounds.
\end{lemma}
\begin{proofof}{\cref{wacc:tools:messageDoublingOnTrees}}
On the first round, $s$ sends to its children in $T_s$, $s_{\ell}$ and $s_r$, some set $Q_1 \subseteq M$, where $|Q_1| = c/2$. On the second round, $s_{\ell}$ and $s_r$ forward $Q_1$ to their children, while $s$ sends them some other such set $Q_2$. This continues for $|M|/c + \tilde O(1)$ rounds. Notice that every node sends and receives at most $c$ messages per round.

To solve $k$ aggregation functions, reversing the flow of messages in the above algorithm suffices.
\end{proofof}

\begin{lemma}[Synchronization]
\label{wacc:tools:synchronize}
In the \wacc model, given a communication tree $T$ rooted at some node $s$ and assuming that every node $v$ has a value $val(v)$, it is possible in $\tilde O(1)$ rounds to ensure that $v$ knows the sum $prev(v)$ of all the values $val(u)$ of the nodes $u$ which come before it in the \emph{in-order} traversal of $T$.
Further, it is possible to solve $k$ instances of this problem in parallel in $\tilde O(k/c + 1)$ rounds.
\end{lemma}

\begin{proofof}{\cref{wacc:tools:synchronize}}
We treat only a single value, noting that allowing $k$ values follows as in the single value case, each node sends and receives only a constant number of messages per round. For a node $v$, denote by $v_{\ell}$, $v_r$ its left and right children in $T$, respectively, and by $v_p$ its parent.

Start from the leaves of $T$ and sum the total of the values upwards till the root. To clarify, a leaf $v$ sends $val(v)$ to $v_p$. Denote by $subtreeVal(v)$ the sum of all of the values of the nodes in the subtree of $T$ rooted at $v$. Once $v$ receives $subtreeVal(v_{\ell})$ and $subtreeVal(v_{r})$ from its children, it sends up to $v_p$ the sum $subtreeVal(v_{\ell}) + val(v) + subtreeVal(v_r)$. 

Then, the root $s$ of $T$ sets $prev(s) = subtreeVal(s_{\ell})$. Further, $s$ sends to $s_{\ell}$ the value zero, and sends to $s_r$ the sum $prev(s) + val(s)$. Then, every node $v$, upon receiving a value $i$ from $v_p$, sets $prev(v) = i + subtreeVal(v_{\ell})$, forwards the value $i$ to $v_{\ell}$, and sends $prev(v) + val(v)$ to $v_r$. This algorithm takes $\tilde O(1)$ rounds and achieves the desired result.
\end{proofof}

\begin{lemma}[Broadcasting]
\label{wacc:tools:broadcasting}
Let $M$ be a set of messages distributed across the nodes arbitrarily. It is possible to broadcast this set of messages to all nodes in $\tilde O(|M|/c + 1)$ rounds of the \wacc model, \whp
\end{lemma}
\begin{proofof}{\cref{wacc:tools:broadcasting}}
Construct a communication tree $T$ from the node with ID 1, and then send down $T$ the communication token of node 1, which implies that from now on every node can communicate with node 1. Using \fullref{wacc:tools:routing}, node 1 receives all of $M$ in $\tilde O(|M|/c + 1)$ rounds.

Once node 1 knows all of $M$, we send $M$ down along $T$ in $\tilde O(|M|/c + 1)$ rounds, using \cref{wacc:tools:messageDoublingOnTrees}.
\end{proofof}

Reversing the flow of messages in the proof of \cref{wacc:tools:broadcasting} proves \cref{wacc:tools:aggregation}. 

\begin{corollary}[Aggregation]
\label{wacc:tools:aggregation}
It is possible to solve $c$ aggregation problems in $\tildeBigO{1}$ rounds in the \wacc model $c$. That is, if every node $v$ has a vector of values $\{v_1, \dots, v_c\}$, denote $i\in[c]:\ S_i = \{v_i\ |\ v \in V\}$, and there are $c$ aggregation functions, $f_1, \dots, f_c$, it is possible to ensure within $\tilde O(1)$ rounds that all the nodes know the values $f_1(S_1), \dots, f_c(S_c)$.
\end{corollary}

\subsubsection{Communication Tools Within Groups of Nodes}
We show the following communication tools related to allowing subsets of nodes in the graph to communicate with one another.

\fullref{wacc:tools:grouping} allows \emph{grouping} together disjoint sets of nodes such that each node in a given set knows all the communication tokens of the other nodes in the set. \fullref{wacc:tools:groupBroadcasting} allows a single node in the set to quickly broadcast messages to or perform aggregation operations on the set.

\begin{lemma}[Grouping]
\label{wacc:tools:grouping}
Let $V_1, \dots, V_k \subseteq V$ be disjoint sets
where $1 \leq |V_i| \leq p$, for some $p$, and where every node $v$ knows if and to which set it belongs. It is possible in $\tildeBigO{k/c + p/c + 1}$ rounds of the \wacc model to ensure that, for each $i$, every node $v \in V_i$ knows $Tokens(V_i)$
\whp
\end{lemma}

\begin{proofof}{\cref{wacc:tools:grouping}}
Due to the definition of the sets, we know that $k \leq n$. Thus, the nodes $[k]$ (with identifiers $1, \dots, k$) broadcast $Tokens([k])$ using \fullref{wacc:tools:broadcasting} in $\tildeBigO{k/c + 1}$ rounds. Using \fullref{wacc:tools:routing}, for each $i \in [k]$, the nodes in $V_i$ send $Tokens(V_i)$ to node $i$, in $\tilde O(p/c + 1)$ rounds.

Fix $i \in [k]$. Node $i$ performs \emph{message duplication} to tell all nodes in $V_i$ the sets $Tokens(V_i)$, as follows. Node $i$
chooses some $m_i \in V_i$, and tells it, within $O(p/c + 1)$ rounds, $Tokens(V_i)$. Now, $i$ and $m_i$,
each proceed to each tell another node in $V_i$ all this information, doubling the number of nodes which know $Tokens(V_i)$ to 4. This continues for $\tilde O(1)$ iterations, each taking $O(p/c + 1)$ rounds.
\end{proofof}

\begin{claim}[Group Communication Tree Construction]
\label{wacc:tools:groupCommunicationTreeConstruction}
Given a set of nodes $W$, where every $v \in W$ knows $Tokens(W)$, it is possible to build a communication tree $T$ over $W$ such that the \emph{in-order} traversal of the tree imposes any desired ordering of the nodes $W$. This is done using local computation only and requires no communication.
\end{claim}
\begin{proofof}{\cref{wacc:tools:groupCommunicationTreeConstruction}}
    Since each node in $W$ knows $Tokens(W)$, this simply entails having every node locally decide which other nodes in $W$ are its parent, left, and right children in the output tree, $T$.
\end{proofof}

The following statement follows immediately from \fullref{wacc:tools:groupCommunicationTreeConstruction} and \fullref{wacc:tools:messageDoublingOnTrees}.
\begin{corollary}[Group Broadcasting and Aggregating]
\label{wacc:tools:groupBroadcasting}
Given a set of nodes $W$, where every $v \in W$ knows $Tokens(W)$, it is possible to allow a single node $s \in W$ to broadcast a set of $M$ messages to the entire set $W$ within $\tilde O(|M|/c)$ rounds of the \wacc model, \whp, while utilizing only the communication bandwidth of the nodes in $W$. Likewise, it is possible to compute $k$ aggregation functions on values of the nodes in $W$, in $\tilde O(k/c+ 1)$ rounds.
\end{corollary}

\fullref{wacc:tools:groupMulticasting} extends \fullref{wacc:tools:groupBroadcasting} in order to allow nodes to efficiently send \emph{multicast} messages within their given set.

\begin{lemma}[Group Multicasting]
\label{wacc:tools:groupMulticasting}
Given a set of nodes $W$, where every $v \in W$ knows $Tokens(W)$, and a value $k$ such that every node $v \in W$ desires to multicast at most $k$ messages to $m_v$ nodes in $W$, where each node $u \in W$ is the destination of messages originating in at most one node, it is possible to perform the communication task in $\tilde O(k/c + m/c + 1)$ rounds of the \wacc model, where $m$ is an upper bound for all $m_v$.
\end{lemma}
\begin{proofof}{\cref{wacc:tools:groupMulticasting}}
Assume, w.l.o.g., that the nodes in $W$ have identifiers $[|W|]$. Then, we use \fullref{wacc:tools:groupCommunicationTreeConstruction} to construct a communication tree $T$ over $W$, with the demand that the \emph{in-order} traversal of the tree will output the nodes of $W$ in ascending order from $1$ to $|W|$. Then, execute, in $\tilde O(1)$ rounds, the algorithm given by \fullref{wacc:tools:synchronize} over $T$ in order to ensure that every node $i \in W$ knows the sum $s_i = \sum_{j \in [i-1]} m_j$. 

Now, node $i$ tells node $s_i + 1$ the values $s_i$ and $m_i$. Then, node $i$ tells these values to $s_i + 2$, while node $s_i + 1$ tells them to node $s_i + 3$, and we continue in this fashion for $\tilde O(1)$ rounds until all nodes in $[s_i + 1, s_i + m_i]$ know the values $s_i$ and $m_i$. We now call these nodes the \emph{gateways} of node $i$. Notice that every node has distinct gateways -- that is, no node is a gateway of more than one node. 

Node $i$ now sends all of its $k$ messages to its gateways. To do so, it tells its first gateway all of its $k$ messages in $O(k/c + 1)$ rounds. Then, it tells its second gateway, while the first gateway tells the third gateway. This continues for $\tilde O(1)$ iterations, each taking $O(k/c + 1)$ rounds, until all the gateways of node $i$ know all the messages of $i$. Next, in $\tilde O(m_i/c + 1)$ rounds, node $i$ sequentially tells its $j^{th}$ gateway the identifier of the $j^{th}$ node which is set to receive the multicast messages from $i$. Finally, every $j^{th}$ gateway forwards the $k$ messages to the target which $i$ tells it to send the messages to, in $O(k/c + 1)$ rounds.
\end{proofof}

\subsection{Carrier Configurations}\label{app:wacc:carrierconfiguration}
\label{subsec:waccAppCarconf}
\begin{figure}[H]
\centering
\includegraphics[width=\linewidth]{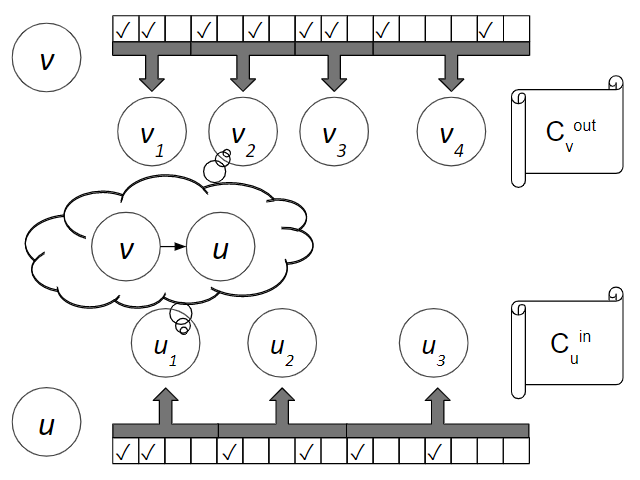}
\caption{The \emph{Carrier Configuration} Distributed Data Structure\\
In this example, $k = 2$, $C_v^{out} = \{v_1, v_2, v_3, v_4\}$, $C_u^{in} = \{u_1, u_2, u_3\}$. The two arrays denote which edges $v$ and $u$ have, with a \emph{checkmark} indicating the existence of an edge. The node $u_1$ holds information about $u$ and the first four nodes. That is, it knows that there are edges from the first two nodes to $u$ and that there are no edges from the following to nodes to $u$. Notice that in this case $v_2$ and $u_1$ both hold the edge $e = (v, u)$ and thus will know its weight, direction, the communication tokens of $v$ and $u$, and the communication tokens of each other ($v_2$ and $u_1$). Further, $v, u$ have communication trees (not shown), which allow them to perform broadcast and aggregate operations on all of $C_v^{out},$ $C_u^{in}$, respectively.}
\label{fig:carrierConfiguration}
\end{figure}

\begin{definition}[Carrier Configuration]\label{definition:carrierConfiguration} Given a set of nodes $V$, a data structure $C$ is a \emph{Carrier Configuration} holding a graph $G=(V, E)$ with average degree $k = |E|/|V|$, if for every node $v \in V$ the following hold:
    ~\\
    ~\\
    \indent\textbf{\emph{Carrier Node Allocations}}
    \begin{enumerate}
        \item{$C_v^{out}, C_v^{in}\subseteq V$ are the outgoing and incoming \emph{carrier nodes} of $v$, where $|C_v^{out}| = \ceil{\deg^{out}(v)/k}$, $|C_v^{in}| = \ceil{\deg^{in}(v)/k}$. \label[propertyCarrierConfiguration]{itm:carrierConfiguration:carrier}}
        
        \item{$v$ is in at most $\zeta\log n$ sets $C_i^{out}$ and $\zeta\log n$ sets $C_j^{in}$, for a constant $\zeta$, and knows which sets it is in.  \label[propertyCarrierConfiguration]{itm:carrierConfiguration:congestion}}
    
    \end{enumerate}
    
    \indent\textbf{\emph{Data Storage}}
    \begin{enumerate}
        \setcounter{enumi}{2}
        \item{An edge $e\in E$ is always stored alongside its weight and direction. \label[propertyCarrierConfiguration]{itm:carrierConfiguration:edge}}

        \item For each $u \in C_v^{out}$, there exists an interval $I_u \subseteq [n]$, such that $u$ knows all of the edges directed away from $v$ and towards nodes with IDs in the interval $I_u$, and there are at most $k$ such edges. It further holds that the intervals $\{I_u\ |\ u \in C_v^{out}\}$ partition $[n]$.
        Similar constraints hold for $C_v^{in}$. \label[propertyCarrierConfiguration]{itm:carrierConfiguration:edges}

        \item Node $v$ knows, for each $u \in C_v^{out} \cup C_v^{in}$, the two delimiters of the interval $I_u$. \label[propertyCarrierConfiguration]{itm:carrierConfiguration:vKnowsEdgeSplit}

    \end{enumerate}
    
    \indent\textbf{\emph{Communication Structure}}
    \begin{enumerate}
        \setcounter{enumi}{5}
        \item{For each $v$, the nodes in $\set{v}\cup C_v^{in}$ are connected by the communication tree $T_v^{out}$, implying that each node knows its parent and children in the tree. The same holds for nodes in $\set{v}\cup C_v^{out}$.
        \label[propertyCarrierConfiguration]{itm:carrierConfiguration:broadcastAggregate}}   
    \end{enumerate}
\end{definition}

The definition of the data structure is compatible with both directed and undirected graphs, where for undirected graphs each edge appears in both directions. We also use carrier configurations for holding matrices, where it can be thought that every \emph{finite} entry at indices $(i, j)$ in a matrix represents an edge from node $i$ to $j$. Each node $i$ stores the finite entries of row $i$ as edges outgoing from $i$, and the finite entries of column $i$ as edges incoming to $i$.

In order to use carrier configurations in the \wacc model, we must slightly extend the definition in order to address the usage of communication tokens. Thus, we present the following definition for \wacc Carrier Configurations, to which we often refer simply as `carrier configurations'.

\begin{definition}[\wacc Carrier Configuration]\label{wacc:def:carrierConfiguration}
Given a set of nodes $V$, a data structure $C$ is a \emph{\wacc carrier configuration} holding a graph $G=(V, E)$ with average degree $k = |E|/|V|$, if it is a \emph{Carrier Configuration} and, additionally, for every node $v \in V$ the following hold:
\begin{enumerate}

    \item{Node $v$ knows $Tokens(C_v^{in})$ and $Tokens(C_v^{out})$.
    \label[propertyWACCCarrierConfiguration]{itm:carrierConfiguration:vKnowsCarriers}}
    
    \item{Node $u$ knows the communication tokens and identifiers of each $v$ such that $u\in C_v^{in}$ or $u\in C_v^{out}$. \label[propertyWACCCarrierConfiguration]{itm:carrierConfiguration:carriersKnowV}}
    
    \item{Node $u$ knows the communication tokens of its parent and children in each communication tree that it belongs to as part of the data structure. \label[propertyWACCCarrierConfiguration]{itm:carrierConfiguration:knowsTree}}

    \item{An edge $e=(u, v)$ is always stored alongside the communication tokens of $u$ and $v$. \label[propertyWACCCarrierConfiguration]{itm:carrierConfiguration:communicationToken}}
    
    \item{Every $u \in C_v^{out}$ knows for every edge $e = (v, w)$ which it holds, the communication token of node $u' \in C_w^{in}$ which also holds $e$. Similarly, $u'$ knows the communication token of $u$. \label[propertyWACCCarrierConfiguration]{itm:carrierConfiguration:endpointsKnow}}
\end{enumerate}
\end{definition}

Throughout this entire section, the term `carrier configuration' refers to \fullref{wacc:def:carrierConfiguration}, unless otherwise specified.

\subsubsection{Initialization}
We show how to construct a carrier configuration, given that the edges of the graph are initially known to the nodes incident to them. As the stages taken during the construction can be partially reused in other algorithms which we show, we break up the construction into two statements -- \fullref{wacc:tools:assignRandomNodes} creates an \emph{empty} carrier configuration by allocating the carrier node sets and creating communication trees spanning them, and \fullref{wacc:tools:populateDS} transfers the data from nodes to their carrier nodes.

\begin{lemma}[Initialize Carriers]
\label{wacc:tools:assignRandomNodes}
    Given a graph $G = (V, E)$, with $k = |E|/|V|$, and $\Delta$ the maximal degree in $G$, where each node initially only knows $\deg_{G}^{in}{(v)},\ \deg_{G}^{out}{(v)}$ (but not even the edges incident to it), it is possible to assign for each node $v \in V$ sets $C_v^{in}$, $C_v^{out}$ which satisfy \cref{itm:carrierConfiguration:carrier,itm:carrierConfiguration:congestion,itm:carrierConfiguration:broadcastAggregate,itm:carrierConfiguration:vKnowsCarriers,itm:carrierConfiguration:carriersKnowV,itm:carrierConfiguration:knowsTree}, in $\tilde O(\Delta / (k \cdot c) + 1)$ rounds, \whp
    
    Note: We do not assume that $k$ and $\Delta$ are originally known to all the nodes.
\end{lemma}

\begin{proofof}{\cref{wacc:tools:assignRandomNodes}}
We perform two operations in this proof. First, we allocate the carrier node sets. Then, we construct communication trees across them. We show the case of outgoing carrier nodes, $C_v^{out}$, and note that the case of $C_v^{in}$ is symmetric.

\textbf{Carrier Allocations:}
We start by computing the values $k=\size{E}/\size{V}$ and $\Delta$, using \fullref{wacc:tools:aggregation}, in $\tilde O(1)$ rounds. Each node $v$ selects  $C_v^{out}$ by sending its communication token and identifier to $\ceil{\deg^{out}(v)/{k}}$ random nodes, and each node which $v$ reaches replies to $v$ with its communication token and identifier. The expected number of times a $u$ node is picked as carrier node is at most ${{1}/{n}}\cdot \sum_{v\in V}{\ceil*{{\deg^{out}{(v)}}/{k}}}=({2\size{E}/k}+n)/{n}=3$, and thus by applying a Chernoff Bound, there exists a constant $\zeta$ such that each node is picked by $\zeta\log{n}$ (not necessarily distinct) nodes in order to be in their carrier node set \whp This concludes the creation of the sets themselves, and satisfies \cref{itm:carrierConfiguration:carrier,itm:carrierConfiguration:congestion,itm:carrierConfiguration:vKnowsCarriers,itm:carrierConfiguration:carriersKnowV}.

The round complexity of this step is $\tilde O(\Delta/(k \cdot c) + 1)$, as each node $v$ initially sends $O(\deg{(v)}/k + 1)$ messages, and then replies to the at most $\tilde O(1)$ nodes which chose it for their carrier configuration sets.

\textbf{Communication Trees:}
Node $v$ locally builds a balanced binary tree $T_v$ which spans $C_v^{out}$, and sends to each $u\in C_v^{out}$ the communication tokens of its parent and children in $T_v$, taking $\tilde O(|C_{v}^{out}|/c + 1) = \tilde{O}(\Delta / (k \cdot c) + 1)$ rounds \whp Notice that $T_v$ is a communication tree (\cref{definition:communicationTree}) as it is of depth $\bigO{\log{n}}$, and thus we satisfy \cref{itm:carrierConfiguration:broadcastAggregate,itm:carrierConfiguration:knowsTree}. 
\end{proofof}

Now, we assume that we are given a carrier configuration which is still incomplete and only satisfies the conditions from the previous statement, and we complete it to a proper carrier configuration.

\begin{lemma}[Populate Carriers]
\label{wacc:tools:populateDS}
Let $G=(V, E)$ be a graph where each node $v$ knows all the edges incident to it and the communication tokens of all of its neighbors. Assume that we have a carrier configuration $C$ which is currently \emph{in-construction} and satisfies all of the properties of a carrier configuration, except for \cref{itm:carrierConfiguration:edge,itm:carrierConfiguration:edges,itm:carrierConfiguration:vKnowsEdgeSplit,itm:carrierConfiguration:communicationToken,itm:carrierConfiguration:endpointsKnow}. Then, it is possible, within $\tilde O(\Delta/c + 1)$ rounds, where $\Delta$ is the maximal degree in the graph, to reach a state where $C$ satisfies all of the properties of a carrier configuration, \whp
\end{lemma}
\begin{proofof}{\cref{wacc:tools:populateDS}}

We show the procedure for, $C_v^{out}$, and note that the case of $C_v^{in}$ is symmetric.

Node $v$ partitions the identifier space $\interval{n}$ into $|C_v^{out}|$ intervals, $I_{v}^{1}, \dots, I_{v}^{|C_v^{out}|}$, such that for every such interval $I_{v}^{i}$, the number of edges directed from $v$ to nodes with identifiers in $I_i$ is at most $k$. Denote by $\delta_{v}^{i}$ the edges from $v$ to nodes with identifiers in $I_{v}^{i}$.
For each node $u \in C_{v}^{out}$, node $v$ assigns $u$ a unique interval $I_{v}^{i_u}$, and sends to $u$ the delimiters of the interval as well as all the edges in $\delta_{v}^{i_u}$. Every edge is sent along with its weight, direction, and the communication tokens and identifiers of both of its endpoints.

The above procedure satisfies \cref{itm:carrierConfiguration:edge,itm:carrierConfiguration:edges,itm:carrierConfiguration:vKnowsEdgeSplit,itm:carrierConfiguration:communicationToken}. We proceed to analyze the round complexity of this step. Clearly, every node $v$ desires to send at most $O(\Delta)$ messages. To bound the number of messages each node receives, recall that by \cref{itm:carrierConfiguration:congestion}, each node $v$ is a carrier node in at most $\tildeBigO{1}$ carried nodes sets, and the number of messages it receives on behalf of each of them is at most $k\leq \Delta$. Thus, each node desires to receive $\tilde O(\Delta)$ messages. Therefore, by \fullref{wacc:tools:routing}, this stage can be executed in \tildeBigO{\Delta/ c + 1} rounds \whp

Finally, we need to satisfy \cref{itm:carrierConfiguration:endpointsKnow}. We assume that every node $v$ followed the above procedure to construct \emph{both} sets $C_v^{out}$, and $C_v^{in}$ which satisfy all properties except for Property~\ref{itm:carrierConfiguration:endpointsKnow}, and now we show how, at once, this property can be satisfied for both $C_v^{out}$, and $C_v^{in}$. For every node $v$, and every edge $e = (v, w)$ for some $w$, let $u \in C_v^{out}$ be the node holding $e$, then node $u$ asks node $w$ which node $x \in C_w^{in}$ holds $e$. Node $w$, which knows this information due to the above, replies to $u$ with the answer. As each node carries $\tildeBigO{k}=\tildeBigO{\Delta}$ edges, and each node receives $\Delta$ requests, one for each of its edges, by \fullref{wacc:tools:routing} it takes an additional $\tildeBigO{\Delta/ c + 1}$ rounds.
\end{proofof}

Applying \fullref{wacc:tools:assignRandomNodes}, followed by \fullref{wacc:tools:populateDS}, directly gives the following.

\begin{lemma}[Initialize Carrier Configuration]
\label{wacc:tools:initDS}
Given a graph $G=(V, E)$, where each node $v$ knows all the edges incident to it and the communication tokens of all of its neighbors in $G$, it is possible, within $\tilde O(\Delta/c + 1)$ rounds, where $\Delta$ is the maximal degree in the graph, to reach a state where $G$ is held in a carrier configuration $C$, \whp
\end{lemma}

\subsubsection{Basic Tools}
We show a basic communication tool within carrier configurations.
\begin{lemma}[Carriers Broadcast and Aggregate]\label{lemma:carrierConfiguration:broadcastAggregate}
    Let $G=(V, E)$ be a graph held in a carrier configuration $C$. In parallel for all nodes, every $v\in V$ can broadcast $c$ messages to all the nodes in $C_v^{in}$ and $C_v^{out}$, as well as solve $c$ aggregation tasks over $C_v^{in}$ and $C_v^{out}$. This requires $\tilde O(1)$ rounds.
\end{lemma}
\begin{proofof}{\cref{lemma:carrierConfiguration:broadcastAggregate}}
    Due to \cref{itm:carrierConfiguration:broadcastAggregate,itm:carrierConfiguration:knowsTree}, there is a communication tree spanning each $C_v^{in}$ and $C_v^{out}$, and every carrier node knows the communication tokens of its parent and children in the tree. Further, since each node is a member of at most $\tildeBigO{1}$ sets of carrier nodes, it is possible to apply \fullref{wacc:tools:messageDoublingOnTrees} simultaneously across all the communication trees in the carrier configuration, in $\tilde O(1)$ rounds, proving the claim.

\end{proofof}

We show the following helpful statement which enables nodes to query a carrier configuration and return to the classical state in which edges are known by the nodes incident to them.

\begin{lemma}[Learn Carried Information]\label{wacc:tools:learn}
Given a graph $G=(V, E)$ with average degree $k=\size{E}/\size{V}$ held in a carrier configuration $C$, it is possible for each node $v$ to learn all edges adjacent to it in $G$ in  $\tildeBigO{\Delta/c + 1}$ rounds \whp, where $\Delta$ is the maximal degree in $G$. It is possible to invoke this procedure for only outgoing or incoming edges separately, requiring $\tildeBigO{\Delta_{out}/c + 1}$, $\tildeBigO{\Delta_{in}/c + 1}$ rounds, respectively, where $\Delta_{out}$ is the maximal out-degree, and $\Delta_{in}$ is the maximal in-degree.
\end{lemma}
\begin{proofof}{\cref{wacc:tools:learn}}
    First, each node $v$ computes $\deg{(v)}$ by summing up the number of edges nodes in $C_{v}^{out}$ and $C_{v}^{in}$ hold. Then, the nodes compute the maximum of their degrees, the value $\Delta$. Every node in $C_{v}^{out}$ and $C_{v}^{in}$ sends to $v$ the edges incident to $v$ which it holds. Node $v$ desires to receive at most $O(\Delta)$ messages, and each node desires to send at most $\tilde O(\Delta)$ messages, as every node is the carrier of at most $\tilde O(1)$ nodes. Thus, due to \fullref{wacc:tools:routing}, this requires $\tilde O(\Delta/c + 1)$ rounds.

\end{proofof}

It is possible to extend \cref{wacc:tools:learn}, and show that if for a node $v$, both $v$ and all the carrier nodes of $v$ know some predicate over edges, then it is possible to send to $v$ only edges incident to it which satisfy the predicate. We formalize this, as follows.

\begin{lemma}[Learn Carried Information with Predicate]\label{wacc:tools:learn:predicate}
Assume that we are given a graph $G=(V, E)$ with average degree $k=\size{E}/\size{V}$ held in a carrier configuration $C$. If each node $v$ has a predicate $p_v$ over the edges incident to $v$, which both $v$ and the nodes $C_v^{out}, C_v^{in}$ know, then it is possible for each node $v$ to learn all edges incident to it in $G$ which satisfy the predicate. The round complexity for this procedure is  $\tildeBigO{\Delta_p/c + 1}$ rounds \whp, where $\Delta_p$ is the maximal number of edges incident to any node $v$ which satisfy $p_v$. 
\end{lemma}

\subsubsection{Merging Carrier Configurations}

We present a useful tool, which shows how to compute the point-wise minimum of two matrices. With respect to graphs, this can be seen as adding edges to a graph, and if an edge exists twice, then setting its weight to the minimum of the two. This tool can be used in order to \emph{merge} two carrier configurations.
\begin{lemma}[Merging]
\label{wacc:tools:matrixMin}
Let $V$ be a set of nodes which hold two $n \times n$ matrices $S, T$ in carrier configurations $A$, $B$, respectively. Denote by $P = \min \{S,\ T\}$ the matrix generated by taking the point-wise minimum of the two given matrices. It is possible within $\tilde O((k_S+k_T+h_A+h_B)/c + 1)$ rounds to output $P$ in a carrier configuration $C$, where the values $k_S, k_T$ denote the average number of finite elements per row of $S, T$, respectively, and the values $h_A, h_B$ denote the maximal number of carriers each node has in $A$, $B$, respectively, \whp
\end{lemma}
\begin{proofof}{\cref{wacc:tools:matrixMin}}

We show how to set up $C_v^{out}$, and note that the case of $C_v^{in}$ is symmetric. Thus, we sometimes drop the superscripts and denote $A_v = A_v^{out}, B_v = B_v^{out}, C_v = C_v^{out}$. A critical note is that in the following proof, when we denote $A_v \cup B_v$, if a node appears in both carrier sets, we count it twice in the union. That is, $A_v \cup B_v$ denotes a \emph{multiset}. Further, let $e$ be some edge which is held in $A$ or $B$ by some carrier node $w$. At the onset of the algorithm, $w$ \emph{attaches} its identifier and communication token to $e$ -- that is, whenever $e$ is sent in a message, it is sent along with these values as metadata.

~\\
\textbf{Proof Overview}

\emph{Goal:} Consider a node $v \in V$. Essentially, node $v$ has a sparse array (row $v$ in matrix $S$, denoted as $S[v]$) held, in a distributed fashion, over the nodes in $A_v$, and a sparse array (row $v$ in matrix $T$, denoted as $T[v]$) held over the nodes in $B_v$. Node $v$ wishes to merge these two arrays into one sparse array (the currently not-yet computed $P[v]$), and hold it in some (currently not allocated) carrier set $C_v$. In the case that an entry appears in both $S[v]$ and $T[v]$, it should keep the minimum of the values. 

\emph{Merging:} Initially, $v$ performs some merging mechanism in order to compute the sparse array $P[v]$. At the end of this step, the array $P[v]$ is distributed across the nodes $A_v$ and $B_v$, as we have yet to allocate $C_v$.

\emph{Constructing $C$:} Finally, we allocate the set $C_v$, and move the data of $P[v]$ from its temporary storage in the nodes $A_v$ and $B_v$ to be distributed across $C_v$. Further, several steps are taken to ensure that $C$ is a valid carrier configuration.

~\\
\textbf{Step: Merging}

Observe some node $v$. In this step, our goal is to compute $P[v]$, and store it in a convenient distributed representation across the nodes in $A_v \cup B_v$.

Initially, we desire for all the nodes in $A_v \cup B_v$ to be able to communicate with one another. Node $v$ knows the communication tokens and identifiers of the nodes in $A_v \cup B_v$ (\cref{itm:carrierConfiguration:vKnowsCarriers}), and broadcasts all of them to all the nodes $A_v \cup B_v$ in $\tilde O(|A_v|/c + |B_v|/c + 1) = \tilde O(h_A/c + h_B/c + 1)$ rounds using \fullref{lemma:carrierConfiguration:broadcastAggregate}.

Due to \cref{itm:carrierConfiguration:edges}, $S[v]$ is distributed across $A_v$ such that an interval $I_w = [w_b, w_e]$ corresponds to each $w \in A_v$, where $w$ holds all the finite elements in $S[v][I_w]$ (entries from index $w_b$ to index $w_e$ of $S[v]$). The same holds for $B_v$. For a set of carrier nodes $X$, denote $I(X) = \{I_w | w \in X\}$, and for a set of intervals $J$, denote by $D(J) = \{x, y | [x, y] \in J\}$ the set of delimiters of $J$. Due to Property~\ref{itm:carrierConfiguration:vKnowsEdgeSplit}, node $v$ knows $D(I(A_v) \cup I(B_v))$ and $D(I(B_v))$. We now perform the following steps.

\begin{enumerate}
    \item Node $v$ computes $J = \{[x, y]\ |\ x < y \in D(I(A_v) \cup I(B_v)) \land \nexists z \in D(I(A_v) \cup I(B_v))$, where $x < z < y \}$, that is, the partition of $[n]$ into intervals using all of the delimiters in $D(I(A_v) \cup I(B_v))$. Notice that $|J| \leq 2(|I(A_v)| + |I(B_v)|) = 2(|A_v| + |B_v|)$, and every $I \in J$ is contained in exactly one interval in $I(A_v)$ and in one interval in $I(B_v)$. Further, $\bigcup_{K\in J} = [n]$, since $\bigcup_{K\in I(A_v)} = [n]$ and $\bigcup_{K\in I(B_v)} = [n]$.
    \item Node $v$ broadcasts $D(J)$ to $A_v \cup B_v$ in $\tilde O(|J|/c + 1) = \tilde O(|A_v|/c + |B_v|/c + 1) = \tilde O(h_A/c + h_B/c + 1)$ rounds using \cref{lemma:carrierConfiguration:broadcastAggregate}.
\end{enumerate}

Notice that all the nodes in $A_v \cup B_v$ know the identifiers of one another (guaranteed above), and also all of $D(J)$. Thus, it is possible for the nodes in $A_v \cup B_v$ to perform local computation which allocates to each $u \in A_v \cup B_v$ two intervals, $K_{u}^{1}, K_{u}^{2} \in J$, and every node in $A_v \cup B_v$ knows that $u$ is assigned $K_{u}^{1}, K_{u}^{2}$. 

Now, we wish for $u$ to learn the finite entries in $S[v][K_{u}^{1}],\ T[v][K_{u}^{1}],\ S[v][K_{u}^{2}],\ T[v][K_{u}^{2}]$, and compute $P[v][K_{u}^{1}],\ P[v][K_{u}^{2}]$. To do so, we need to route the finite entries which $u$ requires from their current storage in the nodes $A_v \cup B_v$ to $u$. We bound the amount of information $u$ receives. For any interval $L_a \in I(A_v)$ and $L_b \in I(B_v)$, there are at most $k_S$ and $k_T$ finite elements in $S[v][L_a]$ and $T[v][L_b]$, respectively. Further, every interval $K \in J$ is contained in exactly one interval in $I(A_v)$ and one interval in $I(B_v)$, and so the number of finite elements in $S[v][K]$ and $T[v][K]$ is at most $k_S$ and $k_T$ respectively. Therefore, node $u$ desires to learn at most $O(k_S + k_T)$ finite elements. We now bound the amount of information node $u$ sends to other nodes in $A_v \cup B_v$ in order to let them learn their desired intervals. Node $u$ originally holds at most $O(k_S + k_T)$ finite elements, and each element is desired by exactly one node. Therefore, node $u$ sends at most $O(k_S + k_T)$ finite elements. Thus, we conclude that every node in $A_v \cup B_v$ sends and receives at most $O(k_S + k_T)$ messages to and from other nodes in $A_v \cup B_v$, showing that this step can be completed in $\tilde O((k_S + k_T)/c + 1)$ rounds, using \fullref{wacc:tools:routing}. 

Finally, we wish for all the nodes in $A_v \cup B_v$ to know, for each $K \in J$, how many finite entries are in $P[v][K]$. Using \cref{wacc:tools:routing}, every node $u \in A_v \cup B_v$ sends to $v$ the number of finite entries in $P[v][K_{u}^{1}]$ and in $P[v][K_{u}^{2}]$, within $\tilde O(|A_v|/c + |B_v|/c + 1) = \tilde O(h_A/c + h_B/c + 1)$ rounds. Then, $v$ broadcasts all of this information to $A_v \cup B_v$ using \fullref{lemma:carrierConfiguration:broadcastAggregate} in $\tilde O(|A_v|/c + |B_v|/c + 1) = \tilde O(h_A/c + h_B/c + 1)$ rounds.

~\\
\textbf{Step: Constructing $C$}

We perform several operations in this step. First, we invoke \fullref{wacc:tools:assignRandomNodes} w.r.t. $P$, in order to create the carrier sets $C_v$, which satisfy all of the properties of a carrier configuration, except for \cref{itm:carrierConfiguration:edge,itm:carrierConfiguration:edges,itm:carrierConfiguration:vKnowsEdgeSplit,itm:carrierConfiguration:communicationToken,itm:carrierConfiguration:endpointsKnow}. These remaining properties relate to populating the sets $C_v$ with data. Therefore, we then populate $C_v$ with the data pertaining to $P[v]$.

\textbf{Sub-step -- Invoking \cref{wacc:tools:assignRandomNodes}:} In order to invoke \cref{wacc:tools:assignRandomNodes} w.r.t. $P$, every node $v$ needs to know the number of finite entries in row $v$ of $P$ and column $v$ of $P$. Notice that $v$ can compute the number of finite entries in $P[v]$ by aggregating over the nodes $A_v \cup B_v$. In order to compute the number of finite entries in column $v$ of $P$, recall that at the beginning of the proof, we say that our analysis follows only the rows of the matrices, thus, inherently one also runs the algorithm up to this point on the columns of the matrices. Therefore, in a symmetric way, $v$ can know the number of finite entries in column $v$ of $P$. Next, we analyze the round complexity of invoking \cref{wacc:tools:assignRandomNodes} w.r.t. $P$. Denote by $k_P$ the average number of finite elements in a row of $P$, and, by aggregation, all of the nodes of the graph compute $k_P$. Notice that $k_P \geq k_S$, and $k_P \geq k_T$, as in each row the number of finite elements could only have increased due to the minimization operation. Further, the maximal number of finite elements in a row of $P$ is at most the maximal number of finite elements in a row of $S$ plus the maximal number of finite elements in a row of $T$. Thus, the round complexity is $\tilde O((k_S \cdot h_A + k_T \cdot h_B) / (k_P \cdot c) + 1) = O((h_A + h_B) / c + 1)$ rounds.

\textbf{Sub-step -- Ensuring \cref{itm:carrierConfiguration:edge,itm:carrierConfiguration:edges,itm:carrierConfiguration:vKnowsEdgeSplit,itm:carrierConfiguration:communicationToken}: }
First, we begin by computing the intervals $I(C_v)$. Notice that $|C_v| \leq |A_v| + |B_v|$, since $k_P \geq k_S,\ k_P \geq k_T$, and $A_v$ and $B_v$ can themselves hold the vectors $S[v],\ T[v]$ with each node in each set carrying at most $k_S,\ k_T$ finite elements, respectively. Now, we partition $[n]$ into $|C_v|$ intervals $I(C_v) = L_{1}, \dots, L_{|C_v|}$, such that the number of finite elements in every $P[v][L_i]$ is at most $k_P$. As the nodes in $A_v \cup B_v$ all know $D(J)$, as well for each $K \in J$, how many finite entries are in $P[v][K]$, every node in $A_v \cup B_v$ knows for every finite element in $P[v]$ which it holds the number of finite elements preceding it in $P[v]$. Thus, for every interval $L_i \in I(C_v)$, there exist some two nodes $x, y \in A_v \cup B_v$ such that $x$ can compute the left endpoint of $L_i$ and $y$ can compute the right. We show this for left endpoints as the proof for right endpoints is symmetric. The left endpoint of $L_i$ is the index of the $[(i - 1) \cdot k_P + 1]$-th finite entry in $P[v]$, and thus the node in $A_v \cup B_v$ which holds this finite element, knows the left endpoint of $L_i$. In $\tilde O(|C_v|/c + 1) = \tilde O(|A_v|/c + |B_v|/c + 1) = \tilde O(h_A/c + h_B/c + 1)$ rounds, the nodes in $A_v \cup B_v$ tell $v$ the contents of $D(I(C_v))$, using \fullref{wacc:tools:routing}. In the same round complexity, $v$ broadcasts $D(I(C_v))$ to $A_v$, $B_v$, and $C_v$, using \fullref{lemma:carrierConfiguration:broadcastAggregate}.

Now, we move the finite entries of $P[v]$ from the nodes in $A_v \cup B_v$ to the nodes in $C_v$. Node $v$ broadcasts the communication tokens and identifiers of all the nodes in $C_v$ to all the nodes in $A_v \cup B_v$, in $\tilde O(|C_v|/c + 1) = \tilde O(|A_v|/c + |B_v|/c + 1) = \tilde O(h_A/c + h_B/c + 1)$ rounds. The nodes in $A_v \cup B_v$ communicate all of the finite entries of $P[v]$ to $C_v$, each node knowing where to send the information which it holds as all the nodes in $A_v \cup B_v$ know $D(I(C_v))$. Each node sends or receives at most $O(k_S+ k_T + k_P) = O(k_S + k_T)$ messages, therefore routing these messages takes $\tilde O(k_S/c + k_T/c + 1)$ rounds using \fullref{wacc:tools:routing}. 

Observe that the above procedure ensures that $C$ satisfies \cref{itm:carrierConfiguration:edge,itm:carrierConfiguration:edges,itm:carrierConfiguration:vKnowsEdgeSplit,itm:carrierConfiguration:communicationToken}.

\textbf{Sub-step -- Ensuring \cref{itm:carrierConfiguration:endpointsKnow}:} Recall that, as stated at the beginning of the proof, we show how to construct $C_v^{out}$, while the case of $C_v^{in}$ is a symmetric algorithm. At this point, we require that all of the above be executed w.r.t. to both $C_v^{out}$ and $C_v^{in}$. This is due to the fact that in order to satisfy \cref{itm:carrierConfiguration:endpointsKnow} for $C_v^{out}$, we query the nodes of $C_v^{in}$ for some information which they compute above. Thus, we now show how \cref{itm:carrierConfiguration:endpointsKnow} is satisfied for $C_v^{out}$. In a symmetric way, it can be shown for $C_v^{in}$.

Let there be some edge $e = (v, w)$, for some $w \in V$, which is now held in $\gamma \in C_v^{out}$. Denote by $u \in A_v^{out} \cup B_v^{out}$, the node which originally held $e$ at the onset of the algorithm,
and recall that at the onset of the algorithm (stated at the beginning of the proof), we attach
to $e$ the communication token and identifier of $u$, and so $\gamma$ knows $u$.
W.l.o.g., assume that $u \in A_v^{out}$. Due to the fact that $A$ is a carrier configuration, node $u$ knows the communication token and identifier of node $w' \in A_w^{in}$ which also holds $e$. Again, assume that at the onset of the algorithm, node $u$ attached to $e$ the communication token and identifier of $w'$. Thus, node $\gamma$ knows the communication token and identifier of $w'$. 

As such, $\gamma$ asks $w'$ which node in $C_w^{in}$ holds $e$. Node $w'$ is able to answer this query, as all the nodes in $A_v^{in} \cup B_v^{in}$ know which intervals are held by which nodes in $C_v^{in}$. The answer to this query is exactly the information which node $\gamma$ needs in order to satisfy \cref{itm:carrierConfiguration:endpointsKnow}. We analyze the round complexity of this routing. Each node in $C_{v}^{out}$ sends queries only w.r.t. edges it holds as part of $C$, and each node in $A_v^{in}, B_v^{in}$ answers queries only w.r.t. edges it holds in $A, B$. Thus, this step can be executed in $\tilde O(k_S/c+ k_T/c + k_P/c + 1) = \tilde O(k_S/c + k_T/c + 1)$ rounds, using \fullref{wacc:tools:routing}.
\end{proofof}

\subsubsection{Partial Carrier Configuration}
We now prove a fundamental tool which can be roughly viewed as computing the transpose of a matrix. Notice that each entry of data is stored twice in a carrier configuration $C$. For instance, an edge $e = (v, w)$ is stored in both $C_{v}^{out}$, and $C_{w}^{in}$. We show that if only the \emph{outgoing} carrier sets $C_{v}^{out}$ are stored, one can complete the data structure to contain also the incoming carrier sets $C_{v}^{in}$. 

This is a very useful tool, as sometimes nodes can only compute the edges directed away from them, and not the edges directed towards them. For instance, in \fullref{theorem:wacc:sparseMM}, we reach a state where there are few edges directed away from every node, but potentially $\Theta(n)$ edges directed towards some nodes. If one were to simply invoke \fullref{wacc:tools:initDS}, this would require every node to learn all of the edges directed both away and towards it, which would incur a high round complexity. Instead, \fullref{wacc:tools:transpose} shows that given that every node $v$ has a partial carrier set holding edges directed away from it, the matching carrier set for edges directed towards $v$ can be allocated and directly populated with these edges without the edges ever being known to $v$ itself.

\begin{definition}[Partial Carrier Configuration]
\label{wacc:def:partialCarrierConfiguration}
Given a set of nodes $V$, a data structure $C$ is a \emph{partial carrier configuration} holding a graph $G = (V, E)$ if all the conditions of \fullref{wacc:def:carrierConfiguration} hold, yet, only for the outgoing edges. That is, each node $v \in V$ only has $C_v^{out}$.

Notice that \cref{itm:carrierConfiguration:endpointsKnow} is not demanded, as it requires the existence of both $C_v^{out}$ and $C_v^{in}$.

\end{definition}

\begin{lemma}[Partial Configuration Completion]
\label{wacc:tools:transpose}
Given a graph $G=(V, E)$ which is held in a partial carrier configuration $C$, there exists an algorithm which runs in $\tilde O(\sqrt{nk}/c + n/(k\cdot c) + 1)$ rounds, where $k = |E|/|V|$, and outputs a carrier configuration $D$ holding $G$, \whp
\end{lemma}

\begin{proofof}{\cref{wacc:tools:transpose}}
We assign $D_v^{out} = C_v^{out}$ for every $v \in V$, and thus we are required to show two items in this proof: how to allocate the sets $D_v^{in}$, and how to populate them with data.

~\\
\textbf{Allocating $D_{v}^{in}$:}

In order to allocate $D_{v}^{in}$, node $v$ needs to know $\deg_{G}^{in}{(v)}$. Denote $m = |E|$, $t = \sqrt{m}$, and $L = \{v \in V\ |\ \deg_{G}^{in}{(v)} \leq 2t\}$, $H = V \setminus L$. The sets $L$ and $H$ contain \emph{light} and \emph{heavy} nodes, respectively, yet, notice that at the current stage in the algorithm, no node knows whether it itself is in $L$ or in $H$, as it does not know $\deg_{G}^{in}{(v)}$. Our goal is to satisfy an even stronger condition -- to make \emph{every} node $v \in V$ know for \emph{every} $u \in V$ whether $u$ is in $L$ or $H$.

For a set of nodes $X \subseteq V$, denote by $\deg_{G}^{in}{(X)} = \sum_{v \in X} \deg_{G}^{in}{(v)}$. Let $\mathbb{V} = \{V_1, \dots, V_t\}$ be an arbitrary, hardcoded, globally known partition of $V$, where all the parts are of roughly equal size.
Using \fullref{wacc:tools:aggregation}, all nodes compute the values $\deg_{G}^{in}{(V_1)}, \dots, \deg_{G}^{in}{(V_t)}$ in $\tilde O(t/c + 1)$ rounds. Denote the set of parts in $\mathbb{V}$ which have low in-degree by $\mathbb{L} = \{V_i \in \mathbb{V} |  \deg_{G}^{in}{(V_i)} \leq 2t\}$, and the high in-degree ones by $\mathbb{H} = \mathbb{V} \setminus \mathbb{L}$. Given $v \in V$, if $v$ belongs to some set in $\mathbb{L}$, that is, $v \in V_i,\ V_i \in \mathbb{L}$, then \emph{certainly} $v \in L$. As $\mathbb{V}$ is hardcoded and globally known, and all the nodes know $\mathbb{L}$, all such nodes $v$ know that they are in $L$, and further all the nodes in the graph know this as well.

As $\deg_{G}^{in}{(V_1)} + \dots + \deg_{G}^{in}{(V_t)} = \deg_{G}^{in}{(V)} = m = t^2$, it holds that $|\mathbb{H}| < t/2$, implying, $\mathbb{L} \geq t/2$. Since the sets in $\mathbb{V}$ are of equal sizes, then we have guaranteed at that least half of the nodes in $V$ are now identified as belonging to $L$. These nodes are set aside, and we iterate over this procedure. In each iteration, at least half of the nodes remaining are marked as belonging to $L$, up until the final iteration where only nodes in $H$ remain. Thus, every node in the graph knows which nodes belong to $L$ and which to $H$.

Fix $v \in L$. It holds that $\deg_{G}^{in}{(v)} = O(t)$. Each node $u$ which holds an edge directed towards $v$ now sends that edge to $v$. This is done using \fullref{wacc:tools:routing}. We must show that $u$ knows the communication token of $v$. This follows from
\cref{itm:carrierConfiguration:communicationToken}, which states that every edge is stored alongside the communication tokens of both of its endpoints.
The execution of \cref{wacc:tools:routing} completes in $\tilde O(t/c + 1)$ rounds, as
every node receives at most $\tilde O(t)$ messages,
and
sends at most $O(k) = O(\sqrt{nk}) = O(\sqrt{m}) = O(t)$ messages. Thus, $v$ computes $\deg_{G}^{in}{(v)}$, as it knows all of the edges which are directed towards itself.

Observe $H$ -- we show that every $v \in H$ also computes $\deg_{G}^{in}{(v)}$. As the minimum in-degree of a node in $H$ is $2t$, and $t^2 = m$, we get $|H| = O(t)$. Further, recall that every node knows which nodes are in $H$. Therefore, using \fullref{wacc:tools:aggregation}, within $\tilde O(|H|/c + 1) = \tilde O(t/c + 1)$ rounds, every node in the graph knows the in-degree of every node in $H$.

Finally, we allocate $D_{v}^{in}$. We are given as input the partial carrier configuration $C$, allowing each node $v \in V$ to compute $\deg_{G}^{out}{(v)}$ in $\tilde O(1)$ rounds, using \fullref{lemma:carrierConfiguration:broadcastAggregate}. Thus, we invoke \fullref{wacc:tools:assignRandomNodes}, in $\tilde O(n/(k \cdot c) + 1)$ rounds, to create the sets $E_v^{out}, E_v^{in}$, which satisfy all of the properties of a carrier configuration, except for \cref{itm:carrierConfiguration:edge,itm:carrierConfiguration:edges,itm:carrierConfiguration:vKnowsEdgeSplit,itm:carrierConfiguration:communicationToken,itm:carrierConfiguration:endpointsKnow}. These remaining properties relate to populating the carrier node sets with input data. 
We throw away $E_v^{out}$ and set $D_v^{in} = E_v^{in}$.

~\\
\textbf{Populating $D_{v}^{in}$:}

Fix $v \in L$. As $v$ knows all the edges directed towards it, it sends these edges to its carrier in $\tilde O(\deg_{G}^{in}{(v)}/c + 1) = \tilde O(t/c + 1)$ rounds, using \fullref{lemma:carrierConfiguration:broadcastAggregate} and trivially completes \cref{itm:carrierConfiguration:edge,itm:carrierConfiguration:edges,itm:carrierConfiguration:vKnowsEdgeSplit,itm:carrierConfiguration:communicationToken}, by sending at most $\tilde O(1)$ messages to each node in $D_{v}^{in}$, requiring $\tilde O(\deg_{G}^{in}{(v)}/(k \cdot c) + 1) = \tilde O(t/(k \cdot c) + 1)$ rounds. The only challenging task is ensuring \cref{itm:carrierConfiguration:endpointsKnow}, which requires that for every edge $e = (w, v)$, the node $w \in D_{v}^{in}$ knows the communication token and identifier of $w' \in D_{w}^{out}$ which holds $e$. However, since every edge in this algorithm is sent directly from the carrier node which holds it (node $w'$ sends $e$), that carrier node can attach its own communication token and identifier to the edge itself when sending it, thus providing the information which the nodes in $D_{v}^{in}$ need in order to satisfy \cref{itm:carrierConfiguration:endpointsKnow}.

Partition $V$ into $t/k$ sets, $W_1 = [1, nk/t], \dots, W_{t/k} = [n - nk/t + 1, n]$, and use \fullref{wacc:tools:grouping} to ensure that for each $W_i$, every node in $W_i$ knows $Tokens(W_i)$, within $\tilde O(t/(k \cdot c) + (nk)/(t \cdot c) + 1) = \tilde O(t/c + 1)$ rounds. For each set $W_i$, denote some arbitrary, hardcoded $w_i \in W_i$ as the \emph{leader} of $W_i$, and in $\tilde O(t/(k \cdot c) + 1)$ rounds, broadcast the communication tokens and identifiers of all the leaders, using \fullref{wacc:tools:broadcasting}.

Fix $v \in H$.
Observe that $|D_{v}^{in}| \geq \deg_{G}^{in}{(v)}/k \geq t/k$. For each set $W_i$, the $i$-th carrier in $D_{v}^{in}$, sends its communication token and identifier to $w_i$, using \fullref{wacc:tools:routing}, in $\tilde O(t/c + 1)$ rounds, as every carrier sends at most one message and every leader receives at most $O(|H|) = O(t)$ messages. Then, leader $w_i$ broadcasts to $W_i$ all the communication token and identifiers of carrier nodes which it received, taking $\tilde O(|H|/c + 1) = \tilde O(t/c + 1)$ rounds, using \fullref{wacc:tools:groupBroadcasting}.

Finally, every node in $W_i$ which holds an edge directed towards $v$, tells the $i$-th carrier in $D_{v}^{in}$ about this edge using \cref{wacc:tools:routing}, in $\tilde O(nk/(t\cdot t) + 1) = \tilde O(t/c + 1)$ rounds, as each carrier node receives at most $|W_i| = nk/t$ messages, and each node sends at most $O(k) = O(\sqrt{nk}) = O(t)$ messages since $C$ is a partial carrier configuration. Now, the nodes $D_{v}^{in}$ know all of the edges directed towards $v$. Since $i < j$ implies that every $x \in W_i,\ y \in W_j$ hold $x < y$, within $\tilde O(nk/(t \cdot c) + |D_{v}^{in}|/c + 1) = \tilde O(t/c + n/(k \cdot c) + 1)$ rounds, the nodes in $D_{v}^{in}$ can rearrange the information stored in them, as well as communicate with $v$, in order to satisfy \cref{itm:carrierConfiguration:edge,itm:carrierConfiguration:edges,itm:carrierConfiguration:vKnowsEdgeSplit,itm:carrierConfiguration:communicationToken}.
To satisfy \cref{itm:carrierConfiguration:endpointsKnow}, an identical claim to the case of $v \in L$ can be used. 
\end{proofof}

\subsection{Efficient Hopset Construction}
\label{subsec:waccAppHopset}

We efficiently compute, in the \wacc model, a \emph{hopset}  \cite{Cohen00}, which allows approximating distance-related problems quickly.
We follow the general outline of \cite{SparseMMPODC2019}, and solve \knearest and \sdk, defined below, in order to construct the hopset. We solve these problems mainly using \fullref{theorem:wacc:sparseMM}.

However, in contrast to the \clique implementation, in the \wacc model we are met with additional challenges, as many operations which are trivial in the \clique model become highly complex.
For instance, upon computing the edges $E_H$ of a hopset $H$, one must add the edges to the graph -- an operation which is straightforward in the \clique model, yet requires \fullref{wacc:tools:matrixMin} in the \wacc model. Further,
in the \clique model, when we consider undirected graphs, once a node $v$ adds an edge $(v, u)$ to the graph then the edge $(u, v)$ is added as well, or updated to the minimum cost, if it exists already. To accomplish this in the \wacc model, one should invoke the algorithm in \fullref{wacc:tools:matrixMin} on the matrix and the transpose of the matrix. However, transposing a matrix is not trivial and we accomplish it due to the definition of the carrier configuration, which implies that whenever nodes hold a matrix $A$, they also implicitly hold $A^T$. This goes to show why various new tools are required in the \wacc model for this problem.

\begin{definition}[$(\beta, \eps)$-Hopset]
For a given weighted graph $G=(V, E)$, a $(\beta, \eps)$-hopset, $H=(V, E_H)$ is a set of edges such that paths of length at most $\beta$ hops in $G \cup H$ approximate distances in $G$ by a multiplicative factor of at most $(1+\eps)$. That is, for each $u, v \in V$, $d_G^\infty(u, v) \leq d_{G\cup H}^\beta(u, v) \leq (1+\eps)d_G^\infty(u, v)$, where $d_G^h(u, v)$ is the weight of the shortest path with at most $h$ hops between $u, v$ in $G$.
\end{definition}

We demonstrate how to construct the $(\log n / \eps, \eps)$-hopset $H$
over the input graph, where the number of edges in $H$ is $\tilde O(n^{3/2})$. This is done using \fullref{theorem:wacc:hopsets}.

\begin{theorem}[Hopset Construction]
\label{theorem:wacc:hopsets}
There exists an algorithm in the \wacc model, such that given a weighted undirected input graph $G=(V, E)$ with $n = |V|$ and $m = |E| = \Omega(n^{3/2})$, held in a carrier configuration $C$, and given some $0 < \eps < 1$, computes a $(\log n / \eps, \eps)$-hopset $H$, with $|H| = \tilde O(n^{3/2})$, and outputs $G'=(V, E \cup H)$ in a carrier configuration $C'$. The round complexity of this algorithm is 
$\tilde O((n^{5/6} / c + m / (n^{2/3}\cdot c) + 1)/\eps)$, \whp

\end{theorem}

Before proving \fullref{theorem:wacc:hopsets}, we prove several theorems related to the following two problems.

\begin{definition}[\knearest]
Given a graph $G=(V, E)$ and a value $k \in [n]$, in the \knearest problem, each node $v$ must learn $k$ of its closest neighbors in $G$, breaking ties arbitrarily.
\end{definition}

\begin{definition}[\sdk]
Given a graph $G=(V, E)$, a set $S \subseteq V$, a value $d \in [n]$, and a value $k \leq |S|$, in the \sdk problem, each node $v$ is required to learn its $k$ closest neighbors in $S$, while considering paths of up to $d$ hops only.
\end{definition}

We solve the \knearest and \knearest problems for the case where $k = \Omega(n^{1/3})$, as the round complexity of our solutions does not improve for $k = o(n^{1/3})$ due to pre-processing costs.

\begin{lemma}[\knearest Algorithm]
\label{wacc:distTools:knearest}
Given a graph $G=(V, E)$, where $n = |V|$, held in a carrier configuration $C$, and some value $k = \Omega(n^{1/3})$, it is possible in the \wacc model, within $\tilde O(k \cdot n^{1/3}/c + n^{2/3}/c + 1)$ rounds, \whp, to output a \emph{directed} graph $G'=(V, E')$ held in a carrier configuration $C'$, where $E'$ contains an edge \emph{from} every node $v \in V$ \emph{to} every node $u$ which is one of the closest $k$ nodes to $v$ (with ties broken arbitrarily), with weight $d_G(v, u)$. Notice that it can be the case that $E \not \subset E'$.
\end{lemma}
\begin{proofof}{\cref{wacc:distTools:knearest}}
    As shown in \cite{SparseMMPODC2019},
    the following process solves the problem. Take the adjacency matrix $A$ of $G$, and in each row keep the $k$ smallest entries (breaking ties arbitrarily), to create \emph{some}\footnote{Given $A$, there potentially are many options for $A'$, since ties can be broken arbitrarily.} matrix $A'$. Then, the matrix $A'$ is iteratively squared, for at most $\log n$ iterations, while after each product only the $k$ smallest entries in each row are preserved.
    We create \emph{some} matrix $A'$ from $A$. Fix $v\in V$. Node $v$ computes $\deg_{out}{(v)}$, in $\tilde O(1)$ rounds, using \fullref{lemma:carrierConfiguration:broadcastAggregate}. If $\deg_{out}{(v)} \leq k$, then $v$ uses \fullref{wacc:tools:learn} in $\tilde O(k/c + 1)$ rounds to learn all of the edges outgoing from it. Otherwise $\deg_{out}{(v)} > k$, and denote by $f(v, p, i)$ the edges directed away from $v$ with weight at most $p$ and towards nodes with identifiers at most $i$. Node $v$ computes two values, $p_v$ and $i_v$, such that $p_v$ is the maximal value where there exists an $i_v$ such that $|f(v, p_v, i_v)| = k$. Given any value $p$, node $v$ can compute $|f(v, p, \infty)|$ within $\tilde O(1)$ rounds using \fullref{lemma:carrierConfiguration:broadcastAggregate}. Thus, in $\tilde O(1)$ rounds, $v$ can compute $p_v$ using binary search. Likewise, $|f(v, p_v, i)|$ can be computed in $\tilde O(1)$ rounds for any $i$, and thus using binary search $v$ computes $i_v$.
    Then, node $v$ broadcasts $p_v$ and $i_v$ to $C_{v}^{out}$, and using \fullref{wacc:tools:learn:predicate}, within $\tilde O(k/c + 1)$ rounds, learns all of the edges in $f(v, p_v, i_v)$.
    We need to hold $A'$ in a carrier configuration in order to use it for matrix multiplication. Each node $v$ with $\deg_{out}{(v)} > k$ knows the entries of row $v$ in $A'$ -- they are $f(v, p_v, i_v)$. Each node $v$ with $\deg_{out}{(v)} < k$, locally adds arbitrary edges directed away from it with infinite weight, to have exactly $k$ edges directed away from it. We denote the new matrix created by this process by $A''$, and notice that $A''$ has the same properties w.r.t. distances as $A'$, since edges of infinite weight do not affect shortest paths. As each node holds \emph{exactly} $k$ edges directed away from it, the nodes themselves are a \emph{partial carrier configuration}, $D$, holding $A''$. That is, for each node $v$, we set $D_v^{out} = \{v\}$. We invoke \fullref{wacc:tools:transpose}, within $\tilde O(\sqrt{nk}/c + n/(k\cdot c) + 1) = \tilde O(k \cdot n^{1/3}/c + n^{2/3}/c + 1)$ rounds, since $k = \Omega(n^{1/3})$, in order to get a carrier configuration $D'$ which holds $A''$.
    
    Finally, we iteratively square $A''$ by applying \fullref{theorem:wacc:sparseMM}. That is, we compute $h(A'' \times A'')$, where $h$ takes a matrix and leaves only the $k$ smallest entries in each row. Then, we compute $h(h((A'')^2) \times h((A'')^2))$, and so forth. Repeating this procedure for $O(\log n)$ iterations results in an output matrix which holds in row $v$ edges only to some $k$ closest nodes to $v$. We perform $O(\log n)$ matrix multiplication, with each taking $\tilde O(k\cdot n^{1/3}/c + n/(k\cdot c) + 1) = \tilde O(k\cdot n^{1/3}/c + n^{2/3}/c + 1)$ rounds, since $k = \Omega(n^{1/3})$ and we always multiply two matrices with at most $k$ elements per row, due to \fullref{theorem:wacc:sparseMM}.
\end{proofof}

\begin{lemma}[\sdk Algorithm]
\label{wacc:distTools:sdk}
Given a graph $G=(V, E)$, where $n = |V|$ and $m = |E|$, held in a carrier configuration $C$, and given $S, d, k$, where $k = |S| = \Omega(n^{1/3})$ and $d \in [n]$, it is possible in the \wacc model, within $\tilde O(([k + m/n] \cdot n^{1/3} / c + n^{2/3}/c + 1) \cdot (d+1))$ rounds, \whp, to output a \emph{directed} graph $G'=(V, E')$ held in a carrier configuration $C'$, where $E'$ contains an edge \emph{from} every node $v$ \emph{to} every node $s \in S$ which is at most $d$ hops away from $v$, with weight $d_G^d(v, s)$.
Notice that it can be the case that $E \not \subset E'$.

It is assumed that the IDs of the nodes in $S$ are known to all of $V$.
\end{lemma}
\begin{proofof}{\cref{wacc:distTools:sdk}}
    In \cite{SparseMMPODC2019}, 
    it is shown that the following process solves the problem. Denote by $A$ the adjacency matrix of $G$. Denote by $A'$ the \emph{sparsified} adjacency matrix with edges only entering nodes in $S$. The matrix $B = A^{d-1} \times A'$ is the solution to the problem.
    
    We construct a carrier configuration which holds $A'$. Fix $v \in V$. Denote by $E(v, S)$ the edges from $v$ directed towards nodes in $S$. Node $v$ uses \fullref{wacc:tools:learn:predicate} to learn $E(v, S)$, in $\tilde O(|S|/c + 1) = \tilde O(k/c + 1)$ rounds. We construct a \emph{partial} carrier configuration $F$ which contains for each $v$ the edges $E(v, S)$, by setting $F_v^{out} = \{v\}$, since the average degree in $F$ is exactly $|S|$.\footnote{Assuming that if a node $v$ does not have an edge to node in $S$, it inserts a dummy edge with infinite weight}. Using \fullref{wacc:tools:transpose}, we turn $F$ into a carrier configuration $D$ holding $A'$, in $\tilde O(\sqrt{nk}/c + n/(k \cdot c) + 1) = \tilde O(k\cdot n^{1/3}/c + n^{2/3}/c + 1)$ rounds, since $k = \Omega(n^{1/3})$.
    Finally, we perform $d-1$ multiplications. We compute the product $B \times A'$ by invoking \fullref{theorem:wacc:sparseMM} with the carrier configurations $C, D$, to get a carrier configuration $E$, which holds $B \times A'$, in $\tilde O([k + m/n] \cdot n^{1/3} / c + n/(k\cdot c) + 1) = \tilde O([k + m/n] \cdot n^{1/3} / c + n^{2/3}/c + 1)$ rounds, since the average number of finite elements per row in $B$ is at most $O(m/n)$, in $A'$ it is at most $k$, and $k = \Omega(n^{1/3})$. Notice that while this invocation of \fullref{theorem:wacc:sparseMM} only computes the $k$ smallest entries in each row of $B \times A'$, there are only at most $k$ entries in $B \times A'$ which are finite -- all the columns not corresponding to nodes in $S$ do not contain finite values. Thus, it turns out that the invocation of \fullref{theorem:wacc:sparseMM} actually computes $B \times A'$ exactly. Thus, we now multiply $C$ by $E$ and repeat $d-2$ times until achieving the final result, taking $\tilde O(([k + m/n] \cdot n^{1/3} / c + n^{2/3}/c + 1) \cdot (d+1))$ rounds.
\end{proofof}

We turn our attention to proving \fullref{theorem:wacc:hopsets}.

\begin{proofof}{\cref{theorem:wacc:hopsets}} In the \wacc model, some preparation is necessary before constructing the desired hopset.
    ~\\
    
    \textbf{Initialization:}
    We initialize the hopset $H$ and denote by $D$ the carrier configuration holding it. We initialize $H$ with $\tilde \Theta(n^{3/2})$ arbitrary, hardcoded edges all with infinite weights. While adding arbitrary edges of infinite weight does not affect distances, it ensures that throughout the entire algorithm $H$ will contain $\tilde \Omega(n^{3/2})$ edges. Further, as no more than $\tilde O(n^{3/2})$ are added in the algorithm which follows, $H$ will always contain $\tilde \Theta(n^{3/2})$ edges, ensuring that the average degree in $H$ is always $\tilde \Theta(\sqrt{n})$, and thus the maximal number of carrier nodes that each node has is at most $\tilde O(n/\sqrt{n}) = \tilde O(\sqrt{n})$.
    
    Whenever a set of edges is added to $H$, it is assumed that if an edge is added from node $v$ to node $u$, then also an edge is added in the opposite direction. As such, assume that whenever we add sets of edges to $H$, we then reset $H$ to be $\min (H, H^T)$, by invoking \fullref{wacc:tools:matrixMin} on $D$. Notice that due to \fullref{wacc:def:carrierConfiguration}, if we set ${D'}_{v}^{in} = {D}_{v}^{out}$ and ${D'}_{v}^{out} = {D}_{v}^{in}$, we get that $D'$ is a carrier configuration holding $H^T$. As the maximal number of carrier nodes each node has in $D$ is $\tilde O(\sqrt{n})$, these invocations of \fullref{wacc:tools:matrixMin} take $\tilde O(n^{1/2}/c + 1)$ rounds. 
    ~\\
    
    \textbf{Construction:}
    We now begin computing the edges of $H$. Initially, we use \fullref{wacc:distTools:knearest} on $C$, with $k = \tilde \Theta(\sqrt{n})$ in order to get a carrier configuration $K$ with an edge from each node $v \in V$ to its $\tilde \Theta(\sqrt{n})$ nearest neighbors. This takes $\tilde O(n^{5/6}/c + 1)$ rounds. We add the edges from $K$ to $D$ using \fullref{wacc:tools:matrixMin} in $\tilde O(n^{1/2}/c + 1)$ rounds.
    
    Next, we sample nodes $S$, where $|S| = \tilde \Theta(\sqrt{n})$, by letting every node join $S$ independently with probability $\tilde \Theta(n^{-1/2})$, ensuring, \whp, that each node $v \in V$ holds in $D$ its distance to at least one node of $S$. We use \fullref{wacc:tools:broadcasting}, in $\tilde O(n^{1/2}/c + 1)$ rounds, in order to let every node $v \in V$ know all of $S$.
    
    We solve the \sdk problem with $S$ over the graph $G \cup H$, and add the resulting edges to $H$. We need to do this iteratively for $\tilde O(1)$ iterations. In each iteration, we invoke \fullref{wacc:distTools:sdk} with $S, d = O(1/\eps), k = |S| = \tilde \Theta(\sqrt{n})$ in $\tilde O(((\sqrt{n} + m/n) \cdot n^{1/3} / c + n^{2/3}/c + 1) \cdot (1/\eps + 1)) = \tilde O((n^{5/6} / c + m / (n^{2/3}\cdot c)+1)/\eps)$ rounds.
    We have now constructed the hopset $H$,
    and therefore can create $C'$ by executing \fullref{wacc:tools:matrixMin} on $C$ and $D$, taking $\tilde O((n^{1/2} + m/n)/c + 1)$ rounds, since we assume that $m = \Omega(n^{3/2})$, completing the proof.
\end{proofof}

\subsection{SSSP}
\label{subsec:waccAppSSSP}

We begin by showing how to perform Bellman-Ford iterations \cite{CLRS} in the \wacc model using carrier configurations. Given a source node $s$ in a graph $G$, in a Bellman-Ford iteration $i$, every node in $v \in G$ broadcasts to its neighbors $d^{i-1}_G(s, v)$, its distance to $s$ with at most $i-1$ hops, and then calculates $d^{i}_G(s, v)$ by taking the minimal distance to $s$ which it receives from its neighbors in this iteration.

\begin{lemma}[Bellman-Ford Iterations in \wacc]
\label{lemma:wacc:bellmanFord}
Given a (directed or undirected) weighted graph $G=(V, E)$ with average degree $k$ held in a carrier configuration $C$, and a source node $s \in V$, it is possible in the \wacc model, within $\tilde O((k/c + 1) \cdot i)$ rounds, \whp, to perform $i$ iterations of the Bellman-Ford algorithm on $G$ with $s$ as the source.
\end{lemma}

\begin{proofof}{\cref{lemma:wacc:bellmanFord}}

Fix $v \in V$. Node $v$ computes $d_{G}^{1}(s, v)$, within $\tilde O(1)$ rounds, using \fullref{lemma:carrierConfiguration:broadcastAggregate}. Then, to simulate the $j$-th iteration, node $v$ broadcasts to $C_{v}^{out}$ the value $d_{G}^{j-1}(s, v)$, in $\tilde O(1)$ rounds. Each node $u \in C_{v}^{out}$, for every edge $e = \{v, w\}$ which $u$ stores, sends to the node $u' \in C_{w}^{in}$ which stores $e$, the value $d_{G}^{j-1}(s, v) + w(e)$. Since the average degree in $G$ is $k$, and due to \fullref{wacc:def:carrierConfiguration}, it holds that each $u \in V$ sends and receives at most $k$ messages in this step, thus taking $\tilde O(k/c + 1)$ rounds, using \fullref{wacc:tools:routing}. Finally, $v$ sets $d_{G}^{j}(s, v)$ to be the minimum over all the values which nodes $C_{v}^{in}$ received in this iteration, within $\tilde O(1)$ rounds.
We repeat the above process $i-1$ times.
\end{proofof}

We show how to compute exact SSSP  in the \wacc model.

\begin{theorem}[Exact SSSP in \wacc]
\label{theorem:wacc:exactSSSP}
Given a weighted undirected graph $G=(V, E)$ with $n = |V|$ and $m = |E|$, held in a carrier configuration $C$, and a source node $s \in V$, it is possible in the \wacc model, within $\tilde O(m^{1/2}n^{1/6}/c + n/c + n^{7/6}/m^{1/2})$ rounds, \whp, to ensure that every node $v \in G$ knows the value $d_G(s, v)$.
\end{theorem}
\begin{proofof}{\cref{theorem:wacc:exactSSSP}}
It was proven in \cite{Nanongkai2014DistributedAA} that it is possible to solve exact SSSP on a weighted, undirected graph $G$ with a source $s$ by using the following steps. First, one solves the \knearest problem, for some $k$, and creates the graph $G'$ by starting with $G$ and adding weighted edges from each node to $k$ of its closest neighbors, with the weights equal to the weighted distance between the nodes in $G$. Then, one performs $O(n/k)$ Bellman-Ford iterations on $G'$ with $s$ as the source, and it is guaranteed that for each $v \in V$, it holds that $d^{O(n/k)}_{G'}(s, v) = d^n_G(s, v)$.

Thus, in order to solve exact SSSP, we choose the value $k' = m^{1/2}/n^{1/6}$ which balances the number of rounds required for our \knearest\footnote{\fullref{wacc:distTools:knearest} requires $k' = \Omega(n^{1/3})$, and this holds since we assume that the graph is connected, implying, $m \geq n$.} and Bellman-Ford implementations. 

Due to \fullref{wacc:distTools:knearest}, we can solve the $k'$-nearest problem in $\tilde O(k' \cdot n^{1/3}/c + n^{2/3}/c + 1) = \tilde O(m^{1/2}n^{1/6}/c + n/c + 1)$ rounds. This gives us a carrier configuration $D$, which for every node holds edges directed away from it to $k'$ of its nearest nodes. We need to add these edges from $D$ to the carrier configuration $C$ which holds $G$. Thus, we invoke \fullref{wacc:tools:matrixMin}, in order to get a carrier configuration $E$ which includes the edges from $C$ and from $D$. Notice that \fullref{wacc:tools:matrixMin} always takes at most $\tilde O(n/c + 1)$ rounds, and so this fits within our desired complexity.

Finally, we perform $O(n/k')$ Bellman-Ford iterations on $E$. Notice that the average degree in $E$ is $\Theta(m/n + k')$. Therefore, due to \fullref{lemma:wacc:bellmanFord}, this completes within $\tilde O(((m/n + k')/c + 1) \cdot n/k') = \tilde O(((m/n + k')/c)\cdot(n/k') + n/k') = \tilde O(((m/n + k')/c)\cdot(n/k') + n^{7/6}/m^{1/2}) = \tilde O((m/k' + n)/c + n^{7/6}/m^{1/2}) = \tilde O(m^{1/2}n^{1/6}/c + n/c + n^{7/6}/m^{1/2})$ rounds.
\end{proofof}

We now proceed to showing how to compute an approximation of SSSP in the \wacc model.

\begin{theorem}[$(1+\eps)$-Approximation for SSSP in \wacc]
\label{theorem:wacc:approxSSSP:noWrapper}
There exists an algorithm in the \wacc model, such that given a weighted, undirected input graph $G=(V, E)$, with $n = |V|$ and $m = |E| = \Omega(n^{3/2})$, held in some carrier configuration $C$, some $0 < \eps < 1$, and a source $s \in V$, ensures that each node knows a $(1+\eps)$-approximation to its distance from $s$. The round complexity of this algorithm is
$\tilde O((n^{5/6} / c + m / (n^{2/3}\cdot c) + 1)/\eps)$, \whp
\end{theorem}
\begin{proofof}{\cref{theorem:wacc:approxSSSP:noWrapper}}
We construct a $(\log n / \eps, \eps)$-hopset $H$ by using \fullref{theorem:wacc:hopsets} on $C$, in $\tilde O((n^{5/6} / c + m / (n^{2/3}\cdot c) + 1)/\eps)$ rounds, and obtain $G'=(V, E \cup H)$ held in a carrier configuration $D$. Due to the definition of $H$, for every $v \in V$, it holds that $d_G(s, v) \leq d_{G'}^{\log n / \eps} (s, v) \leq (1 + \eps) \cdot d_G(s, v)$. Notice that since $G$, $H$, and $G'$ are undirected, for each $v \in V$, the sets $D_v^{out}$, $D_v^{in}$ hold the same edges, and so it is irrelevant which one of them we use. Thus, we denote $D_v = D_v^{out}$ from here on.

We now perform $O(\log n / \eps)$ Bellman-Ford iterations on $G'$, in order to ensure that every node $v$ knows $d_{G'}^{\log n / \eps} (s, v)$ and as such a $(1+\eps)$-approximation to $d_G(s, v)$. To do so, we invoke \fullref{lemma:wacc:bellmanFord} on $G'$, which is held in $D$, and the source $s$, requiring $\tilde O((\log n/\eps) \cdot (\sqrt{n} + m/n)/c + 1) = \tilde O((n^{1/2} / c + m / (n\cdot c) + 1)/\eps)$ rounds, since $|H| = \tilde O(n^{3/2})$.
\end{proofof}

Finally, as our goal is to simulate our SSSP approximation algorithm in other distributed models directly,
we provide the following \emph{wrapper} statement which receives as input a graph where each node knows its neighbors, instead of a graph held in a carrier configuration.

\begin{theorem}[$(1+\eps)$-Approximation for SSSP in \wacc (Wrapper)]
\label{theorem:wacc:approxSSSP}
There exists an algorithm in the \wacc model, such that given a weighted, undirected input graph $G=(V, E)$, with $n = |V|$ and $m = |E|$, where each node $v$ knows all the edges incident to it, and the communication tokens of all of its neighbors in $G$, some $0 < \eps < 1$, and a source $s \in V$, ensures that each node knows a $(1+\eps)$-approximation to its distance from $s$. The round complexity of this algorithm is
$\tilde O((n^{5/6} / c + m / (n^{2/3}\cdot c) + 1)/\eps + \Delta/c)$, where $\Delta$ is the maximal degree in the graph, \whp
\end{theorem}

\begin{proofof}{\cref{theorem:wacc:approxSSSP}}
We wish to invoke \fullref{theorem:wacc:approxSSSP:noWrapper}, yet the main hurdle in our way is that it requires a graph with at least $\Omega(n^{3/2})$ edges. Therefore, we build $G'=(V, E')$
such that $\size{E'}=\bigOmega{n^{3/2}}$, $E\subseteq E'$ and for each $u, v\in V$ $d_{G'}(u, v)=d_{G}(u, v)$. We call a node $u$ with degree less than $\bigO{n^{1/2}}$ a \emph{low degree node}. Each low degree node $u$ \emph{meets} $\Omega(n^{1/2})$ nodes which are not its neighbors in $G$, and adds edges with infinite weight to those nodes. To meet new nodes, each low degree node $u$ sends its identifier and communication token to $\tildeBigO{n^{1/2}}$ random nodes, and each node which received the communication token of $u$, responds with its identifier and communication token. By \fullref{lemma:wacc:samplingUnique}, a low degree node $u$ meets at least $\bigOmega{n^{1/2}}$ unique nodes which are not its original neighbors in $G$, \whp. As such, $u$ connects itself with edges to these nodes which it meets. Notice, that each node was sampled at most $\tildeBigO{n^{1/2}}$ times \whp. Thus, the maximum degree, $\Delta'$, in $G'$ is $\Delta+\tildeBigO{n^{1/2}}$. The number of edges added is $\tildeTheta{n^{3/2}}$, \whp, implying that the number of edges in $G'$ is $m'=m + \tildeTheta{n^{3/2}}=\tildeBigOmega{n^{3/2}}$.

We initialize a carrier configuration $C$ from $G'$, using \fullref{wacc:tools:initDS} in $\tilde O(\Delta'/c + 1)=\tildeBigO{\Delta/c + n^{1/2}/c + 1}$ rounds. Then, we invoke \fullref{theorem:wacc:approxSSSP:noWrapper} in $\tilde O((n^{5/6} / c + m' / (n^{2/3}\cdot c) + 1)/\eps) 
=
\tilde O((n^{5/6} / c + m/ (n^{2/3}\cdot c) + 1)/\eps) 
$ rounds \whp.
\end{proofof}

\begin{lemma}[Sampling Unique Elements]{\label{lemma:wacc:samplingUnique}}
    Let $c$ be some constant. Let $S$ be a set of $n$ elements. Let $B\subseteq S$ be a set of at most $b\sqrt{n}$ bad elements. Denote by $G=S\setminus B$ the set of good elements. Let $r\sqrt{n}$ be a number of required good elements. Let $D$ be a sequence of length at least $d\geq\sqrt{n}r\frac{0.5\ln{n} - \ln{r} + 1}{0.5\ln{n} - \ln{(b + r)}} + \frac{c\ln{n}}{0.5\log{n} - \log{(b + r)}}$ elements, where each element is sampled independently uniformly and randomly from the set $S$. There are more than $r$ unique good elements in the sequence $D$ with probability at least $n^{-c}$. 
\end{lemma}
\begin{proofof}{\cref{lemma:wacc:samplingUnique}}
    We upper bound the number of sequences which contain $r$ or less different good elements. There are ${\binom{n - b\sqrt{n}}{r\sqrt{n}}}$ subsets of good elements which may appear. For each of these subsets there are $((b + r)\sqrt{n})^d$ sequences. Notice, there are a lot of sequences which are counted multiple times, however since we only need an upper bound it is enough for us. So the number of bad sequences is upper bounded by:

    \begin{align*}
        {\binom{n - b\sqrt{n}}{r\sqrt{n}}}\cdot((b + r)\sqrt{n})^d \leq 
        \left(\frac{(n - b\sqrt{n})\cdot e}{r\sqrt{n}}\right)^{r\sqrt{n}} \cdot \left((b + r)\sqrt{n}\right)^d \leq \\
        \frac{(n\cdot e)^{r\sqrt{n}}}{{(r\sqrt{n})}^{r\sqrt{n}}} \cdot \left((b + r)\sqrt{n}\right)^d \leq 
        n^{d-c},
    \end{align*}
    
    for $d$, which satisfies $\frac{\sqrt{n}^d}{(b + r)^d}\geq \left(\frac{\sqrt{n}e}{r}\right)^{r\sqrt{n}}n^c$. Solving this condition for $d$ results in $d\geq\sqrt{n}r\frac{0.5\ln{n} - \ln{r} + 1}{0.5\ln{n} - \ln{(b + r)}} + \frac{c\ln{n}}{0.5\log{n} - \log{(b + r)}}$ for large enough n.

    Since the total number of sequences is $n^d$, and all sequences are obtained with the same probability, the probability to get bad sequence is upper bounded by $n^{-c}$.
\end{proofof}

\subsection{\texorpdfstring{$k$}{k}-SSP and APSP}
\label{subsec:waccAppSSP}

We further show results pertaining to approximating distances from more than one source.

\begin{theorem}[$(1+\eps)$-Approximation for $k$-SSP in \wacc]
\label{theorem:wacc:approxMSSP}
There exists an algorithm in the \wacc model, such that given a weighted, undirected input graph $G=(V, E)$ with $n = |V|$ and $m = |E| = \Omega(n^{3/2})$, held in a carrier configuration $C$, some $0 < \eps < 1$, and a set of sources $M \subseteq V$, $k = |M| = \Omega(n^{1/3})$, outputs:
\begin{enumerate}
    \item 
        A \emph{directed} graph $G'=(V, E')$ held in a carrier configuration $D$, where $E'$ contains an edge \emph{from} every node $v$ \emph{to} every node $s \in M$, where the weight of the edge maintains $d_G(v, s) \leq w((v, s)) \leq (1+\eps)\cdot d_G(v, s)$.
        Notice that it can be the case that $E \not \subset E'$.
    \item 
        Every node $v \in V$, knows a $(1+\eps)$-approximation for $d_G(s, v)$ for every $s \in M$.
\end{enumerate}
The round complexity of this algorithm is $\tilde O((n^{5/6} / c + m / (n^{2/3}\cdot c) + k\cdot n^{1/3}/c+1)/\eps)$, \whp 

\end{theorem}
\begin{proofof}{\cref{theorem:wacc:approxMSSP}}
This proof is split into three parts.

First, we construct a $(\log n / \eps, \eps)$-hopset $H$ by using \fullref{theorem:wacc:hopsets} on $C$, in $\tilde O((n^{5/6} / c + m / (n^{2/3}\cdot c) + 1)/\eps)$
rounds, and obtain $G'=(V, E \cup H)$ held in a carrier configuration $C'$. Due to definition of $H$, for every $v \in V$, it holds that $d_G(s, v) \leq d_{G'}^{\log n / \eps} (s, v) \leq (1 + \eps) \cdot d_G(s, v)$.

Next, we make the IDs of the nodes in $M$ globally known within $\tilde O(k/c + 1)$ rounds, using \fullref{wacc:tools:broadcasting}. This enables us to invoke \fullref{wacc:distTools:sdk} on $G'$ with $S = M$, $k = |M|$, and $d = O(\log n / \eps)$, creating the carrier configuration $D$ which is described in the statement of this theorem. This requires  

$\tilde O(([k + m/n] \cdot n^{1/3} / c + n^{2/3}/c + 1) \cdot (d+1) = \tilde O(( m/(n^{2/3} \cdot c) + k \cdot n^{1/3}/c + n^{2/3}/c + 1) /\eps)$ rounds.

Finally, to satisfy the second guarantee, node $v$ learns all the edges held in $D_{v}^{out}$, using \fullref{wacc:tools:learn} within $\tilde O(k/c + 1)$ rounds.

\end{proofof}

\begin{theorem}[$(3+\eps)$-Approximation for Scattered APSP in \wacc]
\label{theorem:wacc:approxAPSP}
There exists an algorithm in the \wacc model, such that given a weighted, undirected input graph $G=(V, E)$ with $n = |V|$ and $m = |E| = \Omega(n^{3/2})$, held in a carrier configuration $C$, and some $0 < \eps < 1$, solves the $(3+\eps)$-Approximate Scattered APSP problem (\fullref{def:scatteredAPSP}) on $G$.

That is, the algorithm ensures that for every $u, v \in V$, there exist nodes $w_{uv}$, $w_{vu}$ (potentially $w_{uv} = w_{vu}$), which each know a $(3+\eps)$ approximation to $d_G(u, v)$, and node $u$ knows the identifier and communication token of node $w_{uv}$, while node $v$ knows the identifier and communication token of $w_{vu}$.

Further, for a given node $u$, the following hold:
\begin{enumerate}
    \item The set $W_u = \{w_{uv}\ |\ v \in V\}$ contains at most $\tilde O(n^{1/2})$ unique nodes.
    \item Node $u$ can compute a string of $\tilde O(n^{1/2})$ bits, $s_u$, such that using $s_u$, for any $v \in V$, it is possible to determine $w \in W_u$ such that $w = w_{uv}$. \item Denote $P_u = \{x \in V\ |\ \exists v \in V$ s.t. $u = w_{xv}\}$. It holds that $|P_u| = \tilde O(n^{1/2})$.
\end{enumerate}

The round complexity of this algorithm is 
$\tilde O((n^{5/6} / c + m / (n^{2/3}\cdot c) + 1)/\eps)$, \whp
\end{theorem}

The outline of the proof breaks into two parts -- \emph{initialization} and \emph{reshuffling}.

\textbf{Initialization:} First, every node $v$ computes $\Theta(n^{1/2})$ of its nearest neighbors, $K(v)$.
Next, a $(1+\eps)$-approximation for distances from $V$ to a random set $A$ of $\tilde \Theta(n^{1/2})$ nodes is computed, denoted by $\tilde{d}$. 

It holds that for each $v \in V$, \whp, $A \cap K(v) \neq \emptyset$, and so we denote by $p(v)$ a closest node to $v$ in $A \cap K(v)$.

\textbf{Reshuffling:} Due to \fullref{claim:3approximation}, for every two nodes $v, u$, it holds that $d_G(v, p(v)) + \tilde{d}(p(v), u)$ is a $(3+\eps)$ approximation to $d_G(v, u)$. As $G$ is undirected, both $\tilde{d} (p(v), u)$ and $\tilde{d} (u, p(v))$ can be used, and so we work with $\tilde{d} (u, p(v))$. Thus, we desire a state where for every two nodes $v, u$, there exists a node $w_{vu}$ which knows $d_G(v, p(v))$ and $\tilde{d}(u, p(v))$, and whose identifier is known to $v$, and a node $w_{uv}$ which knows $d_G(u, p(u))$ and $\tilde{d}(v, p(u))$ and whose identifier is known to $u$, concluding the proof.

\begin{proofof}{\cref{theorem:wacc:approxAPSP}}
~\\

\textbf{Initialization:} 
Invoke \fullref{wacc:distTools:knearest} on $G$,
with $k = \Theta(n^{1/2})$, within $\tilde O(k \cdot n^{1/3}/c + n^{2/3}/c + 1) = \tilde O(n^{5/6}/c + 1)$ rounds, to get a \emph{directed} graph $G'=(V, E')$ held in a carrier configuration $C'$, where $E'$ contains an edge \emph{from} every node $v \in V$ \emph{to} every node $u \in K(v)$ with weight $d_G(v, u)$, where $K(v)$ is a set of $k$ closest nodes to $v$. Using \fullref{wacc:tools:learn}, in $\tilde O(k/c) = \tilde O(n^{1/2}/c)$ rounds, every node $v$ \emph{itself} knows all the distances to the nodes in $K(v)$. 

Then, a random set $A$ of $\tilde \Theta(n^{1/2})$ nodes is selected, by letting each node join $A$ with probability $\Theta(n^{-1/2})$, and $Tokens(A)$ are broadcast, using \fullref{wacc:tools:broadcasting} within $\tilde O(|A|/c) = \tilde O(n^{1/2}/c + 1)$ rounds. As seen in \fullref{claim:3approximation}, it holds that for each $v \in V$, \whp, $A \cap K(v) \neq \emptyset$, and so we denote by $p(v)$ a closest node to $v$ in $A \cap K(v)$. Node $v$ knows $p(v)$, since it knows the distances to all the nodes in $K(v)$. Finally, invoke \fullref{theorem:wacc:approxMSSP} on $G$,
using $M = A$ as the source set, to compute a $(1+\eps)$ approximation for distances from $V$ to all of $A$, denoted by $\tilde{d}$, and requiring $O((n^{5/6} / c + m / (n^{2/3}\cdot c) + 1)/\eps)$ rounds. As per the specifications of \cref{theorem:wacc:approxMSSP}, $\tilde{d}$ is stored in a carrier configuration $D$ as edges in a \emph{directed} graph $G''$, where for each $v \in V, a \in A$, there is an edge with weight $w_{G''}(v, a) = \tilde{d}(v, a)$.
~\\

\textbf{Reshuffling:} 
For every node $a \in A$, denote by $C(a) =\{v \in V\ |\ p(v)=a\}$.
Notice that $a$ does \emph{not} know the set $C(a)$.

Primarily,
we
compute all the
values, $\{ |C(a)|\ |\ a \in A\}$, at once and make them known to all the nodes in $V$, using \fullref{wacc:tools:aggregation} within $\tilde O(|A|/c) = \tilde O(n^{1/2}/c + 1)$ rounds.

\textbf{Base Case (The Set $A_0$):} Denote by $A_0$ nodes $a \in A$ with $|C(a)| \leq 2n/|A| = \tilde \Theta(n^{1/2})$. Since the values $\{ |C(a)|\ |\ a \in A\}$ are globally known, every node knows which nodes are in $A_0$. Fix a node $a \in A_0$. Every node $v \in C(a)$ sends to $a$ the following values: (1) the identifier $v$, (2), the value $d_G(v, a)$, and (3), the communication token of $v$. Node $v$ knows all of these values, and also knows the communication token of $a$,\footnote{As $Tokens(A)$ are broadcast initially.} and so node $v$ can send these three messages to node $a$. This requires $\tilde O(|C(a)|/c + 1) = \tilde O(n^{1/2}/c + 1)$ rounds using \fullref{wacc:tools:routing}. Node $a$ broadcasts the information it receives to $D_{a}^{in}$, in $\tilde O(n^{1/2}/c + 1)$ rounds, using \fullref{lemma:carrierConfiguration:broadcastAggregate}. Observe some node $w \in D_{a}^{in}$. Notice that due to \cref{itm:carrierConfiguration:edges}, and due to the fact that there is an edge in $D$ from \emph{every} node in $V$ to $a$, then there exists some interval $I_w = [w_b, w_e] \subseteq [n]$, such that node $w$ knows the values $\{ \tilde{d}(v, a) | v \in I_w  \}$. Node $w$ sends to each $v \in C(a)$ the values $w_b$ and $w_e$, using \fullref{wacc:tools:routing}. This is possible as $w$ knows $Tokens(C(a))$,\footnote{This holds since $a$ broadcast the information it received, including $Tokens(C(a))$, to $D_{a}^{in}$.} and takes $\tilde O(|C(a)|/c + |D_{a}^{in}|/c + 1) = \tilde O(n^{1/2}/c + 1)$ rounds,\footnote{$|D_{a}^{in}| = O(n/(|A|) = \tilde O(n^{1/2})$, as the average degree in $D$ is $|A|$, and each node $a \in A$ has $n$ edges directed towards it in $D$.} as $w$ sends $O(|C(a)|)$ messages and every node in $C(a)$ receives $|D_{a}^{in}|$ messages. 

Fix some $v \in V$ where $p(v) \in A_0$, we claim that the output of the theorem is satisfied for $v$. Observe that given any $u \in V$, there exists some node $w_{vu} \in D_{p(v)}^{in}$ which both knows $d(v, p(v))$ and $\tilde {d}(u, p(v))$, and further $v$ knows which node $w_{vu}$ is, as it is the node such that $u \in I_{w}$. Finally, notice that we also satisfy that $W_u = \{w_{uv}\ |\ v \in V\}$ and $P_u = \{x \in V\ |\ \exists v \in V$ s.t. $u = w_{xv}\}$ contain $\tilde O(n^{1/2})$ \emph{unique} nodes, and that it is possible to condense into $\tilde O(n^{1/2})$ bits the information describing which node in $W_u$ is $w_{uv}$, for any $v \in V$, as this depends only on the intervals which each node in $D_{a}^{in}$ holds. Thus, all the conditions of the statement we are proving are satisfied for the case of $A_0$.

\textbf{Iterative Case (Sets $A_i$):} We proceed in $O(\log n)$ iterations. Fix the iteration counter $i \in [O(\log n)]$. Denote by $A_i$, the set of nodes $a \in A$ with $2^i n/|A| < |C(a)| \leq 2^{i+1}n/|A| = \tilde \Theta(2^{i+1}n^{1/2})$. Notice that $\sum_{a \in A}|C(a)| = n$, and therefore, $|A_i| \leq A/2^i$. Further, the values $\{|C(a)|\ |\ a \in A\}$ are globally known, implying that the contents of $A_i$ are globally known, and so all the nodes locally compute an assignment of $2^i$ unique nodes $H(a) \subseteq A$ to each $a \in A_i$. 

Fix $a \in A_i$. The nodes $D_{a}^{in}$ \emph{duplicate} the information which they hold, so that for each $a' \in H(a)$, the nodes $D_{a'}^{in}$ will contain the information held in $D_{a}^{in}$. We use the following observation: Since the graph is connected, for any $x \in A$, the nodes $D_{x}^{in}$ hold \emph{exactly} $n$ values,  $\{\tilde{d}(v, x)\ |\ v \in V\}$. Combining this with the average degree in $D$ being $\Theta(|A|)$, gives that $|D_{a}^{in}| = \tilde O(n/|A|) = \tilde O(n^{1/2})$. Node $a$ selects some node $h_1 \in H(a)$, and sends to $h_1$ the values $Tokens(D_{a}^{in})$, in $\tilde O(n^{1/2}/c + 1)$ rounds. Then, $h_1$ broadcasts $Tokens(D_{a}^{in})$ to $D_{h_1}^{in}$, in $\tilde O(n^{1/2}/c + 1)$ rounds, using \fullref{lemma:carrierConfiguration:broadcastAggregate}. Due to the observation above, it also holds that $|D_{a}^{in}| = n = |D_{h_1}^{in}|$, and so each carrier node in $D_{h_1}^{in}$ selects a unique carrier node in $D_{a}^{in}$, and within $\tilde O(n^{1/2}/c + 1)$ rounds, and learns all of the information which it holds. Thus, the nodes $D_{h_1}^{in}$ know all of the information which the nodes $D_{a}^{in}$ know. Then, nodes $a$ and $h_1$ selects nodes $h_2, h_3 \in H(a)$, and repeat the process where $a$ sends information to $h_2$ and $h_1$ sends information to $h_3$. This process is repeated for $\log |H(a)| = \log 2^i = i = \tilde O(1)$ iterations, until the information has been spread to all of $H(a)$ and their corresponding carrier nodes. 

Fix $a \in A_i$. The set $C(a)$ is split into $2^i$ roughly equal-sized parts, $C_1(a), \dots, C_{2^i}(a)$.
The main challenge is that no node in the graph knows all of $C(a)$, and therefore partitioning $C(a)$ into $C_1(a), \dots, C_{2^i}(a)$ is not trivial. We overcome this final challenge as follows. Every node in $w \in D_{a}^{in}$ observes $I_w$ (defined above as the interval such that $w$ knows $\tilde{d}(v, a)$ for all $v \in I_w$) and sends each $v \in I_w$ a message asking if it is in $C(a)$. Notice that this is possible since $w$ knows the communication tokens of all of $I_w$, due to \cref{itm:carrierConfiguration:communicationToken}. Since $|I_w| = \tilde O(|A|) = \tilde O(n^{1/2})$, then $w$ sends at most $\tilde O(n^{1/2})$ messages. Further, as $|A| = \tilde O(n^{1/2})$, each node in the graph receives at most $\tilde O(n^{1/2})$ messages, and so all of these messages may be routed in $\tilde O(n^{1/2}/c + 1)$ rounds using \fullref{wacc:tools:routing}. Thus, the identities of all of $C(a)$ are dispersed across the carrier nodes $D_{a}^{in}$. Similarly to the step above, this information is \emph{doubled}, in $\tilde O(n^{1/2}/c + 1)$ rounds, in order to make sure that for each $h \in H(a)$, the carrier nodes $D_{h}^{in}$ hold the identifiers of $C(a)$ as well. Finally, for each $h \in H(a)$ (w.l.o.g. nodes in $H(a)$ are numbered from 1 to $|H(a)|$ -- possible since $H(a)$ is globally known), node $h$ performs a binary search, by querying its carrier nodes $D_{h}^{in}$, in order to find the interval $J_h = [h_b, h_e]$ of nodes, such that the number of nodes in $C(a)$ with identifiers at most $h_b$ is at most $|C(a)|/h$, and the number of nodes in $C(a)$ with identifiers in the interval $J_v$ is $|C(a)|/|H(a)|$. These nodes form the set $C_{h}(a)$. Node $h$ broadcasts $h_b$ and $h_e$ to $D_{h}^{in}$, and then uses \fullref{wacc:tools:learn:predicate} in order to learn the identifiers and communication tokens of $C_{h}(a)$, in $\tilde O(|C_{h}(a)|/c + 1) = \tilde O(n^{1/2}/c + 1)$ rounds. Finally, within $\tilde O(|C_{h}(a)|/c + 1) = \tilde O(n^{1/2}/c + 1)$ rounds, node $h$ messages the nodes $C_{h}(a)$ to notify them that they are in $C_{h}(a)$, which overcomes the challenge. This concludes the proof of the statement.

Fix $a \in A_i$, and $h \in H(a)$. The nodes in $C_{h}(a)$ and repeat the same process as done for the case of $A_0$ by communicating with $h$. That is, each $v \in C_{h}(a)$ sends to $h$ the values: (1) the identifier $v$, (2) the value $d_{G}(v, a)$, and (3) the communication token of $v$. Then, $h$ broadcasts this to $D_{h}^{in}$. As shown for the case of $A_0$, this requires $\tilde O(n^{1/2}/c + 1)$ rounds, and completes the proof.
\end{proofof}

\section{The \hybrid Model -- Missing Proofs}
\label{app:littleO}

\subsection{Preliminaries -- Extended Subsection}\label{app:hybrid:prelim}
\subsubsection{Communication Primitives}

We observe several basic routing claims which are known in the \hybrid model.

In \cite[Theorem 2.2]{AGGHKL19}, the following is shown for the weaker \ncc model (in this model there are only global edges), and trivially holds in the \hybrid model. 
\begin{claim}[Aggregate and Broadcast]\label{thr:aggregateAndBroadcast}
There is an Aggregate-and-Broadcast Algorithm that solves any Aggregate-and-Broadcast Problem in \bigO{\log n} rounds in the \hybrid model.
\end{claim}

In \cite{kuhn2020computing,AHKSS20}, solutions are presented for the \fullref{thr:tokenDissemination} (see \cite[Theorem 2.1]{AHKSS20}) and \fullref{thr:tokenRoutingOriginal} (see \cite[Theorem 2.2]{kuhn2020computing}) problems. Token dissemination is useful for broadcasting, while token routing has the ability to be used in a fashion that is more similar to unicast.

\begin{definition}[Token Dissemination Problem]
    The problem of making $k$ distinct tokens globally known, where each token is initially known to one node, and each node initially knows at most $\ell$ tokens is called the \emph{$(k,\ell)$-Token Dissemination (TD) problem}.
\end{definition}
\begin{claim}[Token Dissemination]
There is an algorithm that solves $(k, \ell)$-TD in the \hybrid model in $\tildeBigO{\sqrt{k}+\ell}$ rounds, \whp
\label{thr:tokenDissemination}
\end{claim}

The following is discussed in \cite{kuhn2020computing} for token routing, and we later redefine the problem and remove the strong assumption which requires that each receiver knows the number of messages each sender sends it. We overcome this limitation later.

\begin{definition}[Token Routing Problem]\label{def:tokenRoutingOriginal}
    The \emph{token routing problem} is defined as follows.
    Let $S\subseteq V$ be a set of sender nodes and $R\subseteq V$ be a set of receiver nodes. Each sender needs to send at most $k_S$ tokens and each receiver needs to receive at most $k_R$ tokens, of size \bigO{\log n} bits each. Each token has a dedicated receiver node $r\in R$, and each receiver $r\in R$ knows the senders it must receive a token from and how many token it needs to receive from each sender. The token routing problem is solved when all nodes in $R$ know all tokens they are the receivers of.
\end{definition}
\begin{claim}[Token Routing]
    $S, R\subseteq V$ be sets of nodes sampled from $V$ with probabilities $p_S=n^{x_S-1}$ and $p_R=n^{x_R-1}$, for constant $x_S, x_R \in (0, 1]$, respectively. Let $k_S$ and $k_R$ be the number of tokens to be sent or received by any node in $S$ and $R$, respectively. 
    Let $K=\size{S}\cdot k_S+\size{R}\cdot k_R$ be the total workload. The token routing problem can be solved in $\tildeBigO{\frac{K}{n}+\sqrt{k_S}+\sqrt{k_R}}$ rounds in the \hybrid model \whp
    \label{thr:tokenRoutingOriginal}
\end{claim}

The following claim enables sending a polynomial number of messages uniformly at random while obeying the constraints of the model.
\begin{claim}[Uniform Sending]{\cite[Lemma 3.1]{AHKSS20}}\label{thr:sendingToUniform}
Presume some \hybrid model algorithm takes at most $p(n)$ rounds for some polynomial $p$. Presume that each round, every node sends at most $\sigma=\bigTheta{\log n}$ messages via global edges to $\sigma$ targets in $V$ sampled independently and uniformly at random. Then there is a $\rho=\bigTheta{\log n}$ such that for sufficiently large $n$, in every round, every node in $V$ receives at most $\rho$ messages per round \whp
\end{claim}

\subsubsection{Skeleton Graph}

We use the notion of the skeleton graph presented in \cite{AHKSS20, kuhn2020computing} and augment it with additional conditions. In particular, its nodes are \emph{well spaced} in the graph and satisfy the properties of marked nodes stated above.

\label{app:skelConst}
\begin{restatable}
[Extended Skeleton Graph]{definition}{Skl}\label{def:skeleton}
Given a graph $G=(V, E)$ and a value $0 < x < 1$, a graph $S_x=(M, E_S)$ is called a skeleton graph in $G$, if all of the following hold.
\begin{enumerate}
    \item{$\{v, u\} \in E_S$ if and only if there is a path of at most $h=\Tilde{\Theta}{(n^{1-x})}$ edges between $v, u$ in $G$.\label{itm:skeleton:edge}}
    
    \item Every node $v \in M$ knows all its incident edges in $E_S$.
    
    \item{$S_x$ is connected. \label{itm:skeleton:connected}}
    
    \item{For any two nodes $v, v'\in M$, $d_S(v, v')=d_G(v,v')$. \label{itm:skeleton:distance}}
    
    \item{For any two nodes $u,v\in V$ with $hop(u, v)\geq h$, there is at least one shortest path $P$ from $u$ to $v$ in $G$, such that any sub-path $Q$ of $P$ with at least $h$ nodes contains a node $w\in M$. \label{itm:skeleton:sp}}
    
    \item{$|M| = \tildeTheta{n^{x}}$.\label{itm:skeleton:size}}
    
    \item{For each $v \in M$ there is a helper set $H_v$ which satisfies: 
        \begin{enumerate}
            \item{$|H_v| = n^{1-x}$. \label{itm:helper:size}}

            \item{$\forall u \in H_v\colon hop(u, v)=\tildeBigO{n^{1-x}}$.\label{itm:helper:close}}

            \item{For each node $u \in V$, there are at most $\tildeBigO{1}$ nodes $V_u \subseteq M$ such that for each $w \in V_u$, $u \in H_w$. \label{itm:helper:own}}
            
            \item{$\size{\bigcup_{v\in M} H_v}=\tildeBigOmega{n}$. \label{itm:helper:many}}
    \end{enumerate} \label{itm:skeleton:helper}}

\end{enumerate}
\end{restatable}

In this definition, we merge the properties used by \cite{AHKSS20, kuhn2020computing}, slightly adjust Property~\ref{itm:helper:size} and prove Property~\ref{itm:helper:many}.

\begin{restatable}[Skeleton From Random Nodes]{claim}{SklFromRand}\label{claim:skeletonOnSampled}
Given a graph $G=(V,E)$, a value $0 < x < 1$, and a set of nodes $M$ marked independently with probability $n^{x-1}$, 

there is an algorithm which
constructs a skeleton graph $S_x=(M, E_S)$ in \tildeBigO{n^{1-x}} rounds \whp
If also given a single node $s \in V$, it is possible to construct $S_x=(M\cup\set{s}, E_S)$, without damaging the properties of $S_x$.
\end{restatable}

\begin{proofof}{\cref{claim:skeletonOnSampled}}
    Similarly to \cite[Algorithm 7]{AHKSS20} and \cite[Algorithm 6]{kuhn2020computing}, the algorithm for constructing the skeleton graph $S_x=(M, E_S)$ is to learn the \tildeTheta{n^{1-x}}-hop neighborhood and to run \cite[Algorithm 1]{kuhn2020computing} to compute the helper sets.

    We group and slightly extend the claims given in \cite{AHKSS20,kuhn2020computing}. 
    Properties~\ref{itm:skeleton:connected}~and~\ref{itm:skeleton:distance} holds \whp since $G$ is connected, see ~\cite[Lemma 4.3]{AHKSS20}~or~\cite[Lemma C.2]{kuhn2020computing}. Property~\ref{itm:skeleton:sp} follow from \cite[Lemma 4.2]{AHKSS20}~or~\cite[LemmaC.1]{kuhn2020computing}. Property~\ref{itm:skeleton:size} follows from Chernoff Bounds.
    The helper sets described in Property~\ref{itm:skeleton:helper} are computed using \cite[Algorithm 1]{kuhn2020computing}, and in \cite[Lemma 2.2]{kuhn2020computing}, their Properties~\ref{itm:helper:close}~and~\ref{itm:helper:own} are proven. It is also shown there that, \whp, for every $v\in M$ it holds that $\size{H_v}\geq n^{1-x}$, and thus in an additional \tildeBigO{n^{1-x}} rounds of local communication, we select exactly $n^{1-x}$ helpers and obtain Property~\ref{itm:helper:size}. The remaining Property~\ref{itm:helper:many} of the helper sets states that almost all of the nodes in the graph help other nodes. This holds since there are $\size{M}=\tildeBigOmega{n^{x}}$ skeleton nodes, each has $n^{1-x}$ helpers and each helper helps \tildeBigO{1} skeleton nodes, so, by the pigeonhole principle, the overall number of helpers is at least $\frac{\tildeBigOmega{n^x}\cdot n^{1-x}}{\tildeBigO{1}}=\tildeBigOmega{n}$.

    Finally, for adding a given node $s$ to the skeleton graph, notice that, as stated in \cite{kuhn2020computing}, this node can take as helpers $n^{1-x}$ closest nodes. For each helper node, this will at most double the number of skeleton nodes it helps  (see Property~\ref{itm:helper:own}).
\end{proofof}

For the sake of formality in the following proofs, as some are stated for a set of marked nodes and some for the skeleton graph, we also show the following \fullref{claim:skeleton}.
\begin{corollary}[Construct Skeleton]\label{claim:skeleton}
Given a graph $G=(V,E)$, and a value $0 < x < 1$, there is an algorithm which constructs a skeleton graph $S_x=(M, E_S)$ in \tildeBigO{n^{1-x}} rounds \whp Further, if also given a single node $s \in V$, it is possible to ensure that $s \in M$ without damaging the properties of $S_x$.
\end{corollary}
\begin{proofof}{\cref{claim:skeleton}}
    First mark each skeleton independently with probability $n^{1-x}$, getting a set of skeleton nodes $M$, then, using the algorithm from \fullref{claim:skeletonOnSampled} it is possible to construct the skeleton graph $S_x=(M, E_S)$ within $\tilde O(n^{1-x})$ rounds \whp

\end{proofof}

We show several primitives related to communication within skeleton graphs.

We show the following claim which, given a skeleton graph $S_x=(M, E_S)$, assigns the nodes $M$ unique IDs from the set $[|M|]$. This is useful, among other uses, for symmetry breaking and synchronization among the skeleton nodes. 

\begin{claim}[Unique IDs]\label{hybrid:treeConstruction}
Given a graph $G=(V, E)$, and a skeleton graph $S_x=(M, E_S)$, it is possible to assign the nodes $M$ unique IDs from the set $[|M|]$ within $\tilde O(1)$ rounds in the \hybrid model, \whp
\end{claim}
\begin{proofof}{\cref{hybrid:treeConstruction}}
    We construct a binary tree of $\tilde O(1)$ depth over the nodes $M$, and then assign each node an ID equal to its index in the \emph{pre-order} traversal of the tree.
    
    The nodes $M$ compute the node $v' \in M$ with minimal initial ID, the ID which it has due to the definition of the \hybrid model. Notice it is possible to identify $v'$ and ensure that all nodes in $M$ know the identifier of $v'$ within $\tilde O(1)$ rounds due to \fullref{thr:aggregateAndBroadcast}. Further, using \fullref{thr:aggregateAndBroadcast}, the nodes compute $|M|$.
    
    Next, node $v'$ chooses two nodes at random from $G$, nodes $a, b$, and sends them each a message. Nodes $a, b$ reply each with a random node $\alpha, \beta$, respectively, where $a \in H_\alpha, b \in H_\beta$. Node $v'$ repeats this process as long as it does not receive two distinct nodes $\alpha, \beta$. Node $v$ then sends messages to both $\alpha, \beta$ and lets them know that they are its children in the tree. The nodes added to the tree continue this process, each of them randomly choosing two nodes as its children until it receives two distinct nodes which are not already in the tree, or until some $\tilde O(1)$ rounds elapsed. Clearly, \whp, this process constructs a binary tree of depth $\tilde O(1)$ within $\tilde O(1)$ rounds.
    
    Finally we would like to assign an ordering to the nodes. Each node tells its parent the size of its subtree. That is, the leaves tell their parents that they are leaves, and whenever a node reaches a state where it has heard from all its children, it tells its parent how many nodes are in its subtree. Then, the root of the tree, $v'$, begins with the ID pallet $[|M|]$, takes the first ID for itself, and passes down two contiguous intervals for possible IDs, broken according to the sizes of the subtrees of its children, to its two children nodes -- with the left child receiving the interval with smaller IDs. Inductively, each node takes the first ID from the pallet it receives from its parent, breaks the pallet into two contiguous parts, according to the sizes of the subtrees of its children, and sends the part with smaller IDs to its left child, and the higher part to its right child. Since the depth of the tree is $\tildeBigO{1}$, this completes in $\tildeBigO{1}$ rounds, \whp
\end{proofof}

We use the following statement from \cite{censorhillel2020distance} to prove \fullref{theorem:hybrid:approxSSSP:highMaximalDegree}.

\begin{lemma}[Reassign Skeletons]\label{thr:reassignSkeletons}\cite[Lemma 29]{censorhillel2020distance}
    Given graph $G=(V, E)$, a skeleton graph $S_x=(M, E_S)$, a value $k$ which is known to all the nodes, and nodes $A\subseteq V$ such that each $u \in A$ has at least $\tildeTheta{k\cdot |A|}$ nodes $M_u \subseteq M$ in its $\tilde \Theta(n^{1-x})$ neighborhood, there is an algorithm that assigns $K_u \subseteq M_u$ nodes to $u$, where $|K_u|=\tildeBigOmega{k}$, such that each node in $M$ is assigned to at most $\tilde O(1)$ nodes in $A$. With respect to the set $A$, it is only required that every node in $G$ must know whether or not it itself is in $A$ -- that is, the entire contents of $A$ do not have to be globally known. The algorithm runs in \tildeBigO{n^{1-x}} rounds in the \hybrid model, \whp
\end{lemma}

The skeleton-based techniques allow us to approximate weighted SSSP fast in the \hybrid model. After we do it, the following well-known simple reduction allows us to compute $(2+\eps)$ approximate weighted diameter. 
\begin{restatable}
[Diameter from SSSP]{claim}{DfromSSSP}\label{thr:ssspToDiameter}(see e.g. \cite[Claim 34]{censorhillel2020distance})
Given a graph $G = (V, E)$, a value $\alpha>0$ and an algorithm which computes an $\alpha$ approximation of weighted SSSP in $T$ rounds of the \hybrid model, there is an algorithm which computes a $2\alpha$-approximation of the weighted diameter in $T+\tildeBigO{1}$ rounds of the \hybrid model.
\end{restatable}

We use the following basic claim regarding usage of skeleton graphs for purposes of distance computations in the \hybrid model. It is proven in \cite{AHKSS20}.

\begin{restatable}
[Extend Distances]{claim}{ExtDist}\label{thr:computeMSSPFromMSSPOnSkeleton}\cite[Theorem 2.7]{AHKSS20}Let $G=(V, E)$, let $S_x=(M, E_S)$ be a skeleton graph, and let $V'\subseteq V$ be the set of source nodes. If for each source node $s\in V'$, each skeleton node $v\in M$ knows the $\left(\alpha, \beta\right)$-approximate distance $\tilde{d}\left(v, s\right)$ such that $d(v, s)\leq \tilde{d}(v, s)\leq \alpha d(v, s) + \beta$, then each node $u\in V$ can compute for all source nodes $s\in V'$, a value $\tilde{d}(u, s)$ such that $d(u, s)\leq \tilde{d}(u, s)\leq \alpha d(u, s)+ \beta$ in $\tildeBigO{n^{1-x}}$ rounds.
\end{restatable}

\subsection{Oblivious Token Routing}
\label{app:ObliviousTR}

In \cite{kuhn2020computing}, they introduce and solve the  \emph{token routing problem} over a skeleton graph, where each receiver $r$ knows the number of tokens each sender $s$ has for $r$. This is insufficient for our purposes since we work in the $\littleO{n^{1/3}}$ complexity realm with $\littleOmega{n^{2/3}}$ skeleton nodes, where we can't make the identifiers of the skeleton nodes globally known, let alone the number of messages between pairs of nodes. Therefore, we define the following routing problem, in which the receivers \emph{do not know neither the identifiers of the senders nor the number of messages each sender intends to send them}.

\begin{definition}[Oblivious Token Routing Problem]\label{def:tokenRouting}
    The \emph{oblivious-token routing problem} is defined as follows. Let $S\subseteq V$ be a set of sender nodes and $R\subseteq V$ be a set of receiver nodes. Each sender needs to send and each receiver needs to receive at most $k$ tokens, of size \bigO{\log n} bits each. Each token has a dedicated receiver node $r\in R$, and each sender $s\in S$ and receiver $r\in R$ know the bound $k$ on number of tokens the receiver is going to receive. The oblivious-token routing problem is solved when all nodes in $R$ know all tokens they are the receivers of.
\end{definition}

Notice that the assumption of the knowledge of a bound on the number $k$ of messages each receiver gets is something which is easy to eliminate by having the receiver \emph{double} its estimate and repeat the algorithm till success for \bigO{\log{k}} iterations. To verify if some particular invocation succeeded, we can can make a node broadcast failure if it sent or received more than half of its global capacity at some point.
\begin{restatable}[Oblivious Token Routing]{lemma}{UnkownTR}\label{thr:tokenRouting}
    Given a graph $G=(V, E)$, and a skeleton graph $S_x=(M, E_S)$, let $k$ be an upper bound on the number of tokens to be sent or received by any node in $M$ and let $K=2\cdot \size{M}\cdot k$ be the total workload. The \emph{oblivious-token routing problem} can be solved in $\tildeBigO{{k/n^{1-x}} + n^{1-x}}$ rounds, \whp, in the \hybrid model.
\end{restatable}

\begin{proofof}{\cref{thr:tokenRouting}}
    The problem overcome in~\cite[Theorem 2.2]{kuhn2020computing} (the non-oblivious case), is  that even though there are enough helpers near each skeleton node to send and receive all the messages, it is not straightforward to connect between senders' and receivers' helpers. So, in~\cite{kuhn2020computing} it is suggested to relay messages via some \emph{intermediate receivers}. This way, a message is sent by a sender to one of its helpers, by the helper to an intermediate receiver, from there to a helper of the receiver, and from there it is sent to the receiver.
    To compute intermediate receivers for the message number $i$ from $s$ to $r$, they apply a pseudo-random hash function $h(s, r, i)$. 
    
    However, the receiver needs to be able to compute $h(s, r, i)$ as well, so it needs to know the number of messages it is to receive from each sender $s$, and we cannot assume this for our purposes.
    
    To overcome this limitation, we assign for helper number $i\in\interval{n^{1-x}}$ of the receiver $r$ the intermediate receiver whose identifier is computed as $w=h(r, i)$, where $h$ is pseudo-random hash function. We deliver the messages in $\ceil*{{k/n^{1-x}}}$ phases. To keep the load balanced between phases, for each message $j$ we independently at random sample $p_j\sim U\interval{\ceil*{{k/n^{1-x}}}}$, which is the phase on which it will be sent. In order to keep the load balanced between the receivers' helpers and intermediate receivers on some phase $p$, for each message $j$ we also independently at random sample receivers' helper index $i_j\sim U\interval{n^{1-x}}$. The intermediate receiver is decided by hash function $h$, i.e. we route the message $j$ with the final receiver $r_j$ via $w=h(r_j, i_j)$. Unlike \cite{kuhn2020computing}, we apply $h$ on arguments that are not necessarily distinct, which could increase the number of conflicts. However, we show that every time all nodes apply $h$, each key $\left(r_j, i_j\right)$ is used at most $\tildeBigO{1}$-times \whp, so due to \fullref{claim:pseudoHashOnMultiSet} the congestion on each intermediate receiver is \tildeBigO{1} \whp
        
    The pseudo-code is provided by \cref{alg:tokenRouting}. 
    \begin{algorithm}
        \caption{\textbf{Oblivious-Token Routing Protocol}}
        \label{alg:tokenRouting}
        The node with the minimum ID samples and broadcasts \tildeBigO{1} bits of seed \label{line:tokenRouting:seed} 
        
        Each node uses the seed to sample a pseudo-random hash function $h\in \mathcal{H}$ \label{line:tokenRouting:hash} 
        
        Each sender $s\in M$ balances tokens to send between its helpers $H_s$  \label{line:tokenRouting:balance} 
        
        Each receiver $r$ enumerates its helpers $u\in H_r$ and informs them about their indices \label{line:tokenRouting:enumerate} 
        
        Each sender's helper $v\in V$, for each message $j$ it is assigned to send, samples \emph{receiver's helper index} $i_j\sim U\interval{n^{1-x}}$ and phase $p_j\in U\interval{\ceil*{{k/n^{1-x}}}}$ \label{line:tokenRouting:sample} 
        
        \For{$p$ from $0$ to $\ceil*{{k/n^{1-x}}}$ \label{loop:tokenRouting:phases} } {
            Each sender's helper $v\in V$ for each message $j$ such that $p_j=p$ sends it to $h(r_j, i_j)$
            \label{line:tokenRouting:sendToIntermediate} 
            
            Each receiver's helper $u\in V$, for each receiver $r$ it helps sends $\Braket{u, r}$ to $w=h(r, i)$, such that $i$ is the index of $u$ in $H_r$ \label{line:tokenRouting:askIntermediate} 
            
            Each intermediate receiver $w\in V$ which receives $\Braket{u, r}$ sends all the messages it received for $r$ on this phase to $u$ \label{line:tokenRouting:respondIntermediate} 
        }
        
        Each receiver $r\in M$ collects messages addressed to it from its helpers $H_r$ \label{line:tokenRouting:collect}
    \end{algorithm}
    Notice that each node can play five different roles: it could be a sender $s$, a receiver $r$, a sender's helper $v$, a receiver's helper $u$ and an intermediate receiver $w$. Moreover, it can be a sender's or a receiver's helper for up to $\tildeBigO{1}$ nodes. We show that it can be an intermediate receiver for $\tildeBigO{1}$ receiver's helpers \whp
    
    First, all nodes sample a globally known pseudo-random hash function $h\colon V \times \interval{n^{1-x}} \mapsto V$ from the family of $\tildeTheta{1}$-wise independent random functions $\mathcal{H}$, which is used to compute the intermediate receivers for each message  (\cref{line:tokenRouting:seed,line:tokenRouting:hash}). For this, by \fullref{lemma:pseudoRandom}, \tildeBigO{1} bits of globally known seed are enough and the node with the minimal identifier samples and broadcasts them using \fullref{thr:aggregateAndBroadcast}.
    Afterwards, each sender $s\in M$ distributes the tokens between its helpers $H_v$ in a balanced manner -- each sender's helper is assigned at most $\ceil*{{k/n^{1-x}}}$ messages to send (\cref{line:tokenRouting:balance}). Each receiver enumerates its helpers by identifiers (\cref{line:tokenRouting:enumerate}). Each sender's helper $v$, for each message $j$ it has to send, samples a random phase $p_j\sim U\interval{\ceil*{{k/n^{1-x}}}}$ and a random receiver's helper index $i_j$ (\cref{line:tokenRouting:sample}).
    
    We then proceed for $\ceil*{{k/n^{1-x}}}$ phases. On phase $p$, each sender's helper $v$ sends each message $j$ for which it sampled $p_j=p$ to the node $h(r_j, i_j)$. Afterwards, in \cref{line:tokenRouting:sendToIntermediate}, each receiver's helper $v$ for each receiver $r$ it helps sends $\Braket{v, r}$ to $h(r, i)$, where $i$ is the index of $v$ in $H_r$ computed in \cref{line:tokenRouting:enumerate}. Each intermediate receiver $w$, sends all messages $j$ it received with destination $r_j$ to the $v$ from which it received $\Braket{v, r_j}$.

    \cref{line:tokenRouting:seed} takes \tildeBigO{1} rounds by \fullref{thr:aggregateAndBroadcast}, and  \cref{line:tokenRouting:balance,line:tokenRouting:enumerate,line:tokenRouting:collect} are implemented using local edges in \tildeBigO{n^{1-x}} rounds. There are \tildeBigO{\ceil*{{k/n^{1-x}}}} iterations of the loop in Line~\ref{loop:tokenRouting:phases}, and we argue that each of them requires \tildeBigO{1} rounds of communications via global edges \whp Overall, the complexity is $\tildeBigO{{k/n^{1-x}} + n^{1-x}}$ rounds.
    
    For each of the $k$ messages designated to some receiver $r$, the phase number $p_j$ is sampled independently with probability $\tildeBigO{\min\set{1, {n^{1-x}}/{k}}}$ , therefore by a Chernoff Bound, there are \tildeBigO{n^{1-x}} messages with $r$ as the final destination, which are sent on the $p$-th phase \whp On the $p$-th phase, some receiver's helper index for each of these messages is sampled with probability $\frac{1}{n^{1-x}}$, therefore by a Chernoff Bound it is sampled $\tildeBigO{1}$ times \whp By a union bound over all phases, receivers and receivers' helper indices, on each phase, for each receiver each receiver's helper index is selected \tildeBigO{1} times \whp Thus, by \fullref{claim:pseudoHashOnMultiSet} each $w\in V$ is selected as an intermediate receiver \tildeBigO{1} times and receives \tildeBigO{1} messages in \tildeBigO{1} rounds \whp This implies that no message is lost during \cref{line:tokenRouting:sendToIntermediate} and that \cref{line:tokenRouting:sendToIntermediate,line:tokenRouting:respondIntermediate} take $\tildeBigO{1}$ rounds.
    
    Since each node helps at most \tildeBigO{1} senders and due to Chernoff Bounds, each helper sends \tildeBigO{1} messages \whp on some phase in  \cref{line:tokenRouting:sendToIntermediate}. Since each node is a helper to at most \tildeBigO{1} receivers, \cref{line:tokenRouting:respondIntermediate} also takes \tildeBigO{1} rounds \whp
    Similarly, by \fullref{claim:pseudoHashOnMultiSet}, since there are $\tildeBigO{n^{x}}\cdot n^{1-x}=\tildeBigO{n}$ distinct pairs of receivers and receiver receiver's helper index, \whp each intermediate receiver is assigned to at most $\tildeBigO{1}$ receiver helpers. Thus, \cref{line:tokenRouting:askIntermediate} also takes $\tildeBigO{1}$ rounds \whp
\end{proofof}

\SklUnicast*
\begin{proofof}{\cref{skeletons:unicast}}
    The claim follows by an invocation of \fullref{thr:tokenRouting} with parameters $x,\ k,\ K=\tildeBigO{n^{x}\cdot k}$  resulting in $\tildeBigO{\frac{k}{n^{1-x}}+\sqrt{k}}=\tildeBigO{n^{1-x}}$ rounds, as required.
\end{proofof}

\subsection{\bcc Simulation}

We use the following claims from \cite{censorhillel2020distance} to improve the simulating of the \bcc model in the \hybrid. 

\begin{restatable}
[\local Simulation in \hybrid]{lemma}{LocalSim}\label{claim:LOCALSimulation}\cite[Lemma 16]{censorhillel2020distance}
Given a graph $G=(V, E)$, and a skeleton graph $S_x=(M, E_S)$, it is possible to simulate one round of the \local model over $S_x$ within $\tilde O(n^{1-x})$ rounds in $G$ in the \hybrid model. That is, within $\tilde O(n^{1-x})$ rounds in $G$ in the \hybrid model, any two adjacent nodes in $S_x$ can communicate any amount of data between each other.
\end{restatable}

\begin{restatable}[Sampled neighbors \protect{\cite[Lemma 3.1]{censorhillel2020distance}}]{lemma}{DegSim}\label{lemma:autoSimulator}
    Given is a graph $G=(V, E)$. For a value $q\leq n$, there is a value $x=\tildeBigO{n / q}$ such that the following holds \whp:
    Let $V'\subseteq V$ be a subset of $\size{V'}=x$ nodes sampled uniformly at random from $M$.
    Then each node $u\in V$ with $\deg{(u)}\geq q$ has a neighbor in $V'$.

\end{restatable}

We show how to simulate the \bcc model using the \wacc model and the \local model together, and it then follows by \fullref{claim:wacc:simulation} and \fullref{claim:LOCALSimulation} that this can be converted into a simulation in the \hybrid model. The intuition behind the simulation follows from observing \fullref{lemma:autoSimulator} -- if every node desires to broadcast a single message to the entire graph, then with relatively little bandwidth it is possible to ensure that all nodes above a certain minimal degree will get these messages from all the nodes in the graph.
We begin with the  simulation of the \bcc model in the combined \wacc and \local models.
\begin{lemma}[\bccshort Simulation in \wacc and \local]
    \label{theorem:bcc:simulationInWACC}
    Given a graph $G=(V, E)$ with average degree $k$, given an algorithm $ALG_{BCC}$ in the \bcc model, which runs on $G$ in $t$ rounds, and given some value $c$, there exists an algorithm which uses $\tilde O(t \cdot n/(\sqrt{k}\cdot c))$ rounds of the \wacc model and $O(t)$ rounds of the \local model on $G$ and simulates $ALG_{BCC}$ on $G$. It is assumed that prior to running $ALG_{BCC}$, each node $v \in V$ has at most $\tilde O(\deg_G(v))$ bits of input used in $ALG_{BCC}$, including, potentially, the incident edges of $v$ in $G$. Further, it is assumed that the output of each node in $ALG_{BCC}$ is at most $O(t \log n)$ bits. 
\end{lemma}
\begin{proofof}{\cref{theorem:bcc:simulationInWACC}}
    The outline of the simulation is as follows. We split the graph into high degree nodes, $H \subseteq V$, and low degree nodes, $L = V \setminus H$, at a certain cut-off. The key idea is that if every node $v \in V$ takes a single message and sends it randomly to $c$ nodes in $V$, then every $u \in H$ will have at least one neighbor, \whp, which hears the message from $v$, for every $v \in V$, due to \fullref{lemma:autoSimulator}. Thus, we choose some subset $F \subset H$ and assign to each node $v \in L$ some node $u \in F$ which \emph{partially simulates} $v$. By partially simulating, we mean that, initially, node $v$ tells node $u$ all of its input to $ALG_{BCC}$, and then for each round, node $u$ tells $v$ what message $v$ wants to send in that round, and $v$ then sends this message (that it wishes to broadcast) to $c$ random nodes. Finally, we are guaranteed that every node in $H$ hears all the messages broadcast in the graph, which allows for $u$ to internally simulate the local computation which $v$ should perform in $ALG_{BCC}$ before the next round. In a sense, when $u$ simulates $v$, after each round node $v$ knows what message it wants to send in that round of $ALG_{BCC}$, yet not necessarily other information that it would have learned from other nodes in the graph during that round of $ALG_{BCC}$. Thus, node $v$ might not know its output in $ALG_{BCC}$. To overcome this, notice that $u$ knows the output of $v$ in $ALG_{BCC}$, and due to our assumption in the statement of this theorem, each node outputs at most $O(t\log n)$ bits, and so we can simulate another $t$ rounds where each $v$ will just \emph{broadcast} its output (ensuring that it itself receives it from $u$). \\

    \noindent\textbf{Initialization} \\
    We begin by showing how to initialize the nodes of high degree which simulate those of low degree. The cut-off for being a high or low degree node is $\Theta{(\sqrt{k})}$. That is, we desire to simulate every node $v \in V$ with $\deg(v) = o(\sqrt{k})$ using a node $u \in V$ with degree at least $\deg(u) = \Omega(\sqrt{k})$. Observe that since $k$ is the average degree, there are $\Theta(nk)$ edges in the graph. Since the maximal degree is at most $n$, there must be at least $k$ nodes with degree at least $\Omega(\sqrt{k})$. Thus, we denote by $F$ the $k$ nodes in $G$ with the highest degree, and are guaranteed that for each $v \in F$, $\deg(v) = \Omega(\sqrt{k})$. Notice that it is possible within $\tilde O(1)$ rounds to count the number of nodes in $V$ with degree above a threshold, using \fullref{wacc:tools:aggregation}, and thus within $\tilde O(1)$ rounds it is possible to do a binary search for the degree of the node with $k^{th}$ highest degree, allowing each node to know whether or not it is in $F$. 
    
    Let $v \in F$. The node $v$ now knows that it is in $F$, and thus randomly sends $\tilde \Theta(n/k)$ messages, using $\tilde \Theta(n/(k\cdot c)) \leq \tilde O(n/(\sqrt{k}\cdot c))$ rounds of the \wacc model, containing its ID and communication token in the \wacc model. Clearly, \whp, every node $u \in V \setminus F$ has received a message from at least one node $v \in F$. Thus, if node $u$ needs simulating, that is, $\deg(u) = o(\sqrt{k})$,  it chooses arbitrarily among the nodes from $F$ which it heard from some node $v$ and tells $v$ that it should simulate $u$. Denote by $J_v$ the set of nodes which choose $v$ to simulate them. Each node $v \in F$, upon receiving $J_v$, chooses and arbitrary order for $J_v$ and sends back to each $u \in J_v$ its index in that order. 
    
    Every node $v \in F$ now attempts to learn all the input to $ALG_{BCC}$ of the nodes which it simulates. Notice that now for every node in $v \in F$ it holds that $|J_v|$ is at most $\tilde O(n/k)$. Notice that each node $u \in J_v$ has degree $\deg(u) = o(\sqrt{k}) = O(\sqrt{k})$, since otherwise it would have opted to not be simulated by any node, implying, by the constraints of this theorem, that $u$ has at most $\tilde O(\sqrt{k})$ bits of input to $ALG_{BCC}$, and therefore all the nodes $J_v$ desire to send to $v$ at most $\tilde O(n/\sqrt{k})$ messages. Since $v$ synchronized all the nodes in $J_v$ by sending them each its index in some order of $J_v$, it is possible to send all this data to $v$ in $\tilde O(n/(\sqrt{k}\cdot c))$ rounds of the \wacc model: To do so, assume that every node $u \in J_v$ wishes to send \emph{exactly} $d = \tilde \Theta(n/\sqrt{k})$ messages to $v$ (we can assume this since $u$ wants to send at most $\tilde O(n/\sqrt{k})$ messages to $v$, and so it can just add extra \emph{empty} messages at the end), therefore, since node $u$ knows its index in $J_v$, for some ordering which $v$ decided on, it is possible to order all the messages from all of $J_v$ to $v$ in such a way that each node $u \in F_v$ knows the indices of its messages and such that at all round  neither $v$ receives more than $c$ messages, nor a node $u \in F_v$ sends more than $c$ messages. \\
    
    \noindent\textbf{Round Simulation} \\
    We now show how to simulate every one of the $t$ rounds of the given $ALG_{BCC}$. That is, we show the two final steps of the simulation: how $v \in F$ tells each node in $J_v$ what value to randomly send to nodes across the graph, and how each node in $F$ gets all the messages which were sent by all the nodes. The first part is simple -- we already saw that node $v$ can send a single, unique message to each $u \in J_v$ within $\tilde O(n/(\sqrt{k}\cdot c))$ rounds of the \wacc model. The second part follows from \fullref{lemma:autoSimulator}. According to \fullref{lemma:autoSimulator}, for every node $v\in F$ (which has $\deg_G{v}\geq q=\sqrt{k}$) to receive a single message from some node $u\in V$, it is enough for this node $u$ to send a message $x=\tildeBigO{n / q}=\tildeBigO{n / \sqrt{k}}$ times to nodes sampled uniformly at random and for each node $v\in F$ to learn received messages from its neighbors in $G$. Sending $x=\tildeBigO{n / \sqrt{k}}$ messages, each one to random node requires $\tildeBigO{n / (\sqrt{k}\cdot c)}$ rounds of the \wacc model. And aggregating messages from neighbors requires single round of the \local model.

    \noindent\textbf{Output} \\
    It is critival that every node will know its output at the end of the simulation of $ALG_{BCC}$. This is ensured since we assume in the statement of this theorem that the output of every node in $ALG_{BCC}$ it at most $O(t\log n)$ bits. Thus, instead of simulating $ALG_{BCC}$ directly, we simulate an algorithm $ALG_{BCC}$ which is just like $ALG_{BCC}$, yet is followed by $t$ rounds where each node broadcasts its output. Since $ALG_{BCC}$ takes $t$ rounds, this simply doubles the round complexity achieved above. Due to the fact that our simulation maintains that each node knows all the messages which it broadcasts during the simulated algorithm, every node will necessarily know its output.
\end{proofof}

Finally, we show how to use \fullref{theorem:bcc:simulationInWACC} for simulating a \bcc algorithm in the \hybrid model.

\BCCSim*
\begin{proofof}{\cref{thr:bccInHybrid}}
    We execute the simulation of  \fullref{theorem:bcc:simulationInWACC} on $G$ and $S_x$ with $c = \tilde \Theta (n^{2-2x})$, in order to obtain an algorithm $ALG_{WACC,\ LOCAL}$ which simulates $ALG_{BCC}$ on $S_x$ within $t_ 1=\tilde O(t \cdot n^x/(\sqrt{k}\cdot c)=\tilde O(t \cdot n^x/(\sqrt{k}\cdot n^{2-2x})=\tilde O(t \cdot n^{3x-2}/\sqrt {k})$ rounds of the \wacc model and $t_2=O(t)$ rounds of the \local model on $S_x$. Due to \fullref{claim:wacc:simulation}, since $c = \tilde \Theta (n^{2-2x})$, it is possible to simulate the \wacc rounds of $ALG_{WACC,\ LOCAL}$ on $S_x$ in $\tilde{O} (t_1 \cdot n^{1-x}) = \tilde O(t \cdot n^{2x-1}/\sqrt {k})$ rounds of the \hybrid model on $G$. Likewise, using \fullref{claim:LOCALSimulation}, it is possible to simulate the \local rounds of $ALG_{WACC,\ LOCAL}$ on $S_x$ in $\tilde{O}(t_2\cdot n^{1-x})=\tilde O(t \cdot n^{1-x})$ rounds of the \hybrid model on $G$.
\end{proofof}

\subsection{A \texorpdfstring{$(1+\eps)$}{1 + epsilon}-Approximation for SSSP}

We show the missing proof of how we compute SSSP with low average degree and high maximal degree.
\SSSPLAHM*
\begin{proofof}{\cref{theorem:hybrid:approxSSSP:highMaximalDegree}}
    Let $v' \in S_x$ be some node with degree $\deg_{S_x}(v') = \omega(n^{2-2x})$. Observe that such a node $v'$ exists and can be found and agreed upon by all the nodes of $G$ using \fullref{thr:aggregateAndBroadcast} within $\tilde O(1)$ rounds. Denote by $N_{v'}$, where $|N_{v'}| = \Theta(n^{2-2x})$, an arbitrary subset of the neighbors of $v'$. Using \fullref{thr:tokenDissemination}, it is possible to ensure within $\tilde O(n^{1-x})$ rounds that every node in the graph knows which nodes are in $N_{v'}$. We strive to have the nodes of $M$ send all the contents of $E_S$ to the nodes $N_{v'}$. 

    We now show how the nodes $N_{v'}$ can learn all of $E_S$. We show that there is a way to do this in which every node in $M$ desires to send and receive at most $\tilde O(n^{2-2x})$ messages, and therefore due to \fullref{skeletons:unicast}, the routing will complete in $\tilde O(n^{1-x})$ rounds. 
    
    We start by showing that the nodes $N_{v'}$ even have the bandwidth to receive $E_S$ within at most $\tilde O(n^{1-x})$ rounds. Due to \fullref{skeletons:unicast}, each node in $N_{v'}$ can receive $\tilde \Theta(n^{2-2x})$ messages in $\tilde O(n^{1-x})$ rounds, implying that in total $N_{v'}$ can receive $\tilde \Theta(|N_{v'}| \cdot n^{2-2x}) = \tilde \Theta(n^{4-4x})$ messages. Since the average degree in $S$ is $\tilde O(n^{x/2})$, then $|E_S| = \tilde O(n^{3x/2})$, and $3x/2 \leq 4-4x$ if and only if $x \leq 8/11$, which holds since we assume $x \leq 12/17 \leq 8 /11$.
    
    We now show that each node $v \in M$ has the bandwidth to send all its incident edges within at most $\tilde O(n^{1-x})$ rounds. Notice that if $\deg_{S_x}{(v)} = \tilde O(n^{2-2x})$, then it can clearly do so. Thus, for all other nodes in $M$ which have higher degrees, we assign some of the other nodes of $M$ in their $\tilde \Theta(n^{1-x})$ neighborhood to assist them. Let $A \subseteq M$ be the set of all nodes in $M$ such that $v \in A$ has $\deg_{S_x}(v) = \Omega(n^{2-2x})$. We strive to invoke \fullref{thr:reassignSkeletons} on $A$ in order to assign each $v \in A$ some $\tilde \Theta(n^{3x-2})$ nodes, denoted $A_v \subseteq M$, where $A_v$ are in the $\tilde O(n^{1-x})$-hop neighborhood of $v$ in $G$, and where each node in $M$ is assigned to at most $\tilde O(1)$ nodes in $A$. Thus, we must show that for each node $v \in A$, it has in its $\tilde O(n^{1-x})$ neighborhood in $G$ at least $\tilde \Theta(|A| \cdot n^{3x-2})$ nodes of $M$. Recall that $\deg_{S_x}(v) = \Omega(n^{2-2x})$, implying that $v$ has at least $\Omega(n^{2-2x})$ nodes of $M$ in its $\tilde O(n^{1-x})$ neighborhood in $G$. We thus strive to show that $\tilde \Theta(|A| \cdot n^{3x-2}) = O(n^{2-2x})$. Notice that $|A| = \tilde O(n^{3x/2} / n^{2-2x}) = \tilde O(n^{7x/2 - 2})$ since the average degree in $S_x$ is at most $\tilde O(n^{x/2})$ and the minimal degree of a node in $A$ is $\Omega(n^{2-2x})$. Thus, $\tilde \Theta(|A| \cdot n^{3x-2}) = \tilde O(n^{13x/2 - 4})$, and since $13x/2 - 4 \leq 2 - 2x$ if and only if $x \leq 12/17$, which is given in the conditions of this statement, we conclude. Therefore, it is possible to invoke \fullref{thr:reassignSkeletons} and thus we can assume that each node $v \in A$ is assigned the nodes $A_v$ defined previously. Now, node $v \in A$ distributes its incident edges in $S_x$ to the nodes $A_v$ uniformly, using the local edges of the \hybrid model, within $\tilde O(n^{1-x})$ rounds. Since $v$ has at most $|M| = \tilde O(n^x)$ incident edges in $S_x$, this means that each node $u \in A_v$ receives at most $\tilde O(n^{x - 3x + 2}) = \tilde O(n^{2 - 2x})$ messages from $v$. Every node $u \in A_v$ takes \emph{responsibility} for the edges it received from $v$ and soon forwards them to the nodes $N_{v'}$. Notice that since each node $u \in M$ is assigned to as most $\tilde O(1)$ nodes in $A$, each $u$ takes responsibility for at most $\tilde O(n^{2 - 2x})$ messages in total.
    
    At last, noticed that we reach a state where every node in $S_x$ wishes to send at most $\tilde O(n^{2-2x})$ messages to the nodes in $N_{v'}$ -- this is since nodes with at most $\tilde O(n^{2-2x})$ neighbors in $S_x$ (nodes in $M \setminus A$) have at most such many messages, and each node $v \in A$ distributed such many messages per each node in $A_v$. Further, as stated above, the total number of messages to send are $\tilde O(n^{3x/2})$. Notice that for each message, it does not matter which node in $N_{v'}$ receives it, and so for each message we select the target in $N_v'$ independently and uniformly. The expected number of messages each node in $N_v'$ receives is \tildeBigO{n^{3x/2 - 2 + 2x}}=\tildeBigO{n^{7x/2-2}}=\tildeBigO{n^{2-2x}}, where the last transition holds since $x \leq 12/17 \leq 8/11$, as seen previously. Since the targets of the messages are independent, by an application of a Chernoff Bound, and applying a union bound over all $N_{v'}$, number of messages each node receives is $\tildeBigO{n^{2-2x}}$ \whp Thus by \cref{skeletons:unicast}, it is possible to route the messages within $\tildeBigO{n^{1-x}}$ rounds \whp
    This, combined with the fact that the contents of $N_{v'}$ were previously make globally known using \fullref{thr:tokenDissemination}, allows every node $v \in M$ to locally compute to which node in $N_{v'}$ it should deliver each of its messages in a way such that every node in $N_{v'}$ receives the same number of messages across all the messages being sent. At this point, since every node in $M$ desires to send and receive at most $\tilde \Theta(n^{2-2x})$ messages, it is possible to invoke \fullref{skeletons:unicast} in order to route all these messages within $\tilde \Theta(n^{1-x})$ rounds.

    Finally, since now nodes in $N_{v'}$ know all of $E_S$, node $v$ can learn all of $E_S$ by learning all the information stored in $N_{v'}$ within $\tilde O(n^{1-x})$ rounds. Node $v'$ can compute the exact distance from the source $s \in M$ to any node $v \in M$. Thus, node $v'$ desires to tell every node $v \in M$ the value of $d_{S_x}(s, v)$. This is possible since $S_x$ is connected and thus every node $v \in M$ sent at least one message to $v$ throughout the above algorithm, and thus it is possible to reverse the direction of the messages sent above in order to ensure that for each $v \in M$, node $v'$ can send a unique message to $v$, in the same round complexity as of the above algorithm. 
\end{proofof}

\section{The \congest Model -- Missing Proofs}\label{app:congest}
We begin with some preliminaries for this section. 

\begin{claim}[\congest Routing]\cite[Theorem 1.2]{Ghaffari2017DistributedMA}\cite[Theorem 2]{Chang2019DistributedTD}\label{claim:hc:routing}
Consider a graph $G=(V,E)$ with an identifier assignment $ID\colon V \mapsto \interval{n}$ such that any node $u$ given $ID(v)$ can compute $\floor{\log\deg{(v)}}$, and a set of point-to-point routing requests, each given by the identifiers of the corresponding source-destination pair. If each node $v$ of $G$ is the source and the destination of at most $\deg_G(v)\cdot \twosqlgnlglgn$ messages, there is a randomized distributed algorithm that delivers all messages in time $\tmixtwosqlgnlglgn$ in the \congest model, \whp
\end{claim}

\begin{corollary}[Identifiers]\label{cor:hc:identifiers}
    In the \congest model in $\bigO{\tmix+\log{n}}$ time we can compute an ID assignment $ID\colon V \mapsto \interval{n}$ and other information such that $ID(u) < ID(v)$ implies $\floor*{\log\deg(u)} \leq \floor*{\log \deg(v)}$, and any vertex $u$, given $ID(v)$, can locally compute $\floor*{\log \deg(v)}$ for any $v$.

\end{corollary}
\begin{proofof}{\cref{cor:hc:identifiers}}
    \cite[Lemma 4.1]{Chang2019DistributedTD} shows how to compute the aforementioned set of identifiers in $\bigO{D+\log{n}}$ rounds in the \congest model, where $D$ is the diameter of the graph. Since
    the diameter is at most the mixing time $\tau_{mix}$, we also have that the identifiers are computable in $\bigO{\tmix+\log{n}}$ rounds \whp
\end{proofof}

\label{app:sec:congest}

The following lemma shows how to build a carrier configuration of the $m'$-supergraph $G'=(V', E', w')$ of the input graph $G=(V, E, w)$ in the \congest model and is rather technical. Thus, its proof is deferred to the \cref{app:sec:congest}. Let $ID_{new}$ be an assignment of identifiers, s.t. any node $u$ can compute $\floor{\log\deg{(v)}}$ using $ID_{new}(v)$. For an added node $v\in V'\setminus V$ supergraph, we assume that old identifier $v$ and new identifier $ID_{new}(v)$  are equal $v=ID_{new}(v)$ and greater than identifier of any original node $v'\in V$. Denote by $\rho\colon\interval{n}\mapsto\interval{n}$ a globally known simulation assignment, which satisfies $\size{\{i\colon \rho(i)=i'\}}=\bigO{\deg_G(i')/k}$, for each new identifier $i'$.

\begin{restatable}[Build Carrier Configurations in \congest]{claim}{hcCarrierConfiguration}\label{lemma:hc:waccSimulation:carrierConfiguration}
Given is a graph $G$, $k=\size{E}/\size{V}$, and an assignment of new identifiers  $ID_{new}\colon V\mapsto \interval{n}$. Let $m'$ be s.t. $0\leq m'\leq n^{2}$. Assume that for each $v\in V$, node $\rho(ID_{new}(v))$ knows the original identifier of $v$ and $ID_{new}(v)$. There is an algorithm that builds a carrier configuration $C$, which holds an $m'$-supergraph $G'$ of $G$. The communication token in $C$ for node $u$ is a concatenation of $u$, $\rho(u)$ and $ID_{new}(u)$. The algorithm runs in $\ceil{m'/m}\cdot\tmixtwosqlgnlglgn$ rounds \whp and ensures that information for the carrier node $i\in\interval{n}$ (carried edges, and communication tokens) is stored in the node $\rho(i)$, which simulates $i$.
\end{restatable}
\begin{proofof}{\cref{lemma:hc:waccSimulation:carrierConfiguration}}
    We show how to build the outgoing carrier configuration $C^{out}$ which holds the $m'$-supergraph $G'$. The incoming carrier configuration is built similarly and simultaneously.

    \textbf{Representation:}
    The \emph{added} $\ceil*{m'/n}$ nodes are \emph{represented} by the first $\ceil*{m'/n}$ nodes with lowest $ID_{new}$ identifiers. So the $i$-th added node is simulated by $\rho(i)$. 
    
    \textbf{Carrier Allocation for Added Nodes:}
    We preallocate outgoing carriers for the added $\ceil*{m'/n}\leq n$ carried nodes $V'\setminus V$. For this we compute $\size{E'}=\size{E}+m'$ using BFS, and set $\size{V'}=\size{V}+\ceil*{m'/n}$ and $k'=\size{E'}/\size{V'}$. Then, we split the outgoing edges of each added node into $\ceil{n/k'}$ batches of size at most $k'$. Notice that there are at most $\ceil*{m'/n}\cdot \ceil*{\size{V'}/k'}=\bigO{n}$ added batches, which we assign to the original nodes $V$ to carry, such that each outgoing carrier node carries a constant number of batches. This assignment is done in terms of $ID_{new}$. Now each node knows locally for each added node $v\in V'\setminus V$ the identifiers $ID_{new}$ of its outgoing carrier nodes $C_{v}^{out}$. In particular, each outgoing carrier $u$ knows which added edges it carries.

    \textbf{Carrier Allocation for Original Nodes:}
    Now, we allocate the part of the outgoing carrier configuration which stores \emph{original} nodes and edges. Each node $v$ samples $\ceil{\deg_{G'}(v)/k'}$ identifiers ($ID_{new}$) randomly independently and uniformly. Those are to become its outgoing carriers $C^{out}_v$ (\cref{itm:carrierConfiguration:carrier}). By Chernoff bounds, there is a constant $\zeta$, such that each node is an outgoing carrier for at most $\zeta \log{n}$ nodes \whp (\cref{itm:carrierConfiguration:congestion}).

    \textbf{Acquainting:}
    For each assigned (for new nodes) or sampled (for original nodes) outgoing carrier identifier $i$, which belongs to some outgoing carrier $u$, 
    carried node $v$ or its representative knows the identifier ($ID_{new}$) $\rho(i)$ of $i$'s simulating node. 
    Node $v$ (or its representative) sends the identifiers $v$, $\rho(v)$ and $ID_{new}(v)$, and the identifier $i$ of the carrier node $u$ directly to the simulating node with the new identifier $\rho(i)$. This requires for each node $v$ to send at most $\tildeBigO{\ceil{\deg_{G'}(v)/k'}}=\tildeBigO{\deg_G(v)}$ messages and to receive $\ceil{\deg_{G}(v)/k}=\tildeBigO{\deg_G(v)}$ \whp. Simulating node $\rho(i)$ responds with the identifiers $u$, $\rho(u)$. Again each node sends and receives at most $\tildeBigO{\deg_G(v)}$ messages \whp.
    Now, each carried node $v$, for each $u\in C^{out}_v$, knows the communication token of $u$, which is the concatenation of $u$, $\rho(u)$ and $ID_{new}(u)$ (\cref{itm:carrierConfiguration:vKnowsCarriers}). Also, for each carrier node $u$, $u$'s simulating node $w$ with identifier $\rho(ID_{new}(u))$ knows the communication tokens of each carried node $v$ whose edges $u$ carries (\cref{itm:carrierConfiguration:carriersKnowV}).

    Each carried node $v$ sorts its outgoing carrier nodes by identifiers. It partitions the interval $\interval{n'}$ into $\ceil{\deg_{G'}(v)/k'}$ continuous sub-intervals with at most $k'$ identifiers of opposite endpoints of outgoing edges. We assign the $j$-th sub-interval to the $j$-th carrier. For each carrier $u$, we send to its simulating node $w$ the boundaries of its interval. Each node sends $\tildeBigO{\ceil{\deg_{G'}(v)/k'}}=\tildeBigO{\deg_{G}(v)}$ and receives  $\ceil{\deg_{G}(v)/k}=\tildeBigO{\deg_G(v)}$ messages \whp (\cref{itm:carrierConfiguration:edges,itm:carrierConfiguration:vKnowsEdgeSplit})
    
    Then, carried node $v$, for each true outgoing edge $e=(v, v')$, sends to its other endpoint $v'$ the communication token of the outgoing carrier which is assigned to carry the edge $e$. Now, \cref{itm:carrierConfiguration:endpointsKnow} is satisfied for original outgoing edges but not for added outgoing edges. This requires sending $\bigO{1}$ messages over edges of $G$.
    
    Consider an added outgoing edge $e=(v, v')$, where $v$ is an added node and $v'$ is the original one. Let $u\in C^{out}_v$ be the carrier of the outgoing part of the edge $e$ and $u'\in C^{out}_{v'}$ be the carrier of the incoming part of the edge $e$. New identifier $ID_{new}(u)$ is globally known by construction given $v=ID_2(v)$. Thus, new identifier of the representative of $u$ is globally known, as well as the identifier of its simulator $w'$. Let $w$ be the simulator of the $u$. Node $w$ sends to $w'$ tuple $u', w, ID_{new}(u')$.  This requires for each simulating node to send or to receive $\tildeBigO{k'\cdot \deg_G(v)/k}$ messages. This makes \cref{itm:carrierConfiguration:endpointsKnow} satisfied for added outgoing edges as well.

    \textbf{Communication Tree:}
    Each carried node $v$ (or its representative) locally builds a Communication Tree on its outgoing carrier nodes and sends to each node which simulates a carrier node, its parent and children in the tree (\cref{itm:carrierConfiguration:broadcastAggregate,itm:carrierConfiguration:knowsTree}). Here each node sends $\bigO{1}$ messages and receives $\bigO{\deg_G(v)/k}=\bigO{\deg_G(v)}$ messages.

    \textbf{Carrier Population:}
    Each node $v$, for each of its outgoing carriers $u$, sends to node $w$ which simulates $u$, the batch of original edges assigned for $u$ to carry along with communication token of carriers of the opposite direction of these edges (\cref{itm:carrierConfiguration:edge,itm:carrierConfiguration:communicationToken}). For this, each node sends $\tildeBigO{\deg_G(v)}$ messages and receives $\tildeBigO{k'\cdot \deg_G(v)/k}$. For the added edges we send the identifiers of the first and last edges they store. To do so, each node sends $\tildeBigO{1}$ message and receives $\tildeBigO{\deg_G(v)/k}$ messages.

    \textbf{Round Complexity:}
    The carrier allocation phase is done locally, thus requires no communication. 
    
    In the acquainting phase, we use the routing algorithm from \fullref{claim:hc:routing} for the problems where each node $v$ sends and receives \tildeBigO{\deg_{G}(v)} messages and  $\tildeBigO{{k'\cdot \deg_G(v)/k}}=\tildeBigO{\ceil{m'/m}\deg_G(v)}$ messages. Thus, it requires $\ceil{m'/m}\cdot \tmixtwosqlgnlglgn$ rounds \whp.
    
    Communication tree building requires only $1$ invocation of the \fullref{claim:hc:routing}, thus terminates in $\ceil{m'/m}\cdot \tmixtwosqlgnlglgn$ rounds \whp.
    
    For the Carrier population phase, each node $v$ sends $\tildeBigO{\deg_G(v)}$ and receives $\tildeBigO{{k'\cdot \deg_G(v)/k}}=\tildeBigO{\ceil{m'/m}\deg_G(v)}$ messages, thus runs in $\ceil{m'/m}\cdot \tmixtwosqlgnlglgn$ rounds \whp.
    
    The overall complexity is $\ceil{m'/m}\cdot  \tmixtwosqlgnlglgn$ rounds \whp.
\end{proofof}

In the proof of \fullref{lemma:hc:waccSimulation} we use the following technical claim.
\begin{restatable}[Assignment]{claim}{hcAssignment}\label{lemma:hc:assignment}
Let $n_1, n_2, m, x_1, \dots, x_{n_1}, y_1, \dots, y_{n_2}$ be integers such that $m<\sum_{i=1}^{n_1}{2^{x_i}}+\sum_{i=1}^{n_2}{2^{y_i}}$, for each $i\in\interval{n_1}:0\leq x_i < \floor{\log{k}} - 2$ and $j\in\interval{n_2}:\floor{\log{k}} - 2\leq y_j$, where $n=n_1+n_2$, $k=m/n$. There is a partition of $\interval{n_1}$ to $n_2$ sets $I_1, \dots, I_{n_2}$, such that for each $j\in\interval{n_2}:\size{I_j}\leq 4\cdot \floor{2^{y_j}/k}$.
\end{restatable}
\begin{proofof}{\cref{lemma:hc:assignment}}
    We construct sets greedily, by adding new elements to the set $I_j$ as long as its size is less than $\floor{4\cdot 2^{y_j}/k}$. We notice that the total capacity of sets
    
    \begin{align*}
        \sum_{j=1}^{n_2}{4\cdot \floor*{\frac{2^{y_j}}{k}}}
        \geq 4\cdot \sum_{j=1}^{n_2}{\frac{2^{y_j}}{k}} - n\\
        >4\cdot \frac{m - \sum_{i=1}^{n_1}{2^{x_i}}}{k} - n
        >4\cdot \frac{m - \sum_{i=1}^{n_1}{2^{\floor{\log{k}}-2}}}{k} - n\\
        \geq 4\cdot \frac{m - n\cdot{2^{\log{k}-2}}}{k} - n
        =4\frac{m - n\frac{m}{4n}}{\frac{m}{n}} - n\\
        =2\cdot n
        \geq n, 
    \end{align*}
    is enough to hold all elements.

\end{proofof}

\label{app:hc:knearest}
\hcKnearest*
\begin{proofof}{\cref{claim:hc:knearest}}
    First, we compute $m$ and $\frac{m}{n}$ using a BFS algorithm in $\bigO{D}=\bigO{\tmix}$ rounds.
    We simulate the algorithm from \fullref{wacc:distTools:knearest} to compute the  $\max\set{k, n^{1/3}}$-nearest (notice that \cref{wacc:distTools:knearest} works for $k\geq n^{1/3}$) problem in a carrier configuration, and then we simulate the algorithm from \fullref{wacc:tools:learn} to learn the edges stored in the output carrier configuration $C^{out}$ by nodes in the \wacc[m/n] model. The simulation in the \congest model is done by the algorithm from \fullref{lemma:hc:waccSimulation}. Notice that the out-degree in the resulting graph is $\max\set{k, n^{1/3}}$, and we truncate the output of each node to $k\log{n}$ bits before the end of the simulation. In the \wacc model, solving \knearest requires $\tilde O(\max\set{k, n^{1/3}} \cdot n^{1/3}/c + n^{2/3}/c + 1)$, thus in the \congest model the simulation round complexity is $((\max\set{k, n^{1/3}} \cdot n^{1/3}/(m/n) + n^{2/3}/(m/n) + 1 + k / c)\cdot(m/n)/(m/n) + k)\cdot\tmixtwosqlgnlglgn={(k \cdot n^{4/3}/m + n^{5/3}/m + 1)}\cdot\tmixtwosqlgnlglgn$ \whp. %

\end{proofof}

\end{document}